\documentclass[aps,pra,twocolumn,floatfix,groupedaddress,footinbib,notitlepage]{revtex4-2}
\usepackage{times,bm,bbm,amssymb,amsthm,amsmath,amsfonts,hyperref,graphicx,graphics,color,xcolor}
\usepackage[T1]{fontenc} 
\usepackage{newtxtext,newtxmath} 

\hypersetup{colorlinks,linkcolor=blue,citecolor=blue,urlcolor=blue}
\newcommand{\ignore}[1]{}
\newtheorem{remark}{Remark}

\DeclareFontFamily{OT1}{pzc}{}
\DeclareFontShape{OT1}{pzc}{m}{it}
              {<-> s * [1.25] pzcmi7t}{}
\DeclareMathAlphabet{\mathpzc}{OT1}{pzc}
                                 {m}{it}

\begin{document}

\title{Enhancing energy transfer in quantum systems via periodic driving: Floquet master equations}

\author{Z. Nafari Qaleh}
\affiliation{Department of Physics, Sharif University of Technology, Tehran 14588, Iran}

\author{A. T. Rezakhani}
\affiliation{Department of Physics, Sharif University of Technology, Tehran 14588, Iran}

\begin{abstract}
\noindent We provide a comprehensive study of the energy transfer phenomenon---populating a given energy level---in $3$- and $4$-level quantum systems coupled to two thermal baths. In particular, we examine the effects of an external periodic driving and the coherence induced by the baths on the efficiency of the energy transfer. We consider the Floquet-Lindblad and the Floquet-Redfield scenarios, which both are in the Born-Markov, weak-coupling regime but differ in the treatment of the secular approximation, and for the latter we develop an appropriate Floquet-type master equation by employing a partial secular approximation. Throughout the whole analysis we keep Lamb-shift corrections in the master equations. We observe that, especially in the Floquet-Redfield scenario, the driving field can enhance the energy transfer efficiency compared to the nondriven scenario. In addition, unlike degenerate systems where Lamb-shift corrections do not contribute significantly on the energy transfer, in the Redfield and the Floquet-Redfield scenarios these corrections have nonnegligible effects.
\end{abstract}

\date{\today}
\maketitle

\section{Introduction} 
\label{sec:intro}

The process of energy transfer is a fundamental subject in physics, chemistry, and biology, which involves energy exchange among parts of physical, chemical, and biological systems. In particular, it is interesting to study such processes in quantum regimes \cite{mohseni_quantum_2014, demtroder_molecular_2005}. There exist a vast literature of pertinent studies in various systems such as crystals with impurities \cite{dexter_theory_1953}, quantum-dot nanostructures \cite{gerardot_photon_2005}, polymer chains \cite{collini_coherent_2009, herz_time-dependent_2004}, and light-harvesting complexes in photosynthesis \cite{cheng_dynamics_2009, plenio_dephasing-assisted_2008, caruso_highly_2009,wu_generic_2013}. Increasing efficiency (or power) of energy transfer processes has been one of the main objectives in these investigations \cite{svidzinsky_enhancing_2011, dorfman_increasing_2011}. Energy transfer usually occurs in two forms: (i) in \textit{real space}, where energy in the form of excitation (or exciton) is to move from one particular ``site'' of a lattice to another target site \cite{caruso_entanglement_2010, campos_venuti_excitation_2011, sarovar_environmental_2011, chen_rerouting_2013}; and (ii) in \textit{energy space}, where energy is to move from an eigenstate of the system Hamiltonian to populate another target eigenstate or energy level \cite{irish_vibration-assisted_2014, roden_probability-current_2016}. The latter event in an open quantum system in the presence of baths is of our interest in this paper.  

In real systems, interaction between the system and its environment (or bath) is inevitable. This results in dissipative dynamics of the system in the sense that the system evolution is given by nonunitary master equations, usually after applying relevant approximations such as the Born-Markov, weak-coupling, and secular approximations \cite{zwanzig_ensemble_1960,10.1143/PTP.20.948, breuer_theory_2002, corr-pic}. Under the first two approximations, the Redfield master equation is obtained; the three approximations yield the Lindblad master equation \cite{breuer_theory_2002}. The Redfield equation may not necessarily guarantee the complete-positivity of the system density matrix (nonpositivity issue), and evolution of populations (diagonal terms of the density matrix) through this equation may involve dependence on coherence (offdiagonal) terms. Applying the secular approximation (if holds) retrieves the complete-positivity in the resulting Lindblad equation. If the Hamiltonian of the system is nondegenerate, the Lindblad equation yields an evolution for populations fully independently from coherence terms \cite{lindblad1975, lindblad1976, doi:10.1063/1.522979}. Although there are some arguments against the use of the Redfield equation \cite{doi:10.1063/1.3142485}, several recent studies have shown the utility of this equation in some applications \cite{Plenio-etal, PhysRevA.100.012107}. In fact, the main nonpositivity issue with the Redfield equation may be partially alleviated without need to apply the full secular approximation, e.g., by applying a suitable ``partial'' secular approximation \cite{tscherbul_partial_2015, PhysRevA.100.012107, Plenio-etal}. This way the Redfield equation does not generate unphysical states yet it maintains its appeal to couple populations and coherences and gives physically relevant results.

\begin{figure*}[tp]
\includegraphics[width=.9\linewidth]{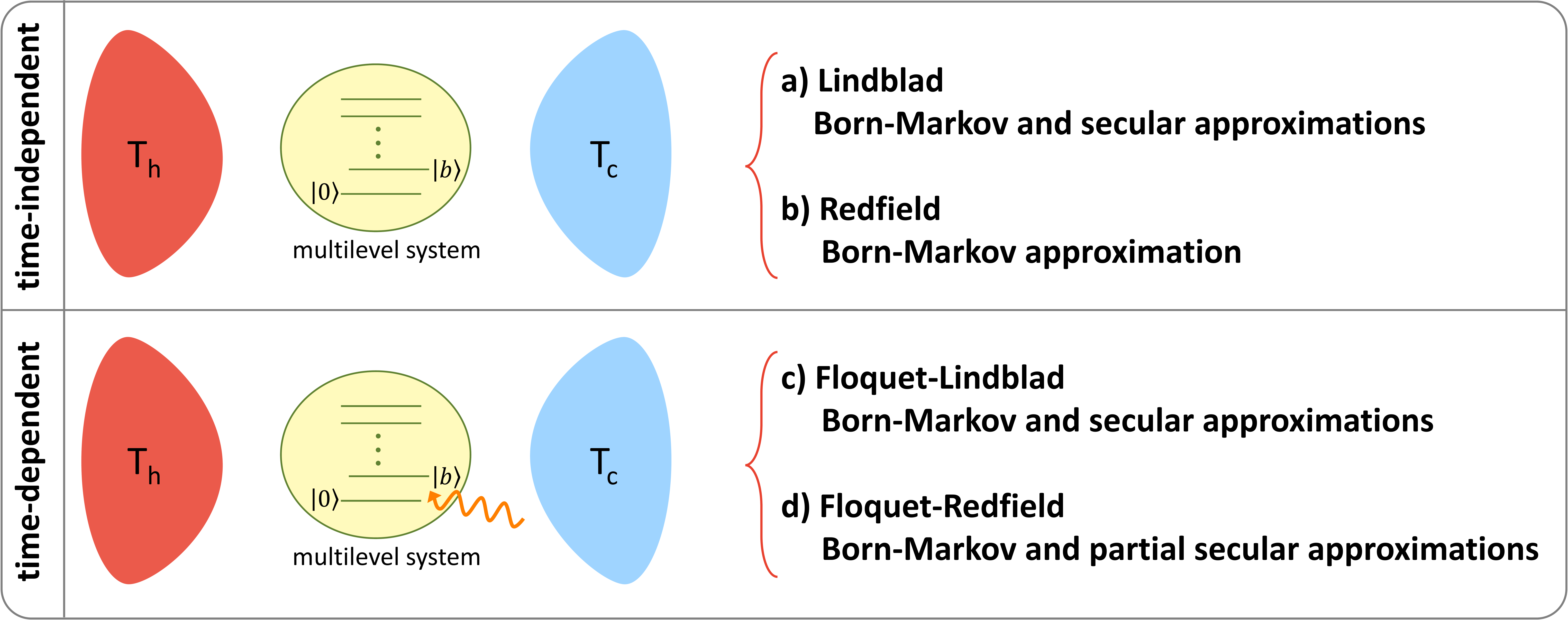}
\caption{Four scenarios of the energy transfer. (a) Lindblad scenario: time-independent system by applying the Born-Markov and secular approximations. Here the standard Lindblad equation [Eq. \eqref{e113}] applies. (b) Redfield scenario: time-independent system and performing only the Born-Markov approximation. Here the Redfield equation [Eq. \eqref{e106}] applies. (c) Floquet-Lindblad scenario: time-dependent system with the Born-Markov and secular approximations. Here the Floquet-Lindblad equation [Eq. \eqref{e29}] applies. (d) Floquet-Redfield scenario: time-dependent system with the Born-Markov approximation and a partial secular approximation. Here the Floquet-Redfield equation [Eq. \eqref{e98}] applies. In time-dependent cases the system is driven by a time-periodic applied field.}
\label{4scenarios}
\end{figure*}

In this paper, we consider a model which includes a multilevel system $\mathsf{S}$ in contact with two thermal baths at two different temperatures. The hot bath (e.g., sunlight) is set to excite the system to its highest energy level(s) and then the exciton moves through the energy space in the presence of another cold bath. We work in the local master equation approximation where the couplings are so weak that the baths can be assumed to act only locally on the part of the system they directly interact with \cite{hofer_markovian_2017, DeChiara-njp}, and thus effectively do not see each other. We consider four scenarios--as in Fig. \ref{4scenarios}---and their impacts on the energy transfer efficiency: (i) Lindblad scenario: assuming the Born-Markov and secular approximations, with no applied driving field on the system, (ii) Floquet-Lindblad scenario: the Born-Markov and secular approximations and driving the system by a time-periodic applied field, (iii) Redfield scenario: the Born-Markov approximation, no secular approximation, and no applied field on the system, and (iv) Floquet-Redfield scenario: the Born-Markov approximation, a partial secular approximation, and driving the system by a time-periodic applied field. 

Each of these scenarios require its own dynamical equation for the system. In particular, in the time-dependent driven cases since the applied field is periodic, one needs to use Floquet theory. Scenarios (i), (ii), and (iii) can be described by, respectively, Lindblad equation, Floquet-Lindblad equation, and Redfield equation \cite{alicki_internal_2006,szczygielski_application_2014}. However, for scenario (iv) we develop an appropriate Floquet-type Redfield equation where we employ a particular partial secular approximation (which is different from that of Ref. \cite{tscherbul_partial_2015}). 

Our objective is to populate a particular energy level of the system as the target state (denoted by $|b\rangle$). Performance of this task can be quantified by the time averaging of the population of the target state \cite{rebentrost_non-markovian_2009, qin_dynamics_2014, rebentrost_environment-assisted_2009, rebentrost_role_2009},
\begin{align}
\eta (t_{\mathrm{f}})=(1/t_{\mathrm{f}})\textstyle{\int_{0}^{t_{\mathrm{f}}}}  ds\, \langle b\vert \varrho_{\mathsf{S}} (s)\vert b\rangle, 
\label{e103}
\end{align}
where $t_{\mathrm{f}}$ is the final time and $\varrho_{\mathsf{S}}(s)$ denotes the state of the system at time $0\leqslant s \leqslant t_{\mathrm{f}}$. Our results show that driving the system by a periodic external field leads to increasing the population of the the target level relative to the case where no external field is applied. In the Redfield and Floquet-Redfield master equations we include the Lamb-shift terms, which are computed by appropriate QED considerations. According to our analysis, it is seen that at least for degenerate systems the Lamb-shift terms do not have a significant impact on the dynamics. In addition, we observe that in the presence of a periodic external field, generating coherence in the system through the dynamics can enhance the energy transfer efficiency (in the Floquet-Lindblad and Floquet-Redfield scenarios).

The structure of this paper is as follows. In Sec. \ref{sec:L-FL} we investigate the Lindblad and Floquet-Lindblad scenarios and their associated dynamical equations. In addition, we consider the case of a $3$-level quantum system and calculate the energy transfer efficiency for each of the scenarios. In Sec. \ref{sec:R-FR} we move to the Redfield and Floquet-Redfield scenarios. In particular, we delineate how one can derive an appropriate dynamical equation which holds for a driven quantum system under a time-periodic external field under the Born-Markov condition. To do so, we introduce a specific partial secular approximation. Next, we calculate the energy transfer efficiency for these scenarios for a $4$-level system, which also enables us to see how coherence may also affect the energy transfer efficiency. We conclude the paper with a summary in Sec. \ref{sec:summary}. 
Several appendixes are also included to elaborate further details and steps of derivations.

\section{Lindblad and Floquet-Lindblad scenarios}
\label{sec:L-FL}

Here we sketch derivation of the quantum master equation for the state of an open quantum quantum system which is driven by a periodic external field. Before presenting the Floquet-Lindblad equation we set up the problem and briefly review the standard Lindblad master equation.

\subsection{Lindblad scenario}
\label{sec:Lscenario}

\begin{figure*}[ht!]
\includegraphics[scale=0.13]{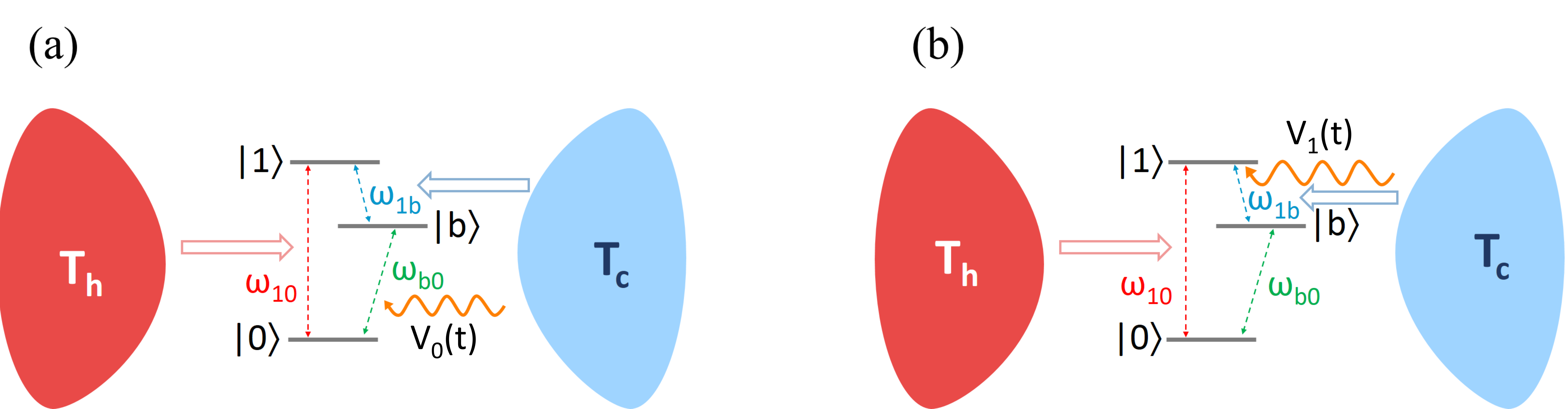}
\caption{A $3$-level system ($\{|0\rangle,|1\rangle,|b\rangle\}$) coupled to two baths and driven by an external periodic field $V(t)$. In (a) the external field couples $\vert 0\rangle$ and $\vert b\rangle$, while in (b) it couples $\vert 1\rangle$ and $\vert b\rangle$.}
\label{fig1}
\end{figure*}

Consider a nondriven, closed bipartite quantum system with the total Hamiltonian  
\begin{align} 
H_{\mathsf{tot}}=H_{\mathsf{S}}+H_{\mathsf{B}}+H_{\mathsf{int}},
\label{e1}
\end{align}
where $H_{\mathsf{S}} =\sum_{\epsilon} \epsilon |\epsilon \rangle \langle \epsilon|$, $H_{\mathsf{B}}$, and $H_{\mathsf{int}}=\sum_{\alpha} S_{\alpha}\otimes B_{\alpha}$ are, respectively, the system, bath, and interaction Hamiltonians. It is more convenient to derive the master equation in the interaction picture which is defined by
\begin{gather}
\bm{\varrho}_{\mathsf{tot}}(t)=  U^{\dagger}_{\mathsf{S}}(t,0)\otimes U^{\dagger}_{\mathsf{B}}(t,0) \varrho_{\mathsf{tot}}(t) U_{\mathsf{S}}(t,0)\otimes U_{\mathsf{B}}(t,0),\\
\ignore{\bm{\varrho}_{\mathsf{S}}(t)= U^{\dagger}_{\mathsf{S}}(t,0) \varrho_{\mathsf{S}}(t) U_{\mathsf{S}}(t,0),\\
\bm{\varrho}_{\mathsf{B}}(t)= U^{\dag}_{\mathsf{B}}(t,0)\varrho_{\mathsf{S}}(t) U_{\mathsf{B}}(t,0),\\}
\bm{H}_{\mathsf{int}}(t)=  U^{\dagger}_{\mathsf{S}}(t,0)\otimes U^{\dagger}_{\mathsf{B}}(t,0) H_{\mathsf{int}} U_{\mathsf{S}}(t,0)\otimes U_{\mathsf{B}}(t,0),
\label{e3}
\end{gather}
where $U_{\mathsf{S}}(t,0)=e^{-itH_{\mathsf{S}}/\hbar}$ and $U_{\mathsf{B}}(t,0)=e^{-itH_{\mathsf{B}}/\hbar}$. We use boldface letters for the interaction picture throughout this paper. The dynamics of the total system in this picture is given by
\begin{align}
\dfrac{d\bm{\varrho}_{\mathsf{tot}}(t)}{dt}=-\dfrac{i}{\hbar}\left[ \bm{H}_{\mathsf{int}}(t) , \bm{\varrho}_{\mathsf{tot}}(t) \right].
\label{e2-}
\end{align}
It has been argued that under the Born-Markov and secular approximations the dynamics of the open quantum system is given by the Lindblad quantum master equation \cite{breuer_theory_2002},
\begin{align}
\dfrac{d\bm{\varrho}_{\mathsf{S}} (t)}{dt}=-\dfrac{i}{\hbar}\left[ H_{\mathrm{lamb}},\bm{\varrho}_{\mathsf{S}}(t)\right] +\mathpzc{D}\left[ \bm{\varrho}_{\mathsf{S}}(t)\right] ,
\label{e113}
\end{align}
where 
\begin{align}
&H_{\mathrm{lamb}}= (1/\hbar) \textstyle{\sum_{\alpha ,\alpha'}} \sum_{{\omega}} \xi_{\alpha \alpha'}(\omega) S_{\alpha}^{\dagger}(\omega )S_{\alpha'}(\omega),
\label{e114}\\
&\mathpzc{D}[\bm{\varrho}_{\mathsf{S}}(t)]=(1/\hbar^{2}) \textstyle{\sum_{\alpha ,\alpha'}} \sum_{{\omega}} \gamma_{\alpha \alpha'}(\omega) \nonumber\\
&\, \times \big[ S_{\alpha'}(\omega)\bm{\varrho}_{\mathsf{S}}(t)S^{\dagger}_{\alpha}(\omega)-\dfrac{1}{2}\left\lbrace S_{\alpha}^{\dagger}(\omega)S_{\alpha'}(\omega),\bm{\varrho}_{\mathsf{S}}(t)\right\rbrace \big],
\label{e115}
\end{align}
are the Lamb-shift Hamiltonian and the dissipator, respectively. Here the $\gamma_{\alpha \alpha'}(\omega)$ and $\xi_{\alpha \alpha'}(\omega)$ coefficients are related to the bath correlation functions (Appendix \ref{app:floquet-lindblad}), and the Lindblad operators $S_{\alpha}(\omega)$ are defined based on the system Hamiltonian as 
\begin{align}
S_{\alpha}(\omega)= \textstyle{\sum_{\epsilon-\epsilon' = \hbar\omega}} \vert \epsilon \rangle \langle \epsilon \vert S_{\alpha} \vert \epsilon ^{'} \rangle \langle \epsilon ^{'} \vert.
\label{Lin-op}
\end{align}

\subsection{Floquet-Lindblad scenario}
\label{sec:FLscenario}

Now we study a driven open quantum system whose Hamiltonian is periodic in time, $H_{\mathsf{S}}(t+\tau)=H_{\mathsf{S}}(t)$ with period $\tau$. It is known from the Floquet theorem \cite{bukov_universal_2015,book:Chicone} that the unitary operator $U_{\mathsf{S}}(t,0)$ satisfying the equation $i\hbar\frac{d}{dt}U_{\mathsf{S}}(t,0)=H_{\mathsf{S}}(t)U_{\mathsf{S}}(t,0)$ can also be represented in the form
\begin{align}
U_{\mathsf{S}}(t,0)=P(t,0) e^{-i\bar{H}t/\hbar},
\label{e11}
\end{align}
where $\bar{H}$ is a Hermitian time-independent operator, referred to as the Floquet Hamiltonian, and $P(t,0)$ is a time-periodic unitary operator, referred to shortly as the periodic operator, such that $P(t+n\tau,n\tau)=P (t+n\tau,0)=P(t,0)$ and $ P (n\tau,0)=\mathbbmss{I}$, $\forall n\in \mathbbmss{N}$.

Starting from the dynamical equation (\ref{e2-}) and employing the Floquet theorem, it has been argued that under appropriate Born-Markov and secular approximations the dynamical equation of an open system $\mathsf{S}$ is given by \cite{breuer_theory_2002,szczygielski_application_2014}
\begin{align}
\dfrac{d\bm{\varrho}_{\mathsf{S}} (t)}{dt}=-\dfrac{i}{\hbar}\big[ H_{\mathrm{lamb}}^{(\mathrm{F})},\bm{\varrho}_{\mathsf{S}}(t)\big] +\mathpzc{D}^{(\mathrm{F})}\left[ \bm{\varrho}_{\mathsf{S}}(t)\right] ,
\label{e29}
\end{align}
where
\begin{align}
& H_{\mathrm{lamb}}^{(\mathrm{F})}= \textstyle{\sum_{\alpha,\alpha'}} \sum_{{\omega}}\sum_{q\in\mathbbmss{Z}} \dfrac{1}{\hbar}\xi_{\alpha \alpha'}(\omega +q\Omega) S_{\alpha}^{\dagger}(q,\omega )S_{\alpha'}(q,\omega) ,
\label{e30-}\\
&\mathpzc{D}^{(\mathrm{F})}[\bm{\varrho}_{\mathsf{S}}(t)]=\dfrac{1}{\hbar^{2}} \textstyle{\sum_{\alpha ,\alpha'}} \sum_{{\omega}}\sum_{q\in\mathbbmss{Z}} \gamma_{\alpha\alpha'}(\omega +q\Omega) \nonumber\\
&\times \big[ S_{\alpha'}(q,\omega)\bm{\varrho}_{\mathsf{S}}(t) S^{\dagger}_{\alpha}(q,\omega)-\dfrac{1}{2}\left\lbrace S_{\alpha}^{\dagger}(q,\omega)S_{\alpha'}(q,\omega),\bm{\varrho}_{\mathsf{S}}(t)\right\rbrace \big] ,
\label{e31-}
\end{align}
are the Lamb-shift Hamiltonian and the dissipator, respectively. Here the Floquet-Lindblad operators $S_{\alpha}(q,\omega)$ are defined as follows. First, note that we can Fourier transform the periodic factor $P ^{\dag}(t,0)S_{\alpha}P(t,0)$ as $P ^{\dag}(t,0)S_{\alpha}P(t,0)= \textstyle{\sum_{q\in \mathbbmss{Z}}} S_{\alpha}(q) e^{iq\Omega t}$, with $\Omega =2\pi/\tau$ being the frequency of the applied field. Next, we define 
\begin{equation}
S_{\alpha}(q,\omega)= \textstyle{\sum_{\bar{\epsilon}-\bar{\epsilon}'=\hbar\omega}} \vert\bar{\epsilon}\rangle\langle\bar{\epsilon}\vert S_{\alpha}(q)|\bar{\epsilon}'\rangle\langle\bar{\epsilon}'\vert,
\end{equation}
where $\bar{H}|\bar{\epsilon}\rangle= \bar{\epsilon}|\bar{\epsilon}\rangle$. In addition, the coefficients $\gamma_{\alpha\alpha'}(\omega +q\Omega)$ and $\xi_{\alpha\alpha'}(\omega +q\Omega)$ are the real and imaginary parts of the one-sided Fourier transformation of the bath correlation functions. For details, see Appendix \ref{app:floquet-lindblad}. An immediate observation is that, unlike the Lindblad scenario, in this case the Lamb-shift Hamiltonian (\ref{e30-}) does not necessarily commute with the system Hamiltonian.

\subsection{Case study: $3$-level system}
\label{sec:F-FL-results}

\begin{figure*}[tp]
\includegraphics[scale=0.64]{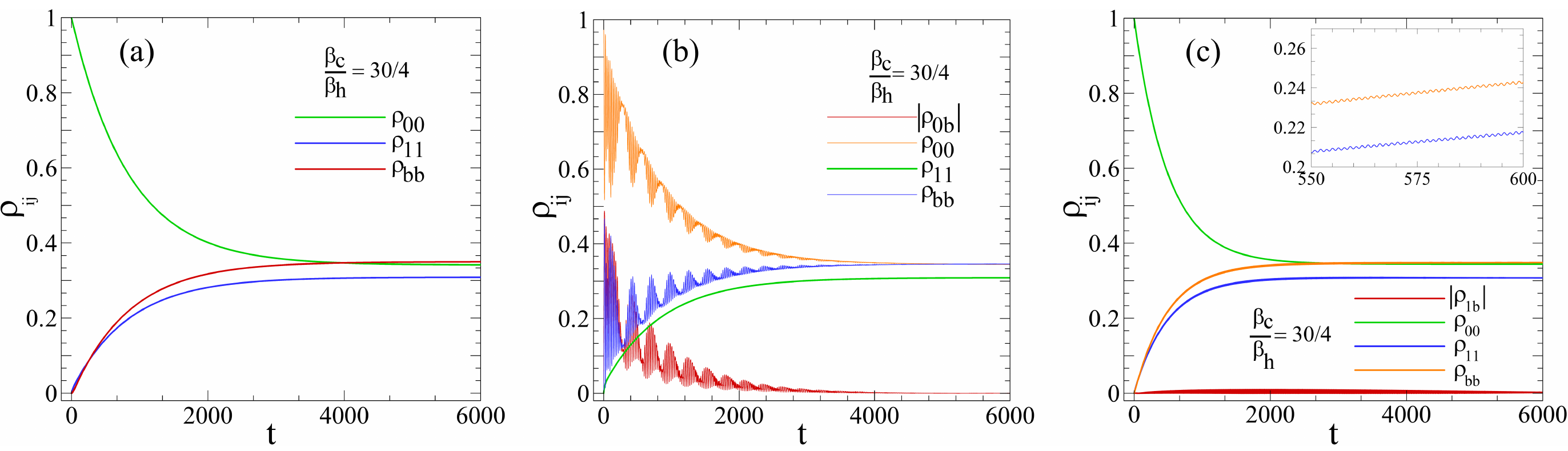}
\caption{Time evolution of the reduced density matrix elements of a $3$-level system (see Sec. \ref{sec:L-FL}). The system is coupled to two baths with $\beta_{\mathsf{c}}/\beta_{\mathsf{h}}=30/4$. (a) Applying no external driving field on the system, (b) Applying an external driving field $V_{0}(t)$. Note that here $\Omega/\omega_{b0}=0.9$. (c) Applying an external driving field $V_{1}(t)$. Note also that $\Omega/\omega_{1b}=4.5$. The properties of the system, baths, and the external fields are listed in Table \ref{tab-2} and the paragraph after Eq. \eqref{h_0}.}
\label{fig2}
\end{figure*}

Consider a driven $3$-level system which is weakly coupled to two bosonic baths $\mathsf{B}_{\mathsf{h}}$ and $\mathsf{B}_{\mathsf{c}}$ with temperatures $T_{\mathsf{h}}$ and $T_{\mathsf{c}}$ ($T_{\mathsf{h}}\geqslant T_{\mathsf{c}}$); see Fig. \ref{fig1}. The baths are assumed sufficiently large and initially in their associated equilibrium states $\varrho_{\mathsf{B}}=e^{-\beta_{\mathsf{B}} H_{\mathsf{B}}} /Z_{\mathsf{B}}$, where $\beta_{\mathsf{B}}=1/(k_{B}T_{\mathsf{B}})$ and $Z_{\mathsf{B}} = \mathrm{Tr}[e^{-\beta_{\mathsf{B}} H_{\mathsf{B}}}]$, with $\mathsf{B}\in\{\mathsf{B}_{\mathsf{c}}, \mathsf{B}_{\mathsf{h}}\}$. One can simulate the baths with electromagnetic radiation fields which interact with a system through dipole interaction. A careful analysis of this interaction is reviewed in Appendix \ref{sec:lm}. As a result, the total Hamiltonian of a driven $3$-level system interacting with a radiation field can be written as
\begin{align}
H_{\mathsf{tot}}(t)=H_{\mathsf{S}}(t)+H_{\mathsf{B}_{\mathsf{h}}}+H_{\mathsf{B}_{\mathsf{c}}}+H_{\mathsf{S}\mathsf{B}_{\mathsf{h}}}+H_{\mathsf{S}\mathsf{B}_{\mathsf{c}}},
\label{e32}
\end{align}
where
\begin{gather}
H_{\mathsf{S}}(t)=H_{0}+V_{i}(t),
\label{e38}\\
H_{0}=\epsilon_{0} \vert \epsilon_{0}\rangle\langle \epsilon_{0}\vert +\epsilon_{b} \vert \epsilon_{b}\rangle\langle \epsilon_{b}\vert +\epsilon_{1} \vert \epsilon_{1}\rangle\langle \epsilon_{1}\vert, 
\label{e33}\\
V_{i}(t)=\mu \cos(\Omega t) \big( \vert \epsilon_{b}\rangle\langle \epsilon_{i}\vert +\vert \epsilon_{i}\rangle\langle \epsilon_{b}\vert\big), \qquad i \in \{0,1\}
\label{e39}\\
H_{\mathsf{B}_{\mathsf{h}}}= \textstyle{\sum_{k}}  \hbar \zeta_{k} \big(\hat{a}_{k}^{\dagger} \hat{a}_{k}+1/2\big),
\label{e34}\\
H_{\mathsf{B}_{\mathsf{c}}}= \textstyle{\sum_{q}}  \hbar \nu_{q} \big(\hat{b}_{q}^{\dagger} \hat{b}_{q}+1/2\big),
\label{e35}\\
H_{\mathsf{S}\mathsf{B}_{\mathsf{h}}}= \textstyle{\sum_{k}}  f_{k}\big(\vert \epsilon_{0}\rangle\langle \epsilon_{1}\vert \otimes \hat{a}^{\dagger}_{k}+\vert \epsilon_{1}\rangle\langle \epsilon_{0}\vert \otimes \hat{a}_{k}\big),
\label{e36}\\
H_{\mathsf{S}\mathsf{B}_{\mathsf{c}}}= \textstyle{\sum_{q}}  g_{q}\big(\vert \epsilon_{b}\rangle\langle \epsilon_{1}\vert \otimes \hat{b}^{\dagger}_{q}+\vert \epsilon_{1}\rangle\langle \epsilon_{b}\vert \otimes \hat{b}_{q}\big) .
\label{e37}
\end{gather}
Here $H_{0}$ denotes the free Hamiltonian of the nondriven system, with energy levels $\epsilon_{0}< \epsilon_{b}<\epsilon_{1}$. For brevity, hereafter and throughout the paper we use the shorthand $|i\rangle$ rather than $|\epsilon_{i}\rangle$. (For an $n$-level system, the free Hamiltonian is extended to 
\begin{equation}
H_{0}=\textstyle{\sum_{i}} \epsilon_{i} \vert i\rangle \langle i\vert,\,\,i\in\{0,b,1,2,\ldots,n-2\},
\label{h_0}
\end{equation}
with $\epsilon_{0}<\epsilon_{b}< \epsilon_{1}<\ldots <\epsilon_{n-2}$.) The external field $V_{0}(t)$ ($V_{1}(t)$) couples the ground level $\vert 0\rangle$ and the target level $\vert b\rangle$ (level $\vert 1\rangle$ to the target level $\vert b\rangle$). The operators $\hat{a}^{\dagger}_{k}$ and $\hat{b}^{\dagger}_{k}$ are the creation operators of the $k$th mode of the hot and the cold baths, respectively, which satisfy the bosonic commutation relations $[\hat{a}_{l},\hat{a}_{k}^{\dagger}]=[\hat{b}_{l},\hat{b}_{k}^{\dagger}]=\delta_{lk}$. Moreover, the interaction Hamiltonian between the system and the hot and the cold baths are given, respectively, by Eqs. \eqref{e36} and \eqref{e37}, where we assume $f_{k}$s and $g_{k}$s to be real-valued parameters. 

\begin{table}[bp]
\caption{Temperatures and constants of the spectral density functions of the cold and hot baths [denoted by $(\mathsf{c})$ and $(\mathsf{h})$], in the form $J(\omega)=J_{0}\omega e^{-\omega^{2}/\omega^{2}_{\mathrm{cutoff}}}$, in natural units.}
\label{tab-2}
\begingroup
\squeezetable
\begin{ruledtabular}
\begin{tabular}{lccccc}
$1/\beta_{\mathsf{h}}$ & $1/\beta_{\mathsf{c}}$ &  $J_{0}^{(\mathsf{h})}$ & $J_{0}^{(\mathsf{c})}$ & $\omega_{\mathrm{cutoff}}^{(\mathsf{h})}$ & $\omega_{\mathrm{cutoff}}^{(\mathsf{c})}$ \\
\hline
$30$ & $4$ & $4 \times 10^{-4}$ & $4 \times 10^{-3}$ & $\sqrt{2} $ & $\sqrt{0.2}$ 
\end{tabular}
\end{ruledtabular}
\endgroup
\end{table}

In the Born-Markov and secular approximations, the evolution of system $\mathsf{S}$ is governed by Eq. \eqref{e29}. Specifically, here we consider $H_{0}$ with $(\epsilon_{0},\epsilon_{b},\epsilon_{1})= (0,2.5,3)$ and $\{|0\rangle=(1,0,0)^{T}, |1\rangle=(0,1,0)^{T}, |b\rangle=(0,0,1)^{T}\}$. The system is assumed to be initially in the state $\varrho_{\mathsf{S}} (0)=\vert 0\rangle_{\mathsf{S}}\langle 0\vert$, and is coupled to two thermal baths with $\beta_{\mathsf{c}} / \beta_{\mathsf{h}}=30/4$. We also assume $(\mu,\Omega)=(0.1,2.25$). In all simulations throughout the paper, we consider natural units where we set $\hbar\equiv c\equiv k_{\mathrm{B}}\equiv 1$, with $c$ being the speed of light and $k_{\mathrm{B}}$ the Boltzmann constant. 
The properties of the baths are listed in Table \ref{tab-2}. We consider three cases: (i) there is no driving field; (ii) there is an external field $V_{0}(t)$ which couples the $\vert 0\rangle$ and $\vert b\rangle$ levels; and (iii) the external field $V_{1}(t)$ couples the $\vert 1\rangle$ and $\vert b\rangle$ levels. (Table \ref{tab-3} in Appendix \ref{sec:analytic} shows the parameters of the Floquet-Lindblad master equation for this example.) We also study the effect of the Lamb-shift in the Lindblad and Floquet-Lindblad scenarios for these cases.
\begin{figure}[bp]
\includegraphics[width=.9\linewidth]{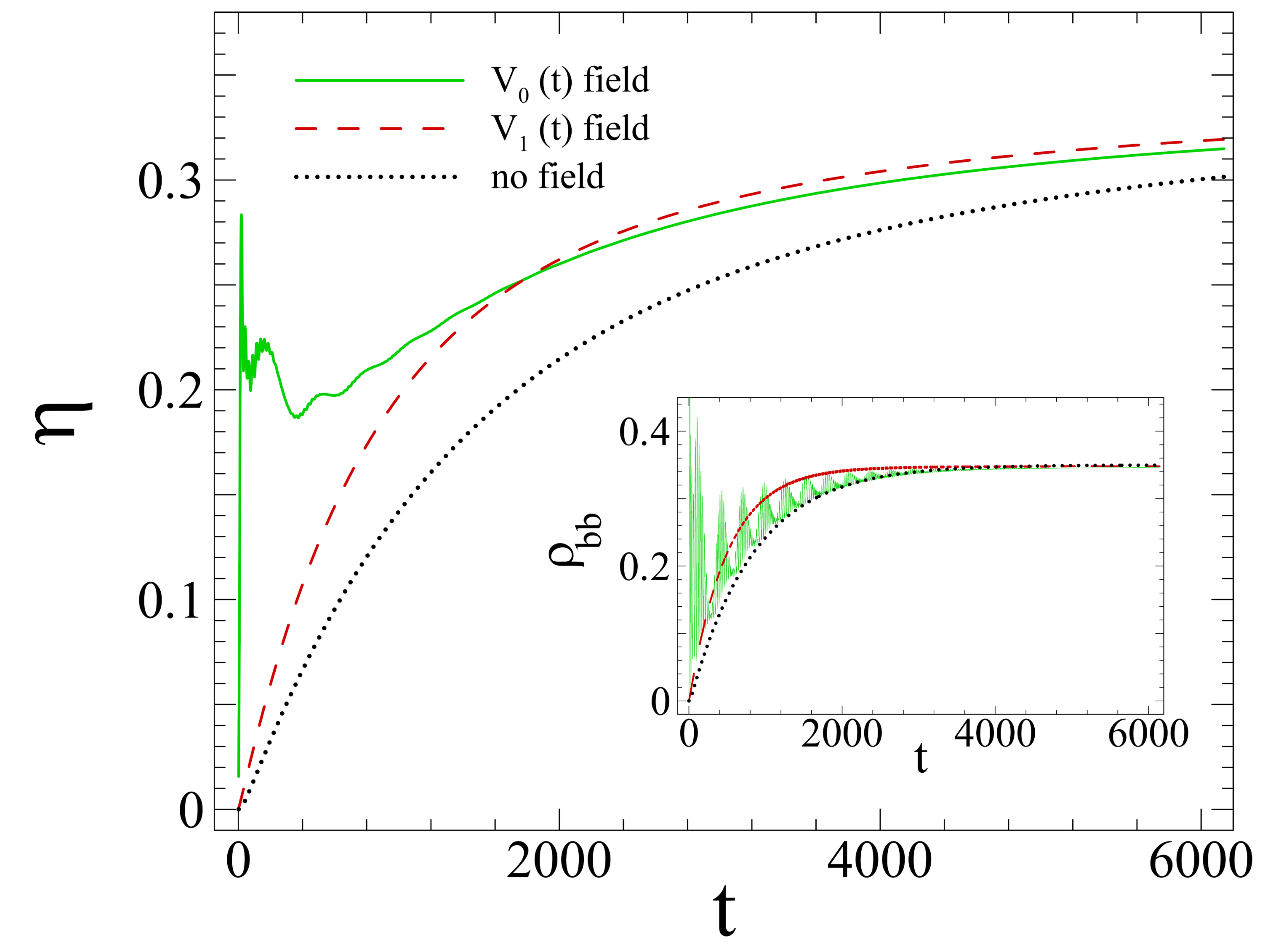}
\caption{The comparison of energy transfer for the three cases of Fig. \ref{fig2}, when $\beta_{\mathsf{c}} / \beta_{\mathsf{h}}=30/4$; see Sec. \ref{sec:L-FL}.}
\label{fig4}
\end{figure}

\begin{figure*}[tp]
\includegraphics[scale=0.125]{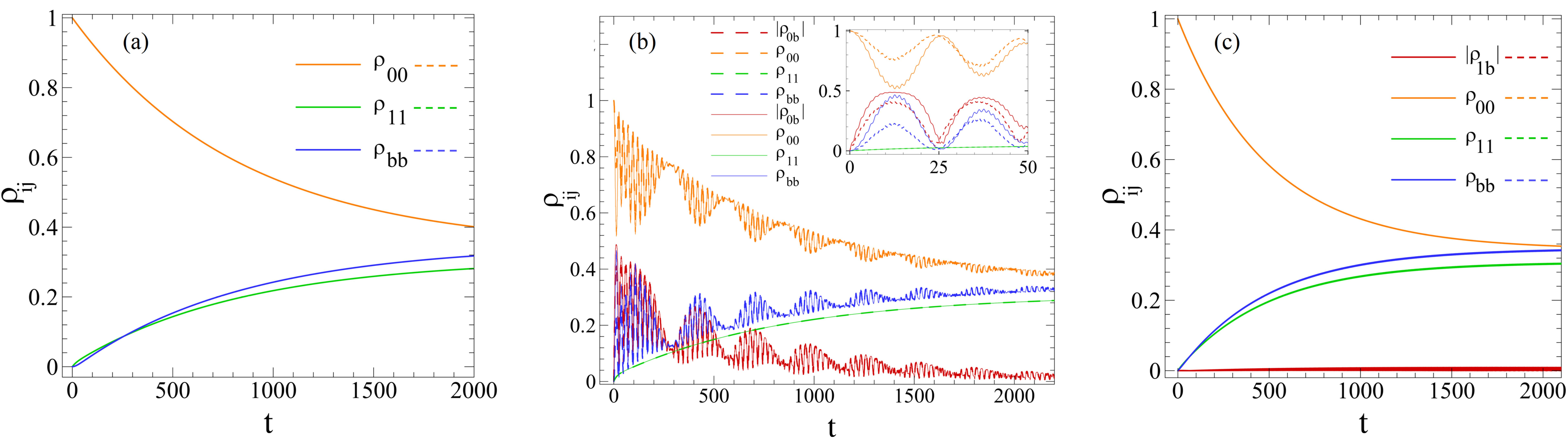}
\caption{Lamb-shift vs. no-Lamb-shift in the dynamics (see Sec. \ref{sec:L-FL}). (a) As Fig. \ref{fig2} a, with no applied field. (b) As Fig. \ref{fig2} b, with a near-resonant applied field $V_{0}(t)$. (c) As Fig. \ref{fig2} c, with an offresonant applied field $V_{1}(t)$. All dashed lines indicate the cases with no-Lamb-shift Hamiltonian.}
\label{fig:lamb}
\end{figure*}

Figure \ref{fig2} shows the evolution of the state of the system. When the system is not driven, it is observed that it reaches its stationary state monotonically; Fig. \ref{fig2} a. Since there is no coherence in the initial state and also noting that the master equation of the system is the Lindblad equation, no coherence is generated during this evolution either. However, when the system is driven by a time-dependent periodic field (Figs. \ref{fig2} b and \ref{fig2} c), it is observed that the system eventually reaches a stationary state, experiencing some fluctuations in the meanwhile. The amplitudes of these transient fluctuations depend on the ratio $\Omega/\omega_{b0}$ (which for Fig. \ref{fig2} b is in the near-resonance regime) and $\Omega/\omega_{1b}$ (which for Fig. \ref{fig2} c is in the offresonance regime). In the former case, at earlier times the amplitude of the fluctuations are considerable, whereas eventually the baths would suppress them. In the latter case, in contrast, the system effectively experiences an effective static potential, which implies negligible fluctuations \cite{bukov_universal_2015}. We note that at long times $|\varrho_{0b}|=0$ [in case (ii)] and $|\varrho_{b1}|=0$ [in case (iii)], which implies a bath-induced decoherence. In addition, by comparing Figs. \ref{fig2} a -- \ref{fig2} c, it is observed that in case (iii) the driven system reaches its stationary state relatively sooner than in the other cases. See Appendix \ref{sec:analytic} for details of calculations. 

Figure \ref{fig4} shows the time evolution of the population of the target level ($\varrho_{bb}$) and the efficiency of energy transfer [as defined in Eq. \eqref{e103}] for the three cases. It is observed that the external field improves this efficiency. However, at earlier times for case (ii) the efficiency is higher compared to the two other cases. At long times, at $t \approx 6000$, when $V_{0}(t)$ is applied, the efficiency grows by $\approx 7\%$ compared to the nondriven case. Similarly, when $V_{1}(t)$ is applied, the efficiency increases by $\approx 8\%$ compared to the nondriven case. As a remark, note that for two baths with equal temperature ratios, their interactions with the corresponding energy levels can be guided through appropriate frequency filters \cite{Scovil-PRL, Scovil-filter, Mahler-filter, Alicki-filter}.

\ignore{
\begin{figure*}[tp]
\includegraphics[scale=0.125]{Lamb-noLamb-Lindblad}
\caption{Lamb-shift vs. no-Lamb-shift in the dynamics (see Sec. \ref{sec:L-FL}). (a) As Fig. \ref{fig2} a, with no applied field. (b) As Fig. \ref{fig2} b, with a near-resonant applied field $V_{0}(t)$. (c) As Fig. \ref{fig2} c, with an offresonant applied field $V_{1}(t)$. All dashed lines indicate the cases with no-Lamb-shift Hamiltonian.}
\label{fig:lamb}
\end{figure*}
}

In practical studies of open quantum systems through master equations, it is often assumed that the Lamb-shift Hamiltonian $H_{\mathrm{lamb}}$ is negligible \cite{noLamb-1,noLamb-2,noLamb-3}. However, a remark is in order. We note that $H_{\mathrm{lamb}}$ in the Lindblad and Floquet-Lindblad scenarios is of the second-order with respect to $H_{\mathsf{int}}$---so is the dissipator \cite{breuer_theory_2002,szczygielski_application_2014}. One can better analyze the effect of $H_{\mathrm{lamb}}$ on the system evolution by removing this term from the master equation and comparing the results with the complete equation. In the Lindblad scenario, since the master equation is a rate-type equation for the population, then the Lamb term does not affect the population evolutions for the nondegenerate case (see Fig. \ref{fig:lamb} a). However, the evolution of the offdiagonal terms of the state of the system $\varrho_{\mathsf{S}}$ (in the eigenbasis of $H_{\mathsf{S}}$) depends on the Lamb-shift Hamiltonian; that is, $H_{\mathrm{lamb}}$ contributes to the coherence effects of $H_{\mathsf{S}}$ in the dynamics. Hence, due to its smallness, neglecting $H_{\mathrm{lamb}}$ compared to $H_{\mathsf{S}}$ seems justifiable.

Nevertheless, note that unlike the Lindblad case, in the Floquet-Lindblad scenario keeping the Lamb-shift Hamiltonian affects the system evolution because $H_{\mathrm{lamb}}^{(\mathrm{F})}$ does not commute with the bare system Hamiltonian $H_{\mathsf{S}}$. Thus, care must be taken when in particular one deals with such cases. From Figs. \ref{fig:lamb} b and \ref{fig:lamb} c we observe that neglecting the Lamb-shift Hamiltonian in the Lindblad scenario as well as the offresonance Floquet-Lindblad scenario are plausible assumptions, because in these cases coherent oscillations are already negligible. Nevertheless, in the near-resonance regime of the Floquet-Lindblad scenario this assumption does not hold unless at the long-time limit of the dynamics. 

\section{Redfield and Floquet-Redfield scenarios}
\label{sec:R-FR}

Here we study the effect of the coherence on the energy transfer efficiency. To do so, we consider two levels $\vert 1\rangle$ and $|2\rangle$ for the system Hamiltonian and investigate the effect of their coherence on the evolution of the population of the target level $\vert b \rangle$. We note that the Lindblad and the Floquet-Lindblad master equations [Eqs. \eqref{e113} and \eqref{e29}] are not appropriate for this particular purpose because neither of them can couple the evolution of the populations to the coherences of the system state. However, if we relax the full secular approximation and retain only the Born-Markov approximation, the resulting Redfield equations for the nondriven and driven cases can offer a physically relevant framework to study the coherence effect. As we explain later, in the driven case (the Floquet-Redfield scenario) we shall still need a suitable \textit{partial} secular approximation to guarantee physicality of the system states.

\subsection{Redfield scenario}
\label{sec:Rscenario}

Consider an open quantum system with a time-independent Hamiltonian $H_{\mathsf{S}}$. When the system and the bath interact weakly so that we can apply the Born-Markov approximation, then the dynamical master equation of the system is given by the Redfield equation, whose general form is as follows \cite{breuer_theory_2002}:
\begin{align}
\dfrac{d\bm{\varrho}_{\mathsf{S}} (t)}{dt}=&\dfrac{1}{\hbar^{2}} \textstyle{\int^{\infty}_{0}}  ds\,\mathrm{Tr}_{\mathsf{B}}\big[ \bm{H}_{\mathsf{int}}(t-s)\bm{\varrho}_{\mathsf{S}}(t)\otimes\bm{\varrho}_{\mathsf{B}}\bm{H}_{\mathsf{int}}(t)\nonumber\\
&-\bm{H}_{\mathsf{int}}(t)\bm{H}_{\mathsf{int}}(t-s)\bm{\varrho}_{\mathsf{S}}(t)\otimes\bm{\varrho}_{\mathsf{B}}+\mathrm{h.c.}\big].
\end{align}
This is to some extent similar to the Lindblad equation but without invoking the secular approximation. To remind the derivation, see Appendix \ref{app:floquet-lindblad} (where we need to consider $H_{\mathsf{S}}$ time-independent and all discussions up to Eq. (\ref{e19}) go through). This equation has been vastly used in various applications to study problems such as light-harvesting \cite{Plenio-etal}. 

\begin{figure*}[tp]
\includegraphics[scale=.13]{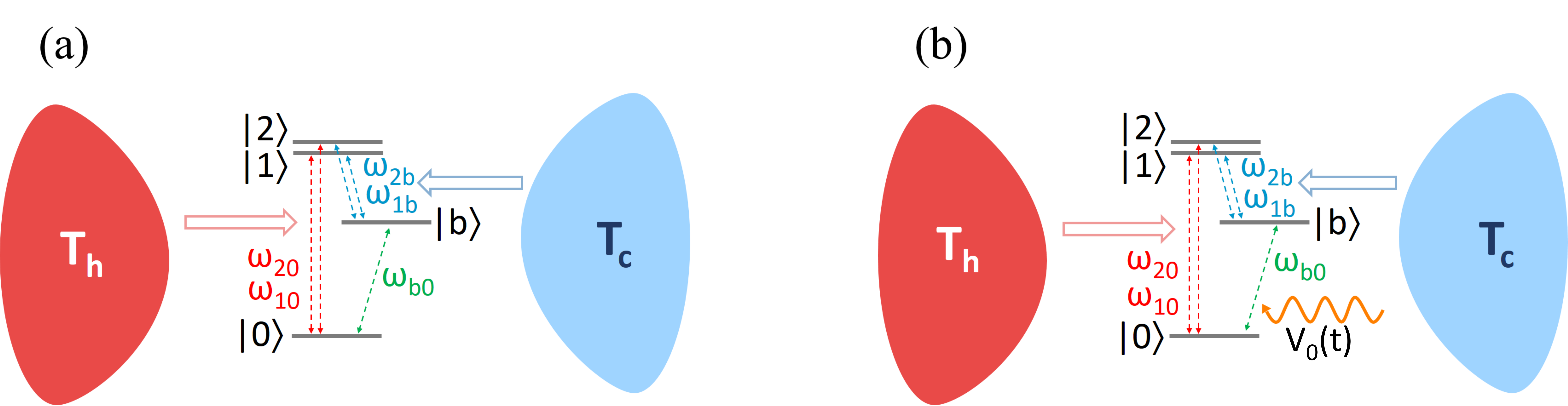}
\caption{A $4$-level system which is in contact with two thermal baths. Two pair of the states are coupled with hot bath, and two other pair with the cold bath. (a) The system is not driven. (b) An external field couples the $\vert 0\rangle$ and $\vert b\rangle$ levels to each other ($V_0$ field).}
\label{fig7}
\end{figure*}

To be specific, let us consider a multilevel system interacting with a bath of radiation field, where the total Hamiltonian is given by Eq. \eqref{e1} with $H_{\mathsf{S}}=H_{0}$ as in Eq. (\ref{h_0}) and the system and the field are coupled through the electric dipole interaction in the ``$\vec{E}\cdot\vec{r}$'' representation \cite{book:Cohen-Tannoudji-QED,book:Cohen-Tannoudji,Scully-book} (see also Appendix \ref{sec:lm}),  
\begin{gather}
H_{\mathsf{B}}=\textstyle{\sum_{k,\lambda}} \hbar \nu_{k} \big( \hat{a}_{k,\lambda}^{\dagger}\hat{a}_{k,\lambda}+1/2\big) ,
\label{e81}\\
H_{\mathsf{int}}= \textstyle{\sum_{i,j }} \textstyle{\sum_{k,\lambda}} g_{k\lambda}^{(ij)} \vert i \rangle\langle j\vert \otimes\hat{a}_{k,\lambda}+\mathrm{h.c.}
\label{e82}
\end{gather}
Here ``h.c.'' denotes Hermitian conjugate, $\hat{a}^{\dag}_{k\lambda}$ is the creation operator of a bath bosonic mode (harmonic oscillator) with frequency $\nu_{k}$, wave vector $\lbrace k\rbrace$, and polarization $\lambda$, and $g_{k\lambda}^{(ij)}$ is the coupling constant of the dipole interaction of the $\vert i\rangle \leftrightarrow \vert j \rangle$ transition of the system with the $(k,\lambda)$th mode of the bath; in particular, $g_{k\lambda}^{(ij)}=\vec{\mu}_{ij}\cdot\hat{e}_{k\lambda}$, with $\hat{e}_{k\lambda}$ being the unit polarization vector, $\vec{\mu}_{ij}=\langle i\vert\vec{\mu}\vert j\rangle$ being the electric-dipole transition matrix elements in the system energy basis \cite{Carmichael:book}. 

After some algebra one obtains (see Appendix \ref{sec:redfield}) the master equation in the Schr\"{o}dinger picture as
\begin{align}
&\dfrac{d\varrho_{\mathsf{S}}(t)}{dt}= -\dfrac{i}{\hbar}\left[ H_{\mathsf{S}},\varrho_{\mathsf{S}}(t)\right] - \dfrac{1}{6\pi c^{3}\hbar \epsilon_{0}}\textstyle{\sum_{i}} \sum_{i'j'} \nonumber\\
&\big[ (\vec{\mu}_{i' i}\cdot\vec{\mu}_{i'j'}) N_{1}(\omega_{i'j'},\beta) \left( \vert i\rangle \langle j'\vert \varrho_{\mathsf{S}}(t)+\varrho_{\mathsf{S}}(t)\vert j'\rangle \langle i\vert \right)+ \nonumber\\
&(\vec{\mu}_{i j'}\cdot\vec{\mu}_{i'j'})N_{2}(\omega_{i'j'},\beta)\left( \vert i\rangle \langle i'\vert \varrho_{\mathsf{S}}(t)+\varrho_{\mathsf{S}}(t)\vert i'\rangle \langle i\vert \right) \big] \nonumber\\
&+\dfrac{1}{6\pi c^{3}\hbar \epsilon_{0}} \textstyle{\sum_{ij}} \sum_{i'j'}\vec{\mu}_{i j}\cdot\vec{\mu}_{i'j'} \times \nonumber\\
& \big[N_{1}(\omega_{i'j'},\beta) \left[ \vert i\rangle \langle j\vert \varrho_{\mathsf{S}}(t)\vert j'\rangle \langle i'\vert +\vert i'\rangle \langle j'\vert \varrho_{\mathsf{S}}(t)\vert j\rangle \langle i\vert \right]+ \nonumber\\
& N_{2}(\omega_{i'j'},\beta) \left[ \vert j\rangle \langle i\vert \varrho_{\mathsf{S}}(t)\vert i'\rangle \langle j'\vert +\vert j'\rangle \langle i'\vert \varrho_{\mathsf{S}}(t)\vert i\rangle \langle j\vert \right] \big] \nonumber\\
&-\dfrac{i}{\hbar}\sum _{ii'j}\dfrac{\vec{\mu}_{ij}\cdot\vec{\mu}_{i'j}}{6\epsilon _{0} \pi ^{2} c^{3}}  C_{2}(\omega _{i'j},\beta)  \big( \vert i\rangle \langle i'\vert \varrho _{\mathsf{S}}(t) -\varrho _{\mathsf{S}}(t)\vert i'\rangle \langle i\vert \big) \nonumber\\
&-\dfrac{i}{\hbar}\sum _{ijj'}\dfrac{\vec{\mu}_{ij}\cdot\vec{\mu}_{ij'}}{6\epsilon _{0} \pi ^{2} c^{3}} C_{1}(\omega _{ij'},\beta) \big( \varrho _{\mathsf{S}}(t)\vert j'\rangle \langle j\vert - \vert j\rangle \langle j'\vert \varrho _{\mathsf{S}}(t)  \big)  \nonumber\\
&+\dfrac{i}{\hbar}\sum _{ii'jj'}\dfrac{\vec{\mu}_{ij}\cdot\vec{\mu}_{i'j'}}{6\epsilon _{0} \pi ^{2} c^{3}} 
\Big[ C_{1}(\omega _{i'j'},\beta) \big( \vert i\rangle \langle j\vert \varrho _{\mathsf{S}}(t)\vert j'\rangle \langle i'\vert - \nonumber\\
& \vert i'\rangle \langle j'\vert \varrho _{\mathsf{S}}(t) \vert j\rangle \langle i\vert  \big) + C_{2}(\omega _{i'j'},\beta) \big( \vert j'\rangle \langle i'\vert \varrho _{\mathsf{S}}(t)\vert i\rangle \langle j\vert - \nonumber\\
& \vert j\rangle \langle i\vert \varrho _{\mathsf{S}}(t) \vert i'\rangle \langle j'\vert  \big) \Big],
\label{e106}
\end{align}
where
\begin{gather}
C_{1}(x ,\beta)=\dfrac{i}{\pi} \mathbbmss{P} \int _{0} ^{\infty} d\nu\, \dfrac{\nu ^{3}\overline{n}(\nu,\beta)}{x-\nu} , \nonumber\\
C_{2}(x ,\beta)=\dfrac{i}{\pi} \mathbbmss{P} \int _{0} ^{\infty} d\nu\, \dfrac{\nu ^{3}[\overline{n}(\nu,\beta)+1]}{x-\nu} , \nonumber\\
N_{1}(x ,\beta)=x^{3} \bar{n}(x,\beta) , \nonumber\\
N_{2}(x ,\beta)=x^{3}  [\bar{n}(x,\beta)+1] .\label{defs-00}
\end{gather}
Here $\bar{n}(\omega,\beta)=(e^{\beta \hbar \omega}-1)^{-1}$. We see that in the time-independent case, there is not any oscillatory terms on the right-hand side (RHS) of the master equation in the Schr\"{o}dinger picture. Hence it seems that this equation [Eq. \eqref{e106}] may generate nonnegative populations. 

Before studying examples, there are two points to take into account. (i) Unlike the Lindblad master equation, in general in the Redfield master equation the spectral density function does not appear. This complicates the comparison of the results of these two approaches. To alleviate this issue and make a fairer comparison, we can study a specific example where the two approaches can yield the similar master equations and then by comparing the equations one can read corresponding parameters. For example, for a qubit interacting with a thermal bath, with an odd spectral density function [$J(-\omega)=-J(\omega)$], comparing the Lindblad and the Redfield equations yield
\begin{gather}
(\mu_{10}/\sqrt{6\pi c^3 \hbar \epsilon_{0}})^{2}=\pi J(\omega_{10})/\omega_{10}^{3},
\label{e100m}
\end{gather}
where $\omega_{10}$ is the qubit Hamiltonian gap. We remind that in all examples in this paper, we choose our parameters according to the explanation after Eq. (\ref{h_0}), Table \ref{tab-2}, and Eq. \eqref{e100m}.

(ii) Note the integrals in Eq. \eqref{defs-00} which appear in the Redfield and Floquet-Redfield master equations. As explained in Appendix \ref{sec:redfield}, these integrals are associated to the Lamb-shift terms. It has been known that the Lamb-shift terms give rise to modifications of the energy levels. Since the corrections of the energy levels due to the Lamb shift are typically small, in some applications they can be safely ignored \cite{noLamb-1, noLamb-2, noLamb-3, tscherbul_partial_2015}. In contrast to the Lindblad scenarios, in the Redfield ones without introducing the spectral density functions, some Lamb-shift terms mathematically diverge. However, by introducing a suitable cutoff frequency (which is compatible with the dipole approximation) and employing appropriate QED considerations \cite{Power-Zienau, PhysRevA.7.1195, book:Cohen-Tannoudji, PhysRevA.87.013804, PhysRevA.89.022128}, these terms take finite values and the divergence issue can be removed. We also follow a similar approach to investigate the Lamb-shift effect in Sec. \ref{sec:lamb-Rdfld}. In some situations such as nondegenerate or time-dependent system Hamiltonians, we observe that the Lamb-shift terms lead to considerable effects on the dynamics.

Now we study a specific example. To see the effect of neglecting the secular approximation, we examine a $4$-level system which is coupled to two thermal baths as in Fig. \ref{fig7} a. The Hamiltonian of the system is 
\begin{equation}
H_{\mathsf{S}}=H_{0}=\textstyle{\sum_{i=0}^{2}} \epsilon_{i}\vert i\rangle\langle i\vert +\epsilon_{b}\vert b\rangle \langle b\vert. 
\end{equation}
We assume that the hot bath leads to the two transitions $\vert 0\rangle \leftrightarrow \vert 1 \rangle$ and $\vert 0\rangle \leftrightarrow \vert 2\rangle$, 
whereas the cold bath causes the $|b\rangle \leftrightarrow |1\rangle$ and $|b\rangle \leftrightarrow |2\rangle$ transitions. The system-bath interaction Hamiltonian reads as
\begin{align}
H_{\mathsf{int}}=H_{\mathsf{S} \mathsf{B}_\mathrm{h}}+H_{\mathsf{S} \mathsf{B}_\mathrm{c}},
\label{e108}
\end{align}
where
\begin{gather}
H_{\mathsf{S}\mathsf{B}_{\mathsf{h}}}= \textstyle{\sum_{i=1,2}} \textstyle{\sum_{k,\lambda}} g_{k\lambda}^{(i 0)} \vert i\rangle\langle 0\vert\otimes  \hat{a}_{k,\lambda}+ \mathrm{h.c.} ,\nonumber\\
H_{\mathsf{S}\mathsf{B}_\mathrm{c}}= \textstyle{\sum_{i=1,2}} \textstyle{\sum_{q,\nu}} f_{q\nu}^{(i b)} \vert i\rangle\langle b\vert \otimes\hat{b}_{q,\nu}+ \mathrm{h.c.}
\label{e109}
\end{gather}

\begin{figure}[tp]
\includegraphics[width=0.9\linewidth]{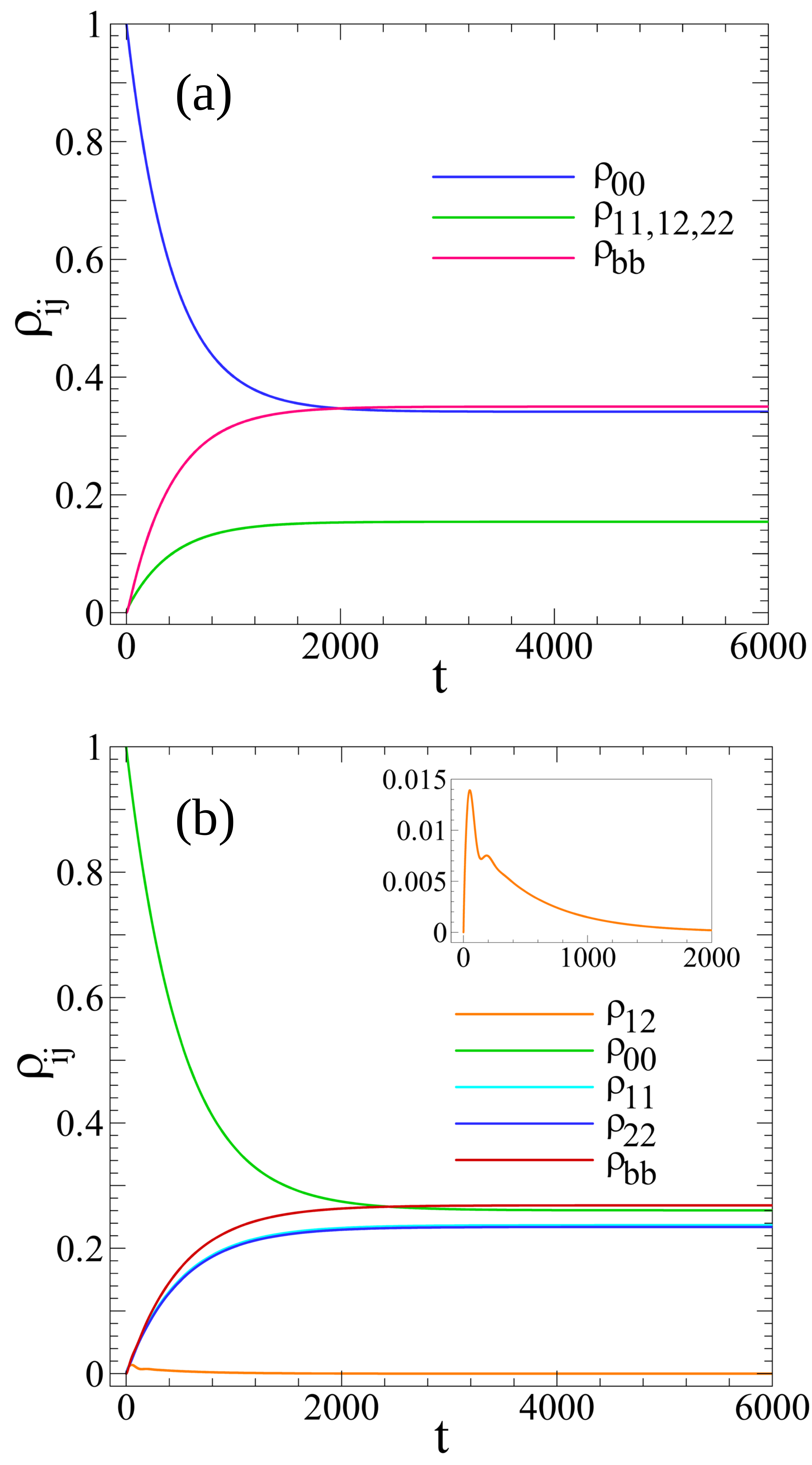}
\caption{Time evolution of a $4$-level nondriven system in contact with two thermal baths with $\beta_{\mathsf{c}}/\beta_{\mathsf{h}}=30/4$. The Lamb-shift term is not included here. (a) The levels $\vert 1\rangle$ and $\vert 2\rangle$ are degenerate. (b) The gap between $\vert 1\rangle$ and $\vert 2\rangle$ is assumed to be $0.05$. All quantities are in natural units.}
\label{fig9}
\end{figure}

We study two cases according to the value of the gap between $\vert 1\rangle$ and $\vert 2\rangle$. First, we consider the nonzero gap case $\omega_{21}=(\epsilon_{2}-\epsilon_{1})/\hbar = 0.05$, and next we focus on the gapless case where $|1\rangle$ and $|2\rangle$ are degenerate states. At first, let us disregard the integrals in Eq. \eqref{e106} (i.e., $C_{1}(x,\beta)$ and $C_{2}(x,\beta)$ coefficients) and analyze the solutions of this dynamical equation. In Fig. \ref{fig9} the evolution of the system in both degenerate and nondegenerate cases are shown (for the corresponding analysis of the dynamics with Lamb-shift terms, see Sec. \ref{sec:lamb-Rdfld}). In Figs. \ref{fig9} a and \ref{fig9} b the temperature ratio of the two baths is $30/4$. When there is a gap between $\vert 1\rangle$ and $\vert 2\rangle$, the final configuration of the system has populations in all levels, which yields a relative reduction of the population of the target level. As expected, according to Eq. \eqref{e106} even in the nondegenerate case a nonvanishing coherence between the levels $\vert 1\rangle$ and $\vert 2\rangle$ is generated because the two $\vert 0\rangle \leftrightarrow \vert 1 \rangle$ and $\vert 0\rangle \leftrightarrow \vert 2\rangle$ ($\vert b\rangle \leftrightarrow \vert 1 \rangle$ and $\vert b\rangle \leftrightarrow \vert 2\rangle$) transitions are due to the same bath, i.e., there are more than one way to populate the levels $|1\rangle$ and $|2\rangle$. The role of coherence in the energy transfer efficiency is studied in Sec. \ref{sec:lamb-Rdfld}, where the Lamb-shift terms are included in the master equation.

\subsection{Floquet-Redfield scenario}
\label{sec:FRscenario}


Now consider a quantum system which interacts with a bath and is driven by a periodic external field. The total system Hamiltonian is 
\begin{align}
H_{\mathsf{tot}}(t)=H_{\mathsf{S}}(t)+H_{\mathsf{B}}+H_{\mathsf{int}} ,
\label{e79}
\end{align}
where the terms on the RHS are given by Eqs. \eqref{e38}, \eqref{h_0}, \eqref{e81}, and \eqref{e82}.

Recalling Sec. \ref{sec:FLscenario}, one can obtain a related master equation for this scenario. However, completely positive of this dynamics is not necessarily guaranteed. To alleviate this issue, we apply a \textit{partial} secular approximation dictated by the assumption $\Omega^{-1}\ll \mathpzc{T}_{R}$, where $\mathpzc{T}_{R}$ is the relaxation time of the driven system. This approximation neglects all terms in the master equation where $q'\neq q$. This removes some fast oscillating terms. Note that this approximation is different from that of Ref. \cite{tscherbul_partial_2015}. After some algebra (see Appendix \ref{sec:redfield} for details), we obtain the following Floquet-Redfield master equation in the Schr\"{o}dinger picture:
\begin{widetext}
\begin{align}
\dfrac{d\varrho_{\mathsf{S}} (t)}{dt}=& -\dfrac{i}{\hbar}\left[ H_{\mathsf{S}}(t),\varrho_{\mathsf{S}}(t)\right] -\dfrac{1}{6\pi c^{3}\hbar \epsilon_{0}}\textstyle{\sum_{ij}} \sum_{i'j'} \sum_{q} \sum_{\omega \omega'}(\vec{\mu}_{ij} \cdot \vec{\mu}_{i'j'}) \times \nonumber\\
&\Big[ \big[N_{2}(\omega'+q \Omega \, ,\beta)+C_{2}(\omega'+q \Omega \, ,\beta) \big]\times \big[ L_{ij}(q,\omega ;t)L^{\dagger}_{i'j'}(q,\omega';t)\varrho_{\mathsf{S}}(t)- L^{\dagger}_{i'j'}(q,\omega';t)\varrho_{\mathsf{S}}(t)L_{ij}(q,\omega ;t)\big]  \nonumber\\      
&+\big[N_{2}(\omega'+q \Omega \, ,\beta)-C_{2}(\omega'+q \Omega \, ,\beta) \big]\times \big[ \varrho_{\mathsf{S}}(t)L_{i'j'}(q,\omega';t)L^{\dagger}_{ij}(q,\omega ;t)- L^{\dagger}_{ij}(q,\omega ;t)\varrho_{\mathsf{S}}(t)L_{i'j'}(q,\omega';t) \big]  \nonumber\\
&+\big[N_{1}(\omega'+q \Omega \, ,\beta)+C_{1}(\omega'+q \Omega \, ,\beta) \big]\times \big[ \varrho_{\mathsf{S}}(t)L^{\dagger}_{i'j'}(q,\omega' ;t)L_{ij}(q,\omega ;t)- L_{ij}(q,\omega ;t)\varrho_{\mathsf{S}}(t)L^{\dagger}_{i'j'}(q,\omega';t) \big]  \nonumber\\
&+\big[N_{1}(\omega'+q \Omega \, ,\beta)-C_{1}(\omega'+q \Omega \, ,\beta) \big]\times \big[  L^{\dagger}_{ij}(q,\omega ;t)L_{i'j'}(q,\omega';t)\varrho_{\mathsf{S}}(t)- L_{i'j'}(q,\omega';t)\varrho_{\mathsf{S}}(t)L^{\dagger}_{ij}(q,\omega ;t)\big] \Big],
\label{e98}
\end{align}
\end{widetext}
where the coefficients are defined in Eqs. \eqref{defs-00} and 
\begin{gather}
L_{ij}(q,\omega ;t)= P(t,0)\sigma_{ij}(q,\omega) P ^{\dagger}(t,0) .
\end{gather}
with $\sigma_{ij}(q,\omega)$ defined through
\begin{gather*}
\sigma_{ij}\equiv \vert i\rangle\langle j\vert,\\
P ^{\dagger}(t,0)\sigma_{ij}P(t,0)= \textstyle{\sum_{q\in\mathbbmss{Z}}} \sigma_{ij}(q)e^{iq\Omega t} ,\\
\sigma_{ij}(q,\omega)= \textstyle{\sum_{\bar{\epsilon}-\bar{\epsilon}'=\hbar \omega}} \vert\bar{\epsilon}\rangle\langle\bar{\epsilon} \vert \sigma_{ij}(q)\vert\bar{\epsilon}'\rangle\langle\bar{\epsilon}'\vert,
\end{gather*}
Note that Eq. \eqref{e98} includes the Lamb-shift terms. 

\begin{figure}[tp]
\includegraphics[width=0.9\linewidth]{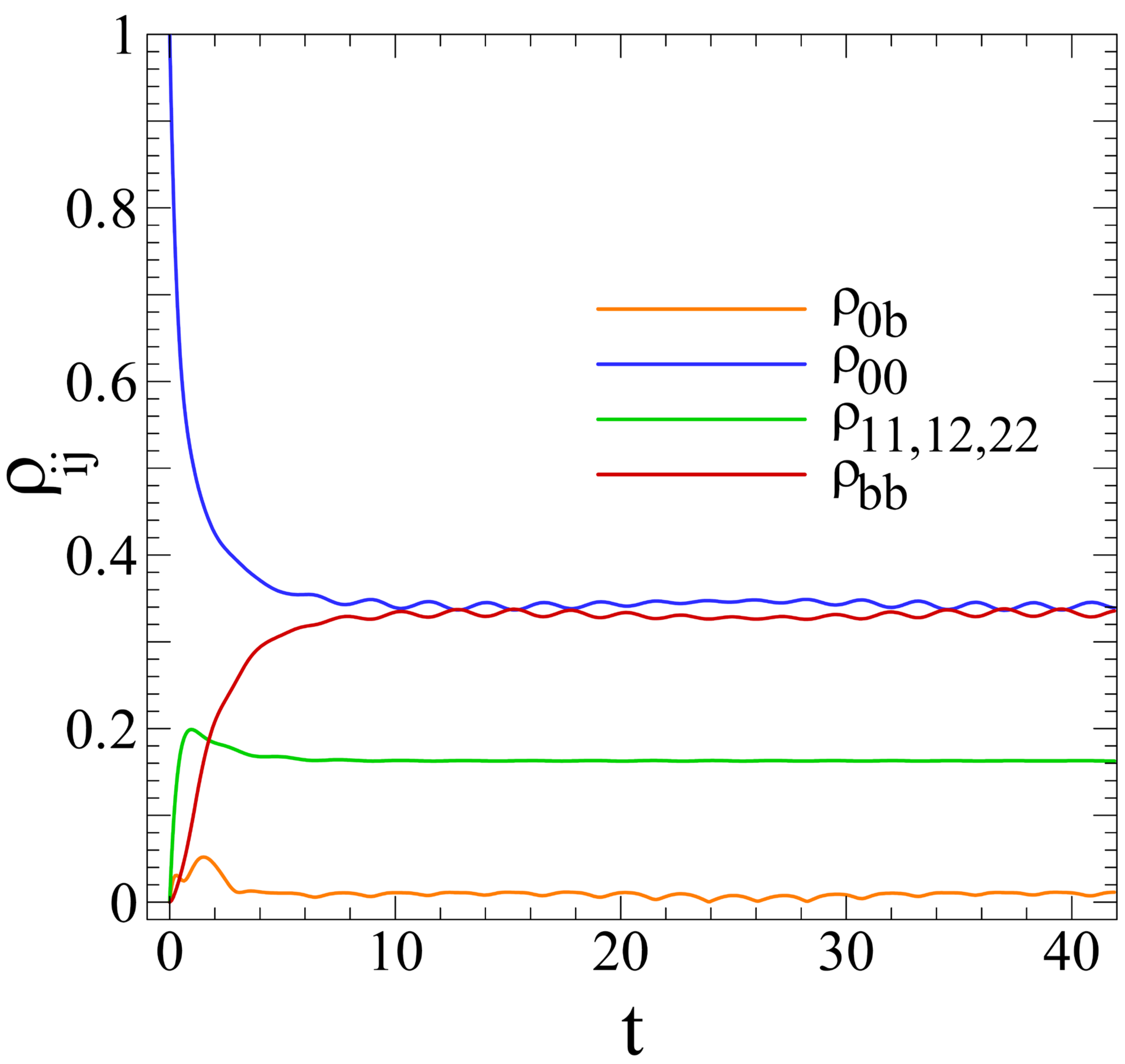}
\caption{Time evolution of the density matrix for the $4$-level driven system in two different temperature ratios. The external field is given by Eq. \eqref{e39}.}
\label{fig11}
\end{figure}

As an example we consider a $4$-level system coupled to two thermal baths and assume that an external field $V_{0}(t)$ couples the $|0\rangle$ and $|b\rangle$ energy levels---see Fig. \ref{fig7} b. The total system Hamiltonian is described by Eq. \eqref{e79}, where 
\begin{align}
&H_{\mathsf{S}}(t)=H_{0}+V_{0}(t) \label{e104}\\
&= \textstyle{\sum_{i=0}^{2}} \epsilon_{i}\vert i\rangle\langle i\vert +\epsilon_{b}\vert b\rangle \langle b\vert +\mu \cos(\Omega t) \left(\vert 0\rangle \langle b\vert +\vert b\rangle \langle 0\vert \right) .\nonumber
\end{align}
The Hamiltonians of the baths and their interaction are also given by Eqs. \eqref{e81} and \eqref{e108}, respectively.

At first, we ignore the contribution of the Lamb-shift terms in the dynamics, i.e., we remove the $C_{1}(x,\beta)$ and $C_{2}(x,\beta)$ coefficients from Eq. \eqref{e98}. In the next section, by considering Lamb-shift terms we will observe that in the Redfield scenario of Sec. \ref{sec:Rscenario}, the energy transfer efficiency in the degenerate case is larger than the nondegenerate case (Fig. \ref{fig10}). Hence in the case of the driven system, we limit ourselves only to the degenerate case, where the two levels $\vert 1\rangle$ and $\vert 2\rangle$ are considered to be degenerate. For the temperature ratio $\beta_{\mathsf{c}} / \beta_{\mathsf{h}}=30/4$ the time evolution of the state of the system is represented in Fig. \ref{fig11}. It is observed that the coherence $\vert \varrho_{12}\vert$, which is also equal to the populations of the degenerate levels, is not destroyed by the bath. However, the coherence $\vert \varrho_{0b}\vert$, which is due to the external field, is affected significantly by the baths. By comparing Figs. \ref{fig11} and \ref{fig9}, it is seen that when the dynamics is driven and given by the Floquet-Redfield equation (\ref{e98}), the system approaches an almost stationary configuration sooner than in the nondriven Redfield dynamics (\ref{e106}). 
\ignore{In addition, it is interesting to see how in this Floquet-Redfield scenario the noise-induced coherence $\varrho_{12}$ in the presence of the field on the energy transfer efficiency compares to the efficiency in the case of the Floquet-Lindblad dynamics (\ref{e29}); Fig.  \ref{fig12} indicated a higher efficiency in the Floquet-Redfield scenario. For example, at $t \approx 42$ we observe an almost $4.4$ times increase in the efficiency compared to the Floquet-Lindblad scenario.}

In the next section, we consider the full master equations \eqref{e106} and \eqref{e98} with the Lamb-shift terms included and investigate the effect of the noise-induced coherence as well as the Lamb-shift terms on the energy transfer efficiency.

\subsection{Effect of the Lamb-shift terms in the Redfield master equations}
\label{sec:lamb-Rdfld}

\begin{figure}[tp]
\includegraphics[width=.9\linewidth]{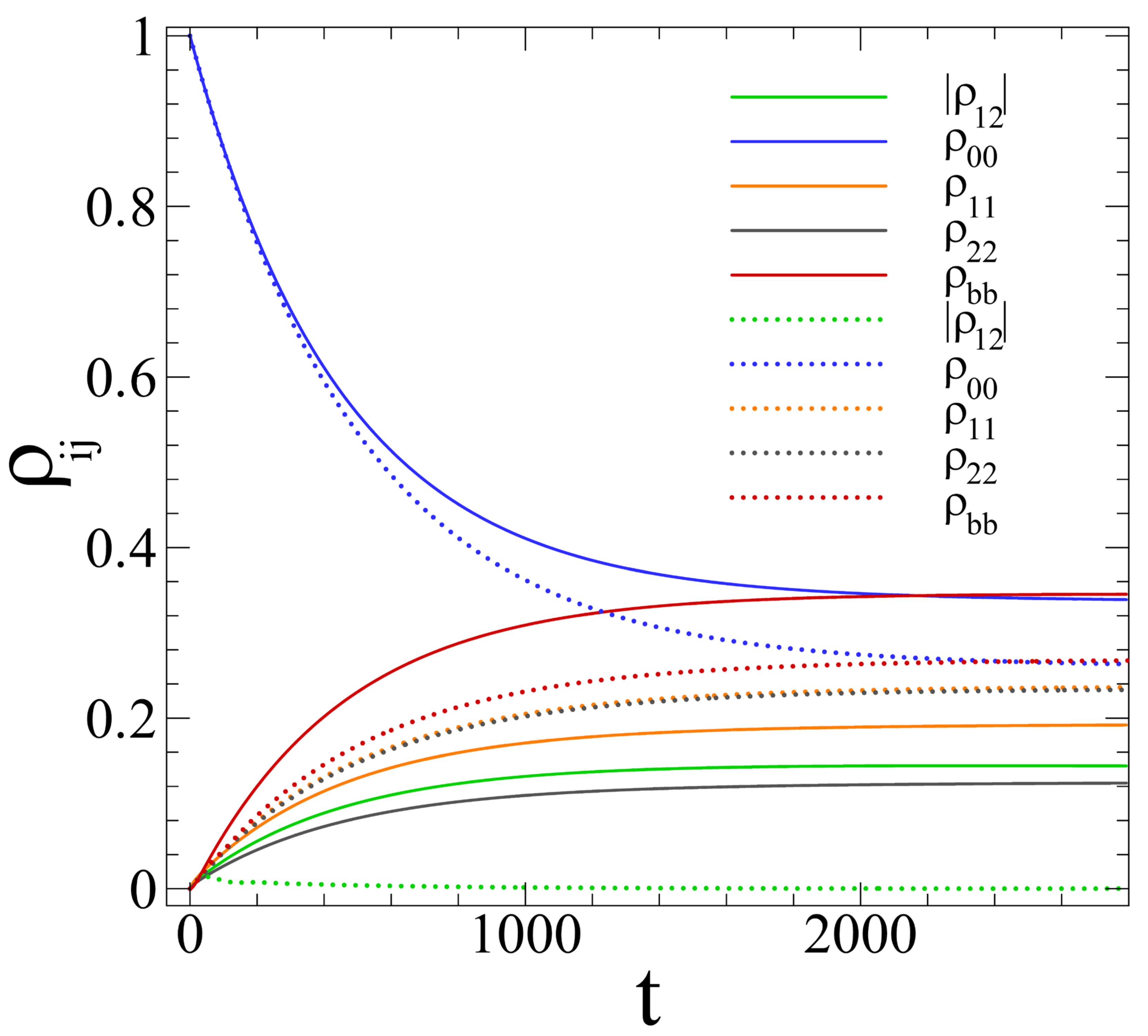}
\caption{Dynamics of a nondriven nondegenerate $4$-level system evolving through Eq. \eqref{e106}. The solid curves correspond to the Redfield scenario with the Lamb-shift correction, while the dotted curves correspond to the dynamics without the Lamb-shift terms. Note that the evolution of the nondriven degenerate system in the presence of the Lamb-shift terms is similar to Fig. \ref{fig9} a.}
\label{fig11lmb}
\end{figure}

In the derivation of the Redfield-type master equations we substitute $\int_{0}^{\infty} ds\,e^{-i(x+\nu)s}=\pi \delta (x+\nu)-i\mathbbmss{P}1/(x+\nu)$---see Appendix \ref{sec:floquet}. The delta function leads to terms which appear in the dissipator and the Cauchy principal value results in the Lamb-shift term, which contributes to the coherent evolution. Although in some studies the principal value part is disregarded (which yield to ignoring the Lamb-shift Hamiltonians in these master equations), a careful analysis of the Lamb-shift term is still in order. To do so, we start from the Hamiltonian of a system interacting with a radiation field. This Hamiltonian is written in the $\vec{E}\cdot\vec{r}$ representation and in the long-wavelength approximation. That is, only the modes with frequency less than a cutoff frequency value $ W <\infty$ are taken into account. Hence the upper limit of all integrals in Eq. \eqref{defs-00} should be replaced with the parameter $ W $. In addition, we assume a low-intensity and low-energy radiation, which respectively yield to disregarding the second-order terms $O(\vec{A}^2)$ term and the interaction of the magnetic field of the radiation and the spin of the system. Now consider the expression for $C_2(\omega ,\beta)$, which includes integrals of the following types:
\begin{align}
\mathbbmss{P} \int_{0}^{W} d\nu\, \dfrac{\nu ^{3}}{\omega -\nu} + \mathbbmss{P} \int_{0}^{W} d\nu\, \dfrac{\nu ^{3} \overline{n}(\nu,\beta)}{\omega -\nu},
\label{divInt}
\end{align}
where
\begin{align}
&\mathbbmss{P} \int_{0}^{W} d\nu\, \dfrac{\nu ^{3}}{\omega -\nu}=-\int_{0}^{W} d\nu\, \nu ^{2} -\omega \int_{0}^{W} d\nu\,  \nu \nonumber\\
& - \omega ^{2}\int_{0}^{W} d\nu + \mathbbmss{P} \int_{0}^{W} d\nu \dfrac{\omega ^{3}}{\omega -\nu} .
\label{divInt1}
\end{align}

By employing appropriate QED arguments (see Appendix \ref {sec:Lamb-app}), one can find that the first term on the RHS of the above relation is completely compensated with another term with opposite sign, which is attributed to a self-energy term. In addition, the second term of the above relation can be concluded to be irrelevant within the low-intensity approximation. Thus, keeping only the relevant terms of Eq. \eqref{divInt}, in the master equation we can effectively replace
\begin{align}
\mathbbmss{P} \int_{0}^{W} d\nu\, \dfrac{\nu ^{3}\big(\overline{n}(\nu,\beta)+1\big)}{\omega -\nu}\to& \mathbbmss{P} \int_{0}^{W} d\nu\, \dfrac{\nu ^{3} \overline{n}(\nu,\beta)}{\omega -\nu}- \omega ^{2} W \nonumber\\
&+ \omega ^{3} \ln (W/\omega).
\label{divInt2}
\end{align}
Note that to establish the validity of the long-wavelength approximation, the cutoff frequency $W$ is chosen large compared to the characteristic frequency of the system and much smaller than the relativistic modes, $\omega _{0}\ll W \ll mc^{2} / \hbar$, where $m$ is the mass of the charged particle \cite{book:Cohen-Tannoudji-QED}. 

\begin{figure}[tp]
\includegraphics[width=.9\linewidth]{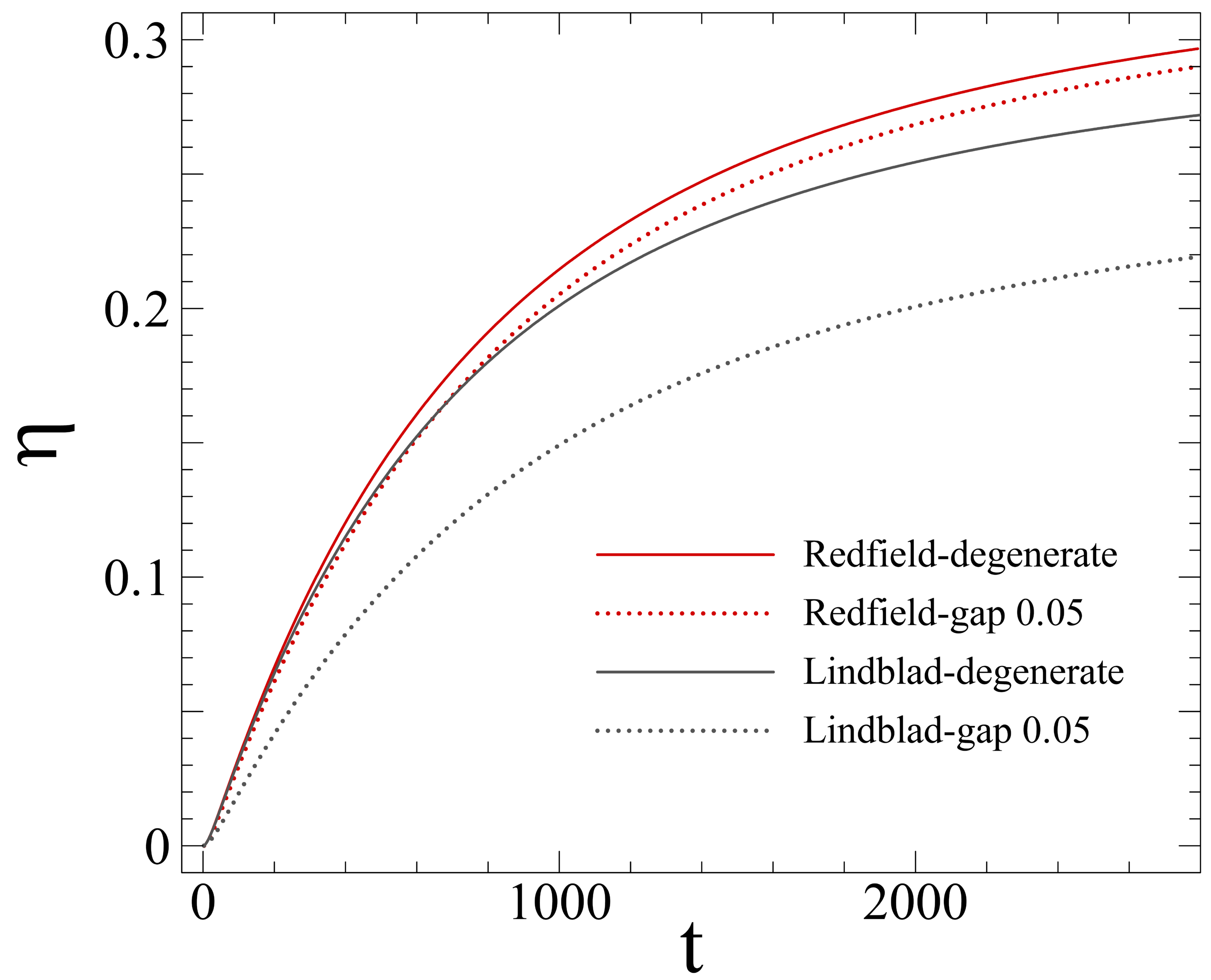}
\caption{Energy transfer efficiency vs. time for the nondriven cases of the Lindblad and the Redfield scenarios. All quantities are in natural units.}
\label{fig10}
\end{figure}

At first we discuss the time-independent Redfield scenario in the presence as well as absence of the Lamb-shift terms. Our simulations show that for the time-independent degenerate $4$-level system (with $\vert 1\rangle$ and $\vert 2\rangle$ degenerate levels) the Lamb-shift terms do not affect the dynamics of the system and the results are the same as Fig. \ref{fig9} a. Thus, for the nondriven scenario (Fig. \ref{fig7} a), we only discuss the nondegenerate system. In all plots of Fig. \ref{fig11lmb}, we consider a nondriven nondegenerate $4$-level system with a $0.05$ gap between the $\vert 1\rangle$ and $\vert 2\rangle$ levels. This figure shows the evolution of the state of the system in the presence as well as absence of the Lamb-shift terms. It is evident that the Lamb-shift terms are nonnegligible. Here we have taken $ W \approx 4\times 10^{4}$ (in natural units). Note that the correct values of the cutoff frequency $W$ in a model depends on its specifics. Here, however, we have estimated this frequency simply by trial and error such that the dynamics generates positive density matrices during the evolution---for a similar analysis see Ref. \cite{PhysRevA.88.042115}. 

To see the effect of coherence on the efficiency of energy transfer see Fig. \ref{fig10}, where we compare the Redfield scenario [Eq. (\ref{e106})] with the Lindblad scenario [Eq. (\ref{e113})] for two degenerate and nondegenerate cases with the temperature ratio $\beta_{\mathsf{c}} / \beta_{\mathsf{h}}=30/4$. At long times the efficiencies of the Redfield scenario for both cases are greater than the similar efficiencies of the Lindblad scenario. For example, at $t \approx 2800$, the Redfield dynamics yields $\approx 8\%$ and $\approx 24\%$ increase in the efficiency over the Lindblad dynamics for the degenerate and nondegenerate ($\omega_{12}=0.05$) cases, respectively. These increases of the efficiency can be attributed to the coupling between the populations and coherence terms in the Redfield scenarios.

\begin{figure}[tp]
\includegraphics[width=0.9\linewidth]{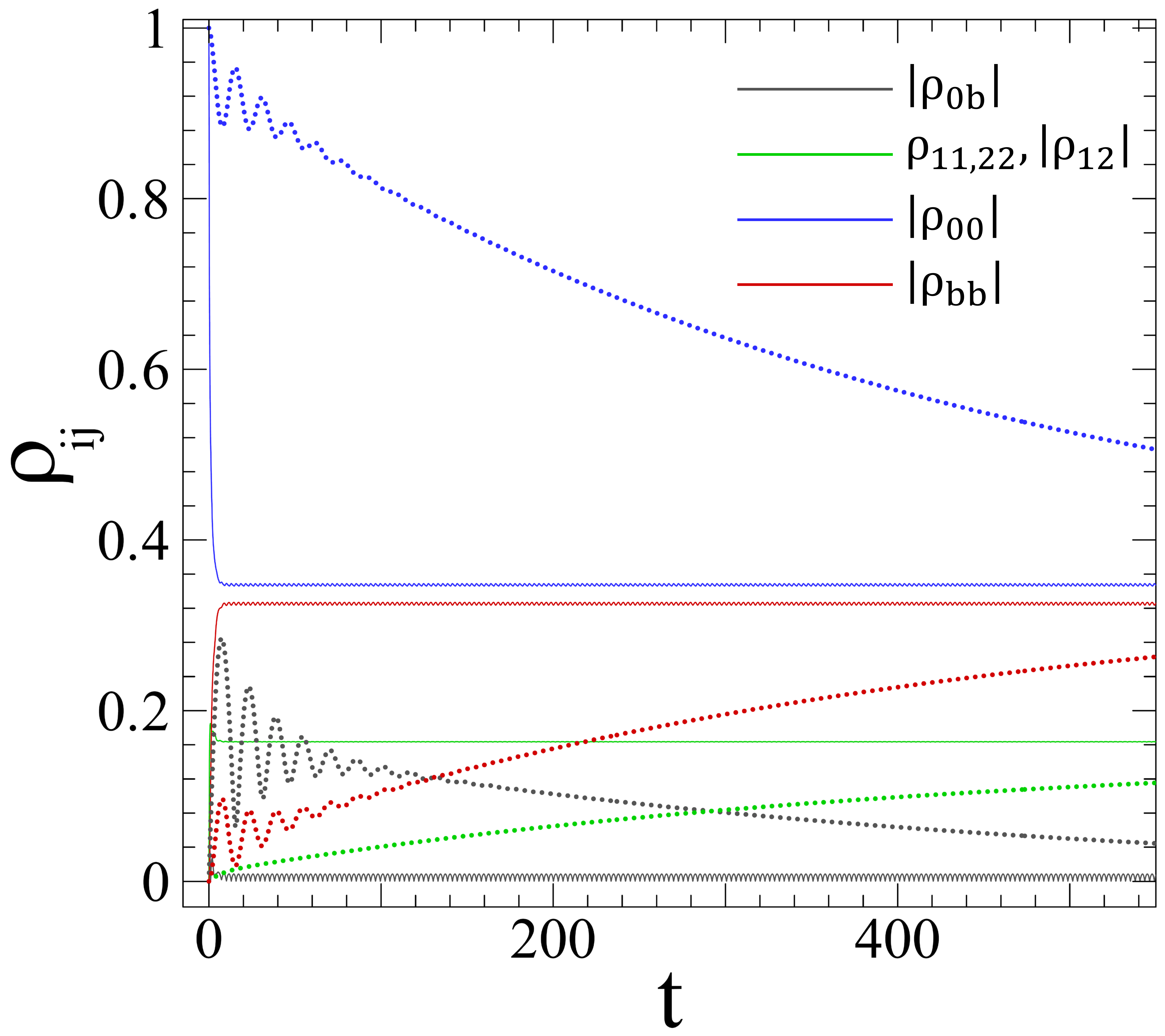}
\caption{State of the $4$-level system. Solid curves are for the Floquet-Redfield dynamics \eqref{e98}, whereas the dotted curves correspond to the Floquet-Lindblad dynamics (\ref{e29}). Both equations include the Lamb-shift terms.}
\label{Lmb-flq-rdfld}
\end{figure}

\begin{figure}[tp]
\includegraphics[width=0.9\linewidth]{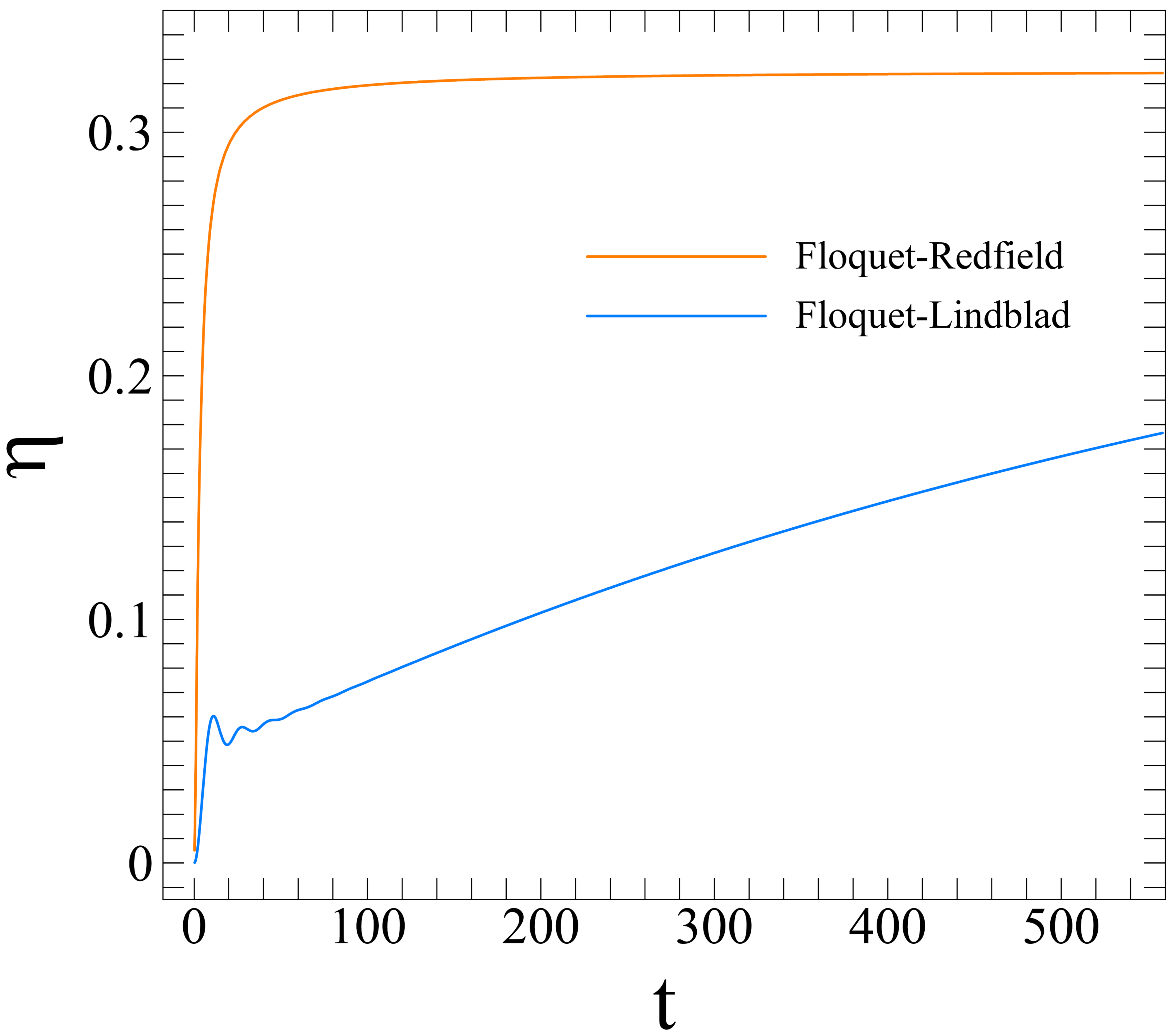}
\caption{Effect of the coherence on the energy transfer efficiency in the presence of the external field. The orange curve corresponds to the Floquet-Redfield master equation (with the lamb-shift terms), and the blue one is related to the Floquet-Lindblad master equation (with the Lamb-shift term).}
\label{fig12}
\end{figure}

\begin{figure}[bp]
\includegraphics[width=.9\linewidth]{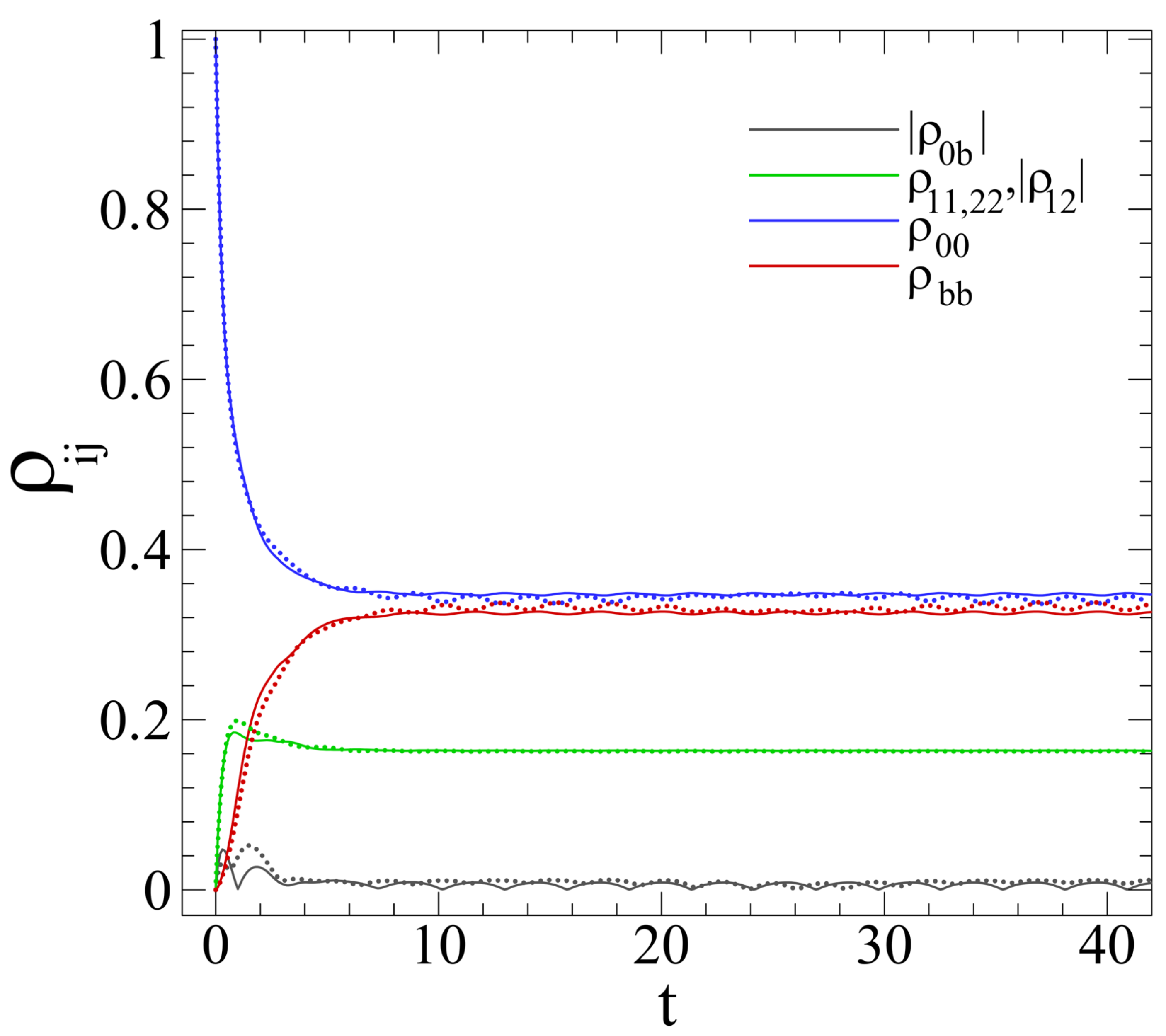}
\caption{Effect of the Lamb-shift terms in the Floquet-Redfield scenario. Solid curves are related to the dynamics with the Lamb-shift terms, and the dotted curves are associated with the no-Lamb-shift dynamics.}
\label{LmbNolmb-flq-rdfld}
\end{figure}

Now let us consider the Floquet-Redfield scenario for a degenerate $4$-level system whose dynamics is given by Eq. \eqref{e98}. Figure \ref{Lmb-flq-rdfld} compares the effects of the full and the partial secular approximations, that is, it shows how the Floquet-Lindblad and Floquet-Redfield master equations compare, note that the Lamb-shift terms are maintained. From this figure, despite remarkable difference between the two sets of plots at short times, it is seems that in the long-time limit these equations yield similar solutions. In Fig. \ref{fig12} the energy transfer efficiency of these two cases are plotted. Since in the Floquet-Redfield scenario the system reaches its stationary state relatively sooner than the Floquet-Lindblad case, from short times the efficiency takes its maximum value and remains so. Figure \ref{LmbNolmb-flq-rdfld} compares the Floquet-Redfield scenarios with and without the Lamb-shift terms. We observe that in the Floquet-Redfield scenario for the degenerate system, the Lamb-shift terms have an insignificant impact on the dynamics. It appears that for the driven nondegenerate system considering the Lamb-shifts terms has nonnegligible impact on the time evolution of the system. This is similar to the time-independent case (Redfield scenario) where including the Lamb-shift terms yielded nonnegligible effect on the time evolution for the nondegenerate system (see Fig. \ref{fig11lmb}),

\section{Summary and conclusions}
\label{sec:summary}

We have studied the problem of energy transfer in a multilevel system in contact with two thermal baths at two different temperatures. The hot bath excites the system to its highest energy level(s) and then the excitation moves through the energy space in the presence of another cold bath. We have considered four different dynamical scenarios within the Born-Markov regime: (i) Lindblad scenario: assuming the secular approximation, with no applied driving field on the system, (ii) Floquet-Lindblad scenario: assuming the secular approximation and driving the system by a time-periodic driving field, (iii) Redfield scenario: assuming no secular approximation and no driving field on the system, and (iv) Floquet-Redfield scenario: assuming a particular partial secular approximation and driving the system by a time-periodic applied field. We have developed a particular dynamical equation for the last scenario. Our overall objective has been to study the efficiency of energy transfer to a particular energy level from other energy levels of the system Hamiltonian in each of the dynamical scenarios. 

We have shown that in the two Lindblad scenarios, Lindblad and Floquet-Lindblad, applying an external periodic field on a $3$-level system which couples the target energy level to another energy level has yielded higher efficiency for the energy transfer phenomenon. We have investigated the effect of the Lamb-shift correction to the system Hamiltonian. In particular, as is known we see that for the no-field Lindblad scenario the Lamb-shit term does not have any effect on the dynamics of the system---and hence on the energy transfer. A similar behavior also has manifested for the case in which an applied field has coupled the target energy level to another level in the offresonance regime. However, when the applied field is in the near-resonance regime, it affects the dynamics at short times, while this effect diminishes as time increases.  

For the two Redfield scenarios, Redfield and Floquet-Redfield, we have considered a $4$-level system, in which a coherence between the two upper levels can be generated through the dynamics. In the Redfield scenario we have considered both nondegenerate and degenerate cases. Since in the Redfield scenario the energy transfer efficiency in the degenerate system grows more than the nondegenerate case, we have only studied the degenerate system in the Floquet-Redfield scenario. Moreover, we have investigate the effect of the Lamb-shift terms in the Redfield scenarios and observed that at least in the degenerate case the lamb-shift terms do not have any significant impact on the dynamics.

Our careful and comprehensive study can be useful in the analysis of energy transfer in synthetic systems, and it offers new outlooks about how one can enhance such processes through suitable driving fields. As a next step, for example, one can study how correlation effects may play a role in enhancing or reducing energy transfer efficiency \cite{unfolding}.

\begin{acknowledgments}
The authors acknowledge helpful discussions with V. Rezvani. This work was partially supported by Sharif University of Technology's Office of Vice President for Research and Technology through Contract No. QA960512.
\end{acknowledgments}

\newpage
\begin{widetext}
\onecolumngrid
\appendix
\section{Derivation of the Floquet-Lindblad master equation}
\label{app:floquet-lindblad}

Here based on Ref. \cite{szczygielski_application_2014}, we reproduce the derivation of the Floquet-Lindblad master equation. Consider a quantum system which is in contact with a bath and is driven by an external periodic field. We want to obtain a master equation which describes the dynamic of the system. The total Hamiltonian is given by
\begin{align}
H_{\mathsf{tot}}(t)=H_{\mathsf{S}}(t)+H_{\mathsf{B}}+H_{\mathsf{int}},
\label{ea1}
\end{align}
where $H_{\mathsf{S}}(t)$ is the periodic driving Hamiltonian of the system ($H_{\mathsf{S}}(t+\tau)=H_{\mathsf{S}}(t)$ with period $\tau$), $H_{\mathsf{B}}$ is the bath Hamiltonian, and $H_{\mathsf{int}}$ describes the interaction between the system and the bath.
Evolution of the total system in the interaction picture is governed by
\begin{align}
\dfrac{d\bm{\varrho}_{\mathsf{tot}}(t)}{dt}=-\dfrac{i}{\hbar}\left[ \bm{H}_{\mathsf{int}}(t) , \bm{\varrho}_{\mathsf{tot}}(t) \right] ,
\label{e2}
\end{align}
where the interaction Hamiltonian in the interaction picture is defined as
\begin{align}
\bm{H}_{\mathsf{int}}(t)= U^{\dagger}_{\mathsf{S}} (t,0)\otimes U^{\dagger}_{\mathsf{B}}(t,0) H_{\mathsf{int}} U_{\mathsf{S}}(t,0)\otimes U_{\mathsf{B}}(t,0).
\label{e3}
\end{align}

From Eq. \eqref{e2} we obtain
\begin{align}
\dfrac{d\bm{\varrho}_{\mathsf{tot}}(t)}{dt}=-\dfrac{i}{\hbar} \left\lbrace \left[ \bm{H}_{\mathsf{int}}(t) , \bm{\varrho}_{\mathsf{tot}} (0) \right] -\dfrac{i}{\hbar} \big[ \bm{H}_{\mathsf{int}}(t) , \textstyle{\int^{t}_{0}}  ds\,\left[\bm{H}_{\mathsf{int}}(s) , \bm{\varrho}_{\mathsf{tot}}(s) \right] \big] \right\rbrace .
\label{e4}
\end{align}

By considering the Markov approximation, the dynamical equation reduces to  
\begin{align}
\dfrac{d\bm{\varrho}_{\mathsf{tot}}(t)}{dt}=-\dfrac{i}{\hbar} \left[ \bm{H}_{\mathsf{int}}(t) , \bm{\varrho}_{\mathsf{tot}}(0) \right] -\dfrac{1}{\hbar^{2}} \textstyle{\int^{t}_{0}}  ds\,\big[ \bm{H}_{\mathsf{int}}(t) ,\left[ \bm{H}_{\mathsf{int}}(s) , \bm{\varrho}_{\mathsf{tot}}(t) \right] \big],
\label{e5}
\end{align}
from whence
\begin{align}
\dfrac{d\bm{\varrho}_{\mathsf{S}} (t)}{dt}=-\dfrac{i}{\hbar} \mathrm{Tr}_{\mathsf{B}} \left[ \bm{H}_{\mathsf{int}}(t) , \bm{\varrho}_{\mathsf{tot}} (0) \right] -\dfrac{1}{\hbar^{2}} \textstyle{\int^{t}_{0}}  ds\,\mathrm{Tr}_{\mathsf{B}} \big[ \bm{H}_{\mathsf{int}}(t) ,\left[ \bm{H}_{\mathsf{int}}(s) ,\bm{\varrho}_{\mathsf{tot}}(t) \right] \big].
\label{e6}
\end{align}

Consider the general form of the interaction Hamiltonian in the Schr\"{o}dinger picture as
\begin{align}
H_{\mathsf{int}}= \textstyle{\sum_{\alpha}} S_{\alpha}\otimes B_{\alpha},
\label{e63}
\end{align}
where $S_{\alpha}^{\dagger}=S_{\alpha}$ and $B_{\alpha}^{\dagger}=B_{\alpha}$. Substituting Eq. \eqref{e63} into Eq. \eqref{e3} yields
\begin{align}
\bm{H}_{\mathsf{int}}(t)=\textstyle{\sum_{\alpha}} U^{\dagger}_{\mathsf{S}}(t,0) S_{\alpha}U_{\mathsf{S}}(t,0)\otimes U^{\dagger}_{\mathsf{B}}(t,0)B_{\alpha}U_{\mathsf{B}}(t,0)=\sum_{\alpha}\bm{S}_{\alpha}(t)\otimes \bm{B}_{\alpha}(t) ,
\label{e7}
\end{align}
where $U_{\mathsf{S}}(t,0)=\mathbbmss{T}e^{(-i/\hbar)\int_{0}^{t}ds\,H_{\mathsf{S}}(s)}$, $U_{\mathsf{B}}(t,0)=e^{(-it/\hbar)H_{\mathsf{B}}}$, $\bm{S}_{\alpha}(t)=U^{\dagger}_{\mathsf{S}}(t,0)S_{\alpha}U_{\mathsf{S}}(t,0)$, and $\bm{B}_{\alpha}(t)=U^{\dagger}_{\mathsf{B}}(t,0)B_{\alpha}U_{\mathsf{B}}(t,0)$. 

If we assume that the initial state of the total system is an uncorrelated state $\varrho_{\mathsf{tot}}(0)=\varrho_{\mathsf{S}}(0)\otimes\varrho_{\mathsf{B}}(0)$ and the bath is prepared in a thermal state, then the first term in Eq. \eqref{e6} vanishes,
\begin{align}
\mathrm{Tr}_{\mathsf{B}} \left[ \bm{H}_{\mathsf{int}}(t) , \bm{\varrho}_{\mathsf{tot}} (0) \right] &=\mathrm{Tr}_{\mathsf{B}} \left[\textstyle{\sum_{\alpha}}\bm{S}_{\alpha}(t)\otimes \bm{B}_{\alpha}(t) , \varrho_{\mathsf{S}}(0)\otimes\varrho_{\mathsf{B}}(0) \right] \nonumber\\
& =\textstyle{\sum_{\alpha}} \left[ \bm{S}_{\alpha}(t), \varrho_{\mathsf{S}}(0) \right] \mathrm{Tr}_{\mathsf{B}}\left[ \bm{B}_{\alpha}(t)\varrho_{\mathsf{B}}(0)\right] =0,
\label{e8}
\end{align}
because $\mathrm{Tr}_{\mathsf{B}}\left[ \bm{B}_{\alpha}(t)\varrho_{\mathsf{B}}(0)\right] =0$. In addition, we assume the Born approximation, where the coupling between the system and the bath is considered to be weak such that one can neglect the effects of the system on the bath. This approximation is often incorporated by writing the instantaneous state of the total system approximately as a tensor product of the system state $\varrho_{\mathsf{S}}(t)$ and the bath state is assumed almost unchanged in the course of time, $\bm{\varrho}_{\mathsf{B}}(t)\approx \bm{\varrho}_{\mathsf{B}}(0) \equiv \bm{\varrho}_{\mathsf{B}}$; that is, $\bm{\varrho}_{\mathsf{tot}}(t) \approx \bm{\varrho}_{\mathsf{S}}(t)\otimes\bm{\varrho}_{\mathsf{B}}$. Thus, by applying this approximation, Eq. \eqref{e6} becomes
\begin{align}
\dfrac{d\bm{\varrho}_{\mathsf{S}} (t)}{dt}=-\dfrac{1}{\hbar^{2}} \textstyle{\int^{t}_{0}}  ds\,\mathrm{Tr}_{\mathsf{B}} \big[ \bm{H}_{\mathsf{int}}(t) ,\left[ \bm{H}_{\mathsf{int}}(s) , \bm{\varrho}_{\mathsf{S}}(t)\otimes \bm{\varrho}_{\mathsf{B}} \right] \big] .
\label{e9}
\end{align}

If we substitute $s \rightarrow t-s$ in Eq. \eqref{e9} and put the upper limit of the integral to be infinity (presuming that the integrand, which is proportional to the bath correlation functions, does not have a substantial contribution for $s\gg \mathpzc{T}_{\mathsf{B}}$) \cite{breuer_theory_2002}, this reduces Eq. \eqref{e9} to the following Markovian master equation:
\begin{align}
\dfrac{d\bm{\varrho}_{\mathsf{S}} (t)}{dt}=\dfrac{1}{\hbar^{2}} \textstyle{\int^{\infty}_{0}}  ds\,\mathrm{Tr}_{\mathsf{B}}\big[ \bm{H}_{\mathsf{int}}(t-s)\bm{\varrho}_{\mathsf{S}}(t)\otimes\bm{\varrho}_{\mathsf{B}}\bm{H}_{\mathsf{int}}(t)-\bm{H}_{\mathsf{int}}(t)\bm{H}_{\mathsf{int}}(t-s)\bm{\varrho}_{\mathsf{S}}(t)\otimes\bm{\varrho}_{\mathsf{B}}+\mathrm{h.c.}\big].
\label{e19}
\end{align}
We now need to incorporate $\bm{H}_{\mathsf{int}}(t)$ in the above equation from Eq. (\ref{e7}). To do so, we need to calculate $U_{\mathsf{S}}(t,0)$, which in general requires a time-ordering as $U_{\mathsf{S}}(t,0)=\mathbbmss{T}e^{-(i/\hbar)\int_{0}^{t} H_{\mathsf{S}}(s)\,ds}$. However, noting that the system Hamiltonian $H_{\mathsf{S}}(t)$ is assumed to be periodic, one can employ the Floquet theorem to partially simplify the form of $U_{\mathsf{S}}(t,0)$.

The Floquet theorem states that \cite{bukov_universal_2015} the time evolution operator associated to a periodic Hamiltonian $H_{\mathsf{S}}(t)$, $i\hbar\frac{d}{dt}U_{\mathsf{S}}(t,t_{0})=H_{\mathsf{S}}(t)U_{\mathsf{S}}(t,t_{0})$, can be factorized as
\begin{align}
U_{\mathsf{S}}(t,t_{0})= P (t,t_{0}) e^{-i\bar{H}(t-t_{0})/\hbar},
\label{e11}
\end{align}
where $\bar{H}$ is a Hermitian \textit{time-independent} operator, referred to as the Floquet Hamiltonian, and $ P (t,t_{0})$ is a unitary and periodic operator such that $ P (t+n\tau,n\tau)=P (t+n\tau,0)=P(t,0)$ and $ P (n\tau,0)=\mathbbmss{I}$,  with $\tau$ being the period of the Hamiltonian and $n\in \mathbbmss{N}$.

The Floquet theorem helps us rewrite $\bm{S}_{\alpha}(t)$ operators in Eq. (\ref{e7}) as
\begin{align}
\bm{S}_{\alpha}(t)=e^{i\bar{H}t/\hbar} P ^{\dagger}(t,0)S_{\alpha}P(t,0)e^{-i\bar{H}t/\hbar}.
\label{e12}
\end{align}
Since the operator $P(t,0)$ is periodic, then $ P ^{\dagger}(t,0)S_{\alpha}P(t,0)$ is also periodic and hence has a Fourier expansion as
\begin{align}
 P ^{\dag}(t,0)S_{\alpha}P(t,0)= \textstyle{\sum_{q\in \mathbbmss{Z}}} S_{\alpha}(q) e^{iq\Omega t},
\label{e13}
\end{align}
where $\Omega =2\pi/\tau$ is the frequency of the periodic Hamiltonian $H_{\mathsf{S}}(t)$. The Fourier coefficients read
\begin{align}
 S_{\alpha}(q) =\dfrac{1}{\tau} \textstyle{\int^{\tau}_{0}}  dt\, e^{-iq\Omega t}P ^{\dag}(t,0)S_{\alpha}P(t,0),
\label{eSq}
\end{align}
whose calculation needs the knowledge of the periodic operator $P(t,0)$ only in the first period $0\leqslant t\leqslant \tau$.

One can construct another set of related operators as 
\begin{align}
S_{\alpha}(q,\omega)= \textstyle{\sum_{\bar{\epsilon}-\bar{\epsilon}'=\hbar\omega}} \vert\bar{\epsilon}\rangle\langle\bar{\epsilon}\vert S_{\alpha}(q)\vert\bar{\epsilon}'\rangle\langle\bar{\epsilon}'\vert ,
\label{e14}
\end{align}
where $\bar{\epsilon}$ and $\vert\bar{\epsilon}\rangle$ are, respectively, the eigenvalues and eigenvectors of the Floquet Hamiltonian, $\bar{H}\vert\bar{\epsilon}\rangle =\bar{\epsilon}\vert\bar{\epsilon}\rangle$. The summation in Eq. \eqref{e14} is over all eigenvalues with the same (quasi) energy difference $\hbar\omega$, and whenever we encounter a $\sum_{\omega}$ it denotes the summation over all possible (quasi) energy gaps. 

It is useful to identify several symmetries here. By comparing Eqs. \eqref{e13} and \eqref{e14} with their Hermitian conjugates, we find $S_{\alpha}^{\dagger}(q)=S_{\alpha}(-q)$ and $S_{\alpha}^{\dagger}(q,\omega)=S_{\alpha}(-q,-\omega)$. By summing $S_{\alpha}(q,\omega)$ over all $\omega$s we have
\begin{align}
\textstyle{\sum_{\omega}} S_{\alpha}(q,\omega)=S_{\alpha}(q).
\label{e15}
\end{align}

Substituting Eq. \eqref{e13} into Eq. \eqref{e12} and then using Eq. \eqref{e14}, gives 
\begin{align}
\bm{S}_{\alpha}(t)= \textstyle{\sum_{q\in\mathbbmss{Z}}} \sum_{\omega} e^{it(\omega +q\Omega)}S_{\alpha}(q,\omega),
\label{e16}
\end{align}
which in turn yields 
\begin{gather}
\bm{H}_{\mathsf{int}}(t)= \textstyle{\sum_{\alpha}} \textstyle{\sum_{q\in\mathbbmss{Z}}} \textstyle{\sum_{\omega}} e^{it(\omega +q\Omega)}S_{\alpha}(q,\omega)\otimes \bm{B}_{\alpha}(t).
\label{e17}
\end{gather}
Inserting this form in Eq. \eqref{e19} gives the following dynamical equation:
\begin{align}
\dfrac{d\bm{\varrho}_{\mathsf{S}} (t)}{dt}=&\dfrac{1}{\hbar^{2}} \textstyle{\sum_{\alpha ,\alpha'}} \sum_{{\omega},{\omega'}}\sum_{q,q'}e^{it\left[ (\omega -\omega')+(q-q')\Omega\right] } \Big[\left( S_{\alpha'}(q,\omega)\bm{\varrho}_{\mathsf{S}}(t)S_{\alpha}^{\dagger}(q',\omega')- S_{\alpha}^{\dagger}(q',\omega')S_{\alpha'}(q,\omega)\bm{\varrho}_{\mathsf{S}}(t)\right)  \nonumber\\
& \times \textstyle{\int^{\infty}_{0}}  ds\,e^{-is(\omega +q\Omega)} \mathrm{Tr}_{\mathsf{B}}\big[\bm{B}_{\alpha}^{\dagger}(t) \bm{B}_{\alpha'}(t-s)\varrho_{\mathsf{B}}\big] \Big]+ \mathrm{h.c.}
\label{e22}
\end{align}
This from can still be simplified further by introducing the one-sided Fourier transformation of the bath correlation functions as
\begin{align}
\Gamma_{\alpha \alpha'}(\omega +q\Omega)\equiv \textstyle{\int^{\infty}_{0}}  ds\,e^{-is(\omega +q\Omega)} \langle \bm{B}_{\alpha}^{\dagger}(t)\bm{B}_{\alpha'}(t-s)\rangle ,
\label{e23}
\end{align}
where $\langle \bm{B}_{\alpha}^{\dagger}(t)\bm{B}_{\alpha'}(t-s)\rangle =\mathrm{Tr}_{\mathrm{B}}\big[\bm{B}_{\alpha}^{\dagger}(t)\bm{B}_{\alpha'}(t-s) \varrho_{\mathsf{B}}\big]$. If we assume the thermal equilibrium state as $\varrho_{\mathsf{B}}=e^{-\beta H_{\mathsf{B}}}/ Z_{\mathsf{B}}$, with $\beta =1/k_{B}T$ and $Z_{\mathsf{B}}=\mathrm{Tr}[e^{-\beta H_{\mathsf{B}}}]$, then one can see that the bath correlation functions become homogeneous in time,
\begin{align}
\left\langle \bm{B}_{\alpha}^{\dagger}(t)\bm{B}_{\alpha'}(t-s)\right\rangle =\left\langle \bm{B}_{\alpha}^{\dagger}(s) \bm{B}_{\alpha'}(0)\right\rangle.
\label{e24}
\end{align}
Thus, we obtain
\begin{align}
\dfrac{d\bm{\varrho}_{\mathsf{S}} (t)}{dt}=&\dfrac{1}{\hbar^{2}} \textstyle{\sum_{\alpha ,\alpha'}}\sum_{{\omega},{\omega'}}\sum_{q,q'} \Gamma_{\alpha \alpha'}(\omega +q\Omega) e^{it\left[ (\omega -\omega')+(q-q')\Omega\right] }\nonumber\\
& \times \left[ S_{\alpha'}(q,\omega)\bm{\varrho}_{\mathsf{S}}(t)S_{\alpha}^{\dagger}(q',\omega')- S_{\alpha}^{\dagger}(q',\omega')S_{\alpha'}(q,\omega)\bm{\varrho}_{\mathsf{S}}(t)\right]+ \mathrm{h.c.}
\label{e25}
\end{align}

If the time scale $\mathpzc{T}_{\mathsf{S}}=\max_{\substack{\omega\neq\omega', m}} |\omega -\omega'+m\Omega|^{-1}$ is considerably small compared to the relaxation time scale of the driven system ($\mathpzc{T}_{R}$), i.e., $\max_{\substack{\omega\neq\omega', m}} |\omega -\omega'+m\Omega|^{-1}\ll \mathpzc{T}_{R}$, then in the summation $\sum_{\alpha,\alpha'}$ and $\sum_{\omega,\omega'}$ only the terms for which $\omega =\omega'$ and $q=q'$ will have a nonvanishing contribution. This condition is called the \textit{secular approximation}, and gives
\begin{align}
\dfrac{d\bm{\varrho}_{\mathsf{S}} (t)}{dt}=&\dfrac{1}{\hbar^{2}} \textstyle{\sum_{\alpha ,\alpha'}} \sum_{{\omega}}\sum_{q} \Gamma_{\alpha \alpha'}(\omega +q\Omega) 
\big[ S_{\alpha'}(q,\omega)\bm{\varrho}_{\mathsf{S}}(t)S_{\alpha}^{\dagger}(q,\omega)- S_{\alpha}^{\dagger}(q,\omega)S_{\alpha'}(q,\omega)\bm{\varrho}_{\mathsf{S}}(t)\big]+ \mathrm{h.c.}
\label{e26}
\end{align}

We can decompose $\Gamma_{\alpha \alpha'}(x)$ to a real part and an imaginary part as follows:
\begin{align}
\Gamma_{\alpha \alpha'}(x)=\dfrac{1}{2}\gamma_{\alpha \alpha'}(x)+i\xi_{\alpha \alpha'}(x).
\label{e28}
\end{align}

By substituting the above expression in Eq. \eqref{e26}, we obtain the master equation of the system in the interaction picture as follows:
\begin{align}
\dfrac{d\bm{\varrho}_{\mathsf{S}} (t)}{dt}=-\dfrac{i}{\hbar}\big[ H_{\mathrm{lamb}} ^{(\mathrm{F})},\bm{\varrho}_{\mathsf{S}}(t)\big] +\mathpzc{D} ^{(\mathrm{F})}\left[ \bm{\varrho}_{\mathsf{S}}(t)\right] ,
\label{ee29}
\end{align}
where 
\begin{gather}
H_{\mathrm{lamb}}^{(\mathrm{F})}= \dfrac{1}{\hbar}\textstyle{\sum_{\alpha ,\alpha'}}\sum_{{\omega}}\sum_{q} \xi_{\alpha \alpha'}(\omega +q\Omega) S_{\alpha}^{\dagger}(q,\omega )S_{\alpha'}(q,\omega) ,
\label{e30}\\
\mathpzc{D}^{(\mathrm{F})}[\bm{\varrho}_{\mathsf{S}}(t)]=\dfrac{1}{\hbar^{2}} \textstyle{\sum_{\alpha ,\alpha'}} \sum_{{\omega}}\sum_{q} \gamma_{\alpha \alpha'}(\omega +q\Omega)\big[ S_{\alpha'}(q,\omega)\bm{\varrho}_{\mathsf{S}}(t)S^{\dagger}_{\alpha}(q,\omega)-\dfrac{1}{2}\big\lbrace S_{\alpha}^{\dagger}(q,\omega)S_{\alpha'}(q,\omega),\bm{\varrho}_{\mathsf{S}}(t)\big\rbrace \big].
\label{e31}
\end{gather}
Here $H_{\mathrm{lamb}}^{(\mathrm{F})}$ is referred to  as the Lamb-shift Hamiltonian, which is due to the interaction between the system and the bath. The second term on the RHS of Eq. \eqref{e29} $\left( \mathpzc{D}^{(\mathrm{F})}[\bm{\varrho}_{\mathsf{S}}(t)] \right)$ is a decoherence term induced by the bath. 

For the sake of comparison, we close this appendix with recalling the standard Lindblad master equation for a \textit{time-independent} system in the interaction picture,
\begin{align}
\dfrac{d\bm{\varrho}_{\mathsf{S}} (t)}{dt}=-\dfrac{i}{\hbar}\left[ H_{\mathrm{lamb}},\bm{\varrho}_{\mathsf{S}}(t)\right] +\mathpzc{D}\left[ \bm{\varrho}_{\mathsf{S}}(t)\right] ,
\label{ea113}
\end{align}
where 
\begin{gather}
H_{\mathrm{lamb}}=  \textstyle{\sum_{\alpha ,\alpha'}} \sum_{{\omega}} \dfrac{1}{\hbar}\xi_{\alpha \alpha'}(\omega) S_{\alpha}^{\dagger}(\omega )S_{\alpha'}(\omega),
\label{e114} \\
\mathpzc{D}[\bm{\varrho}_{\mathsf{S}}(t)]= \dfrac{1}{\hbar^{2}} \textstyle{\sum_{\alpha ,\alpha'}} \sum_{{\omega}} \gamma_{\alpha \alpha'}(\omega)
\big[ S_{\alpha'}(\omega)\bm{\varrho}_{\mathsf{S}}(t)S^{\dagger}_{\alpha}(\omega)-\dfrac{1}{2}\big\lbrace S_{\alpha}^{\dagger}(\omega)S_{\alpha'}(\omega),\bm{\varrho}_{\mathsf{S}}(t)\big\rbrace \big].
\label{e115}
\end{gather}
Unlike the time-dependent case, here the Lindblad operators $S_{\alpha}(\omega)$ are obtained according to the system Hamiltonian, i.e., $H_{\mathsf{S}}=H_{0}$. In addition, here the $\omega$'s are the gaps of the system Hamiltonian. Note that the $\gamma_{\alpha \alpha'}(\omega)$ and $\xi_{\alpha \alpha'}(\omega)$ coefficients are obtained by Eq. \eqref{e28}, too.

\section{Hamiltonian of the charge-field interaction}
\label{sec:lm}

Consider a particle of charge $q$ and mass $m$ (an atom or a molecule) localized in the origin and coupled to a (quantized) electromagnetic radiation field. Let us also assume the long-wavelength approximation, which neglects the spatial variation of the electromagnetic field on the length scale of the charged particle, and the Coulomb gauge for the electromagnetic field. Based on elementary physical principles the total charge-field Hamiltonian is given by \cite{book:Cohen-Tannoudji-QED} 
\begin{align}
H_{\mathrm{tot}}=\dfrac{1}{2m}[\vec{p}-q\vec{A}(0)]^{2}+V_{\mathrm{Coul}}(\vec{r}) + \textstyle{\sum_{j}} \hbar \omega_{j} (\hat{a}^{\dagger}_{j} \hat{a}_{j} +1/2), 
\label{q1}
\end{align}
where $\vec{p}$ is the momentum operator of the charged particle, $\vec{A}$ is the vector potential operator associated with the electromagnetic field, $V_{\mathrm{Coul}}$ is the Coulomb potential, $\{\omega_{j}\}$ are the quantized frequencies or modes of the field, and $\hat{a}_{j}$ is the annihilation operator corresponding to mode $\omega_{j}$ of the field. This form of the Hamiltonian is usually referred to as the ``$\vec{A}\cdot \vec{p}$'' representation. One can rewrite this Hamiltonian such that wherein three separate terms can be discerned, 
\begin{align}
H_{\mathrm{tot}}=H_{\mathsf{P}}+H_{\mathsf{R}}+H_{I},
\label{q2}
\end{align}
where
\begin{gather}
H_{\mathsf{P}}=\dfrac{\vec{p}^2}{2m}+V_{\mathrm{Coul}},
\label{q3}\\
H_{\mathsf{R}}= \textstyle{\sum_{j}} \hbar \omega_{j} (\hat{a}^{\dagger}_{j} \hat{a}_{j} +1/2),
\label{q4}\\
H_{I}=H_{I1}+H_{I2},
\label{q5}
\end{gather}
are the Hamiltonians associated to the particle, the radiation field, and the interaction, respectively. Here $H_{I1}$ is linear in terms of the field $\vec{A}$ and $H_{I2}$ is quadratic, 
\begin{gather}
H_{I1}=-\dfrac{q}{m}\vec{p}\cdot\vec{A}(0),
\label{q6}\\
H_{I2}=\dfrac{q^{2}}{2m}[\vec{A}(0)]^{2}.
\label{q7}
\end{gather}

For practical purposes, it might be more useful to transform the $\vec{A}\cdot \vec{p}$ representation to another equivalent ``$\vec{E}\cdot\vec{r}$'' representation which explicitly depends on the associated electric field $\vec{E}$ rather than the vector potential $\vec{A}$. The transformation between these two representations is carried out by the unitary operator $T=e^{-(i/\hbar)q \vec{r}\cdot \vec{A}(0)}$, which yields 
\begin{align}
H'_{\mathrm{tot}}&=TH_{\mathrm{tot}}T^{\dagger} \nonumber\\
&=\dfrac{\vec{p}^2}{2m}+V_{\mathrm{Coul}}+ \textstyle{\sum_{j}} \hbar \omega_{j} (\hat{a}^{\dagger}_{j} \hat{a}_{j} +1/2)-\vec{d}\cdot\vec{E}(0) +\varepsilon_{\mathrm{dipole}}, \label{q8}
\end{align}
where $\varepsilon_{\mathrm{dipole}}$ is the dipole self-energy related to the charged particle Hamiltonian (which is usually ignored in the literature), and the fourth term is the expected dipole interaction, where $\vec{d}=q\vec{r}$ is the dipole operator of the charged particle. It is seen that another appeal of the $\vec{E}\cdot \vec{r}$ representation is that in this representation the Hamiltonian only depends on the electric field linearly. One can note that the charge-field Hamiltonian (\ref{q8}) can be recast in the well-known form
\begin{equation}
H'_{\mathrm{tot}} = H_{\mathsf{P}} + H_{\mathsf{R}} -\vec{d}\cdot\vec{E}(0),
\end{equation}
where the first, second, and last terms denote the particle, the reservoir, and the particle-reservoir dipole interaction, respectively. In this paper we work in the $\vec{E}\cdot\vec{r}$ representation and simply denote the associated Hamiltonian by $H_{\mathrm{tot}}$ (without prime).

\begin{remark}
\label{remark-}
In some physical applications, one may assume the low-intensity approximation which corresponds to ignoring the $O(\vec{A}^{2})$ term from the Hamiltonian (\ref{q1}). In the $\vec{E}\cdot\vec{r}$ representation this approximation leads to the subtraction of $H_{I2}$ from the Hamiltonian (\ref{q8}).
\end{remark}

\section{Floquet Hamiltonian and periodic operator for a driven system}
\label{sec:analytic}

\subsection{General analysis}

As it was shown in the Born-Markov and secular approximations, the evolution of a driven system with a periodic external field is governed by Eq. \eqref{e29}, and for using this equation we should calculate the Floquet Hamiltonian. Thus, we present a way to obtain the Floquet Hamiltonian approximately. 

On the one hand, we have 
\begin{equation}
U_{\mathsf{S}}(t,0)=\mathbbmss{T}e^{-(i/\hbar)\int_{0}^{t}ds\, H_{\mathsf{S}}(s)},
\label{ec1}
\end{equation}
where $H_{\mathsf{S}}(t)=H_{0}+V(t)$ and $\mathbbmss{T}$ denotes time-ordering. On the other hand, from the Floquet theory, we have 
\begin{align}
U_{\mathsf{S}}(\tau,0)=e^{-i\bar{H}\tau/\hbar}.
\label{e40}
\end{align}

According to Eq. \eqref{e40} to obtain the Floquet Hamiltonian $\bar{H}$ it is better that we have the evolution operator in the exponential form, to do this notice the following equation which relates the evolution operator in the both Schr\"{o}dinger and interaction picture together
\begin{align}
U_{\mathsf{S}}(t,0)=U_{0}(t,0) \bm{U}_{I}(t,0) ,
\label{e42}
\end{align}
where $U_{0}(t,0)=e^{-iH_{0}t/\hbar}$ and $\bm{U}_{I}(t,0)$ is the evolution operator in the interaction picture which evolves as
\begin{align}
\dfrac{d\bm{U}_{I}(t,0)}{dt}=-\dfrac{i}{\hbar} \bm{V}_{I}(t)\bm{U}_{I}(t,0),
\label{e43}
\end{align}
where $\bm{V}_{I}(t)=e^{iH_{0}t/\hbar}V(t)e^{-iH_{0}t/\hbar}$. By straightforward calculation the operator $\bm{V}_{I}(t)$ takes the following form for the operator defined as Eq. \eqref{e39} [see Fig. \ref{fig1}]
\begin{align}
\bm{V}_{I}(t)=\mu \cos(\Omega t) \left[ e^{-it\omega_{b0}}\vert 0\rangle\langle b\vert +e^{it\omega_{b0}}\vert b\rangle\langle 0\vert\right], 
\label{e44}
\end{align}
where $\omega_{b0}\equiv (\epsilon_{b}-\epsilon_{0})/\hbar$ is the gap of the two states which are couple together through the external field with strength $\mu$. Thus the RHS of the differential Eq. \eqref{e43} is proportional to $\mu$ and we can apply the Magnus expansion to obtain the $\bm{U}_{I}(t,0)$ as follows:
\begin{align}
\bm{U}_{I}(t,0)=e^{\bm{\Lambda} (t)}, 
\label{e45}
\end{align}
where the operator $\bm{\Lambda} (t)$ obtains by the following series expansion:
\begin{align}
\bm{\Lambda} (t)= \textstyle{\sum_{n=1}^{\infty}} ( -i\mu/\hbar)^{n}\bm{\Lambda}_{n}(t).
\label{e46}
\end{align}
The three first terms of the series Eq. \eqref{e46} is as follows \citep{blanes_magnus_2009}:
\begin{align}
& \bm{\Lambda}_{1}(t)=\textstyle{\int_{0}^{t}} dt_{1}\,\bm{V}(t_{1}),\nonumber\\
& \bm{\Lambda}_{2}(t)= (1/2!)\textstyle{\int_{0}^{t}} dt_{1}\, \int_{0}^{t_{1}} dt_{2}\left[ \bm{V}(t_{1}),\bm{V}(t_{2})\right] , \nonumber\\
& \bm{\Lambda}_{3}(t)=(1/3!)\textstyle{\int_{0}^{t}} dt_{1}\, \int_{0}^{t_{1}} dt_{2}\int_{0}^{t_{2}}dt_{3} \big\{ \left[\bm{V}(t_{1}), \left[\bm{V}(t_{2}),\bm{V}(t_{3})\right] \right]+ \left[ \left[ \bm{V}(t_{1}),\bm{V}(t_{2})\right] ,\bm{V}(t_{3})\right] \big\},
\label{e47}
\end{align}
where $\bm{V}(t)=\cos(\Omega t)\big( e^{-it\omega_{b0}}\vert 0\rangle\langle b\vert +e^{it\omega_{b0}}\vert b\rangle\langle 0\vert\big)$. Here we keep $\bm{U}_{I}(t)$ up to $O(\mu^{3})$. By direct calculation the evolution operator in the interaction picture is obtained as
\begin{align}
\bm{U}_{I}(t,0)=e^{ x(t)\vert 0\rangle\langle b\vert +y(t)\vert b\rangle\langle 0\vert +h(t)\vert 0\rangle\langle 0\vert -h(t)\vert b\rangle\langle b\vert},
\label{e48}
\end{align}
where
\begin{gather}
x(t)\equiv g(t)+a(t)+b(t)+c(t)+d(t)+e(t)+f(t), \nonumber\\
y(t)\equiv -[g^{*}(t)+a^{*}(t)+b^{*}(t)-c(t)+d^{*}(t)+e^{*}(t)+f^{*}(t)], \nonumber\\
h(t)\equiv \dfrac{\mu^{2}}{\hbar^{2}}\dfrac{i \omega_{b0}}{\omega^{2}_{b0}-\Omega^{2}} \Big[ \dfrac{\Omega \cos (\omega_{b0}t) \sin (\Omega t)-\omega_{b0}\cos (\Omega t) \sin (\omega_{b0}t)}{\omega^{2}_{b0}-\Omega^{2}}+\dfrac{2t\Omega +\sin (2t\Omega)}{4\Omega}\Big] , \nonumber\\
\label{e49}
\end{gather}
and the functions $a(t)$, $b(t)$, $c(t)$, $d(t)$, $e(t)$, $f(t)$, and $g(t)$ are defined as follows:
\begin{align}
a(t)\equiv& \dfrac{\mu^{3}\omega_{b0}}{12 \hbar^{3}\Omega(\Omega^{2}-\omega^{2}_{b0})} \Big[ \dfrac{4\Omega^{2}\sin (t\Omega) e^{-it\omega_{b0}}\left( t\Omega^{2}+2i\omega_{b0}-t\omega^{2}_{b0} \right)-4\Omega \left(\Omega^{2}+\omega^{2}_{b0}\right)}{(\Omega^{2}-\omega^{2}_{b0})^{2}} \nonumber\\
&+\dfrac{e^{-it\omega_{b0}}\left( \Omega \cos (\Omega t)+i\omega_{b0}\sin (\Omega t)\right)-\Omega}{\omega_{b0}^{2}-\Omega^{2}}+\dfrac{e^{-it\omega_{b0}}\big(3\Omega \cos (3\Omega t)+i\omega_{b0}\sin (3\Omega t)\big)-3\Omega}{\omega_{b0}^{2}-9\Omega^{2}}\nonumber\\
&+ \dfrac{4\Omega e^{-it\omega_{b0}}\cos (t\Omega )\left(\Omega^{2}-it\Omega^{2}\omega_{b0}+\omega^{2}_{b0}+it\omega_{b0}^{3}\right)}{(\Omega^{2}-\omega^{2}_{b0})^{2}} \Big] ,\\
 b(t)\equiv & -\dfrac{2\mu^{3}\omega_{b0}}{3\hbar^{3}(\Omega^{2}-\omega^{2}_{b0})^{2}}\times \dfrac{e^{-2it\omega_{b0}}}{8\Omega(\omega^{2}_{b0}-\Omega^{2})} \Big[ \Omega \cos (2t\Omega )\left( \Omega^{2}(1+e^{2it\omega_{b0}})-\omega^{2}_{b0}(e^{2it\omega_{b0}}-1)\right) \nonumber\\
& -\Omega^{3}\left( e^{2it\omega_{b0}}(1+2it\omega_{b0})+1\right) +\Omega \omega^{2}_{b0}\left( 1+e^{2it\omega_{b0}}(2it\omega_{b0}-1)\right) \nonumber\\
&+i\omega_{b0}\sin (2t\Omega )\left( \Omega^{2}(2-e^{2it\omega_{b0}})+\omega_{b0}^{2}e^{2it\omega_{b0}}\right) \Big] , \nonumber \\
c(t) \equiv &\dfrac{2i\mu^{3}\omega_{b0}^{2}}{3\hbar^{3}(\Omega^{2}-\omega^{2}_{b0})^{2}}\Big[ \dfrac{\Omega \cos (t\omega_{b0})\sin (t\Omega) -\omega_{b0}\cos (t\Omega) \sin (t\omega_{b0})}{\Omega^{2}-\omega^{2}_{b0}}-\dfrac{2t\Omega +\sin (2t\Omega)}{4\Omega}\Big] , \nonumber \\
d(t)\equiv& \dfrac{i\mu^{3}}{12 \hbar^{3}}\dfrac{\Omega^{2}-2\omega^{2}_{b0}}{(\Omega^{2}-\omega^{2}_{b0})^{3}} \left[ -i\omega_{b0}+e^{it\omega_{b0}}\big( i\omega_{b0}\cos (t\Omega )+\Omega \sin (t\Omega )\big) \right] ,\\
e(t)\equiv& \dfrac{i\mu^{3}}{12 \hbar^{3}\Omega(\Omega^{2}-\omega^{2}_{b0})^{2}(9\Omega^{4}-10\Omega^{2}\omega^{2}_{b0}+\omega^{4}_{b0})} \Big[ i\Omega \omega_{b0}(9\Omega^{2}-\omega^{2}_{b0})(\omega_{2}+4\omega^{2}_{b0})\nonumber \\
& +\sin (t\Omega) \Bigl(-\omega_{b0}^{6}-2\Omega^{2}\omega_{b0}^{4}(1+it\omega_{b0})+4\Omega^{4}\omega_{b0}^{2}(13+5it\omega_{b0})-9\Omega^{6}(1+2it\omega_{b0}) \nonumber\\
& -\omega_{b0}^{2}(\omega_{b0}^{4}-3\Omega^{4}+2\Omega^{2}\omega_{b0}^{2})\Bigr)+2i\Omega \omega^{3}_{b0}(\Omega^{2}-\omega^{2}_{b0})-2i\Omega \omega^{3}_{b0}\cos (3t\Omega) (\Omega^{2}-\omega^{2}_{b0})\nonumber\\
&+e^{-it\omega_{b0}} \bigl(-\Omega \omega_{b0} \cos (t\Omega) (9\Omega^{2}-\omega^{2}_{b0})(-2\omega^{2}_{b0}(-2i+t\omega_{b0})+\Omega^{2}(i+2t\omega_{b0})\bigr) \Big],\\
f(t)\equiv & \dfrac{-\mu^{3}e^{-it\omega_{b0}}}{24\hbar^{3}(\Omega^{2}-\omega^{2}_{b0})^{2}(9\Omega^{4}-10\Omega^{2}\omega_{b0}^{2}+\omega_{b0}^{4})} \Big[\cos (3t\Omega) \left( \omega_{b0}^{5}-3\Omega^{4}\omega_{b0}+2\Omega^{2}\omega_{b0}^{3}\right) \nonumber\\
& +2\Omega \Bigl(-8\Omega \omega_{b0} e^{it\omega_{b0}}(\omega_{b0}^{2}-3\Omega^{2})-i\Omega^{2}\sin (t\Omega) (e^{2it\omega_{b0}}-1)(\omega_{b0}^{2}-9\Omega^{2}) \nonumber\\
& +2i\omega_{b0}^{2}\sin(3t\Omega) (\omega_{b0}^{2}-\Omega^{2})\Bigr)-\omega_{b0}\cos (t\Omega)\, (\omega_{b0}^{2}-9\omega_{2})\bigl( \omega_{b0}^{2}-\Omega^{2}(3+2 e^{2it\omega_{b0}})\bigr)
\Big],\nonumber \\
g(t)\equiv& \dfrac{-i\mu}{\hbar}\left[ -i\omega_{b0}+ie^{-it\omega_{b0}} \big(\omega_{b0}\cos(t\Omega) +i\Omega \sin(t\Omega)\big)\right] .
\label{e50}
\end{align}

Now we want to calculate the evolution operator in the Schr\"{o}dinger picture by applying the Baker-Campbell-Hausdorff (BCH) formula to the RHS of Eq. \eqref{e42}. By keeping only first twelve terms of this formula, we obtain the operator $\mathpzc{E}$ as \cite{hall_lie_2015}
\begin{align}
\mathpzc{E}(t)=&\log[ e^{\Theta(t)} e^{\bm{\Lambda}(t)}] \approx \Theta(t) +\bm{\Lambda}(t) +\dfrac{1}{2}\left[ \Theta(t) ,\bm{\Lambda}(t) \right]\nonumber\\
&+\dfrac{1}{12}\left( \left[ \Theta(t) ,\left[ \Theta(t) ,\bm{\Lambda}(t) \right] \right] +\left[ \bm{\Lambda}(t) ,\left[ \bm{\Lambda}(t) ,\Theta(t) \right] \right] \right)-\dfrac{1}{24}\left[ \bm{\Lambda}(t) ,\left[ \Theta(t) ,\left[ \Theta(t) ,\bm{\Lambda}(t) \right] \right] \right] \nonumber\\
&-\dfrac{1}{720}\left( \left[ \bm{\Lambda}(t) ,\left[ \bm{\Lambda}(t) ,\left[ \bm{\Lambda}(t) ,\left[ \bm{\Lambda}(t) ,\Theta(t) \right] \right] \right] \right] +\left[ \Theta(t) ,\left[ \Theta(t) ,\left[ \Theta(t) ,\left[ \Theta(t) ,\bm{\Lambda}(t) \right] \right] \right] \right]\right) \nonumber\\
&+\dfrac{1}{360}\left( \left[ \Theta(t) ,\left[ \bm{\Lambda}(t) ,\left[ \bm{\Lambda}(t) ,\left[ \bm{\Lambda}(t) ,\Theta(t) \right] \right] \right] \right] +\left[ \bm{\Lambda}(t) ,\left[ \Theta(t) ,\left[ \Theta(t) ,\left[ \Theta(t) ,\bm{\Lambda}(t) \right] \right] \right] \right]\right) \nonumber\\
&+\dfrac{1}{120}\left( \left[ \bm{\Lambda}(t) ,\left[ \Theta(t) ,\left[ \bm{\Lambda}(t) ,\left[ \Theta(t) ,\bm{\Lambda}(t) \right] \right] \right] \right] +\left[ \Theta(t) ,\left[ \bm{\Lambda}(t) ,\left[ \Theta(t) ,\left[ \bm{\Lambda}(t) ,\Theta(t) \right] \right] \right] \right]\right) + \ldots,
\label{e51}
\end{align}
where
\begin{align}
& \Theta(t)\equiv (-it/\hbar)H_{0}=-it\big(\omega_{0}\vert 0\rangle\langle 0\vert + \omega_{b}\vert b\rangle\langle b\vert +\textstyle{\sum_{j=1}^{n-2}} \omega_{j}\vert j\rangle\langle j\vert\big), \nonumber\\
& \bm{\Lambda} (t) \equiv x(t)\vert 0\rangle\langle b\vert +y(t)\vert b\rangle\langle 0\vert +h(t)\vert 0\rangle\langle 0\vert -h(t)\vert b\rangle\langle b\vert,
\label{e52} \\
&\omega_{j}\equiv \epsilon_{j}/\hbar,\nonumber\\
&U_{\mathsf{S}}(t,0) = e^{\mathpzc{E}(t)}. \label{expon}
\end{align}
Equation \eqref{e52} is related for $n$-level atoms. In the following a general expression for the evolution operator in the Schr\"{o}dinger picture for an $n$-level system which its two levels are coupled together via an external field is obtained,
\begin{align}
\mathpzc{E}(t)=u_{00}(t)\vert 0\rangle\langle 0\vert +u_{0b}(t)\vert 0\rangle\langle b\vert +u_{b0}(t)\vert b\rangle\langle 0\vert +u_{bb}(t)\vert b\rangle\langle b\vert -it \textstyle{\sum_{j=1}^{n-2}} \omega_{j}\vert j\rangle\langle j\vert .
\label{e53}
\end{align}
where the functions $u_{00}(t)$, $u_{0b}(t)$, $u_{b0}(t)$ and $u_{bb}(t)$ have been obtained by BCH expansion as follows:
\begin{align}
u_{00}(t)\equiv & -it\omega_{0}+h(t)+\dfrac{i}{6}tx(t)y(t)\omega_{b0}-\dfrac{i}{90}tx(t)y(t)\omega_{b0}\bigl(h(t)^{2}+x(t)y(t)\bigr) \nonumber\\
&+\dfrac{i}{180}t^{3}\omega_{b0}^{3}x(t)y(t)+\dfrac{1}{30}t^{2}\omega_{b0}^{2}h(t)x(t)y(t) \nonumber\\
u_{0b}(t)\equiv & x(t)+\dfrac{i}{2}tx(t)\omega_{b0}-\dfrac{1}{12}\omega_{b0}^{2}t^{2}x(t)-\dfrac{i}{6}tx(t)h(t)\omega_{b0}+\dfrac{1}{12}h(t)\omega_{b0}^{2}t^{2}x(t)\nonumber\\
&+\dfrac{i}{90}tx(t)h(t)\omega_{b0}\bigl(h^{2}(t)+x(t)y(t)\bigr)-\dfrac{1}{720} t^{4}\omega_{b0}^{4}x(t)+\dfrac{i}{60}t^{3}\omega_{b0}^{3}x(t)h(t)\nonumber\\
&-\dfrac{i}{180}t^{3}\omega_{b0}^{3}x(t)h(t)-\dfrac{1}{30}t^{2}\omega_{b0}^{2}x(t)h^{2}(t)+\dfrac{1}{90}t^{2}\omega_{b0}^{2}x(t)\bigl(h^{2}(t)+x(t)y(t)\bigr)\nonumber\\
u_{b0}(t)\equiv& y(t)-\dfrac{i}{2}ty(t)\omega_{b0}-\dfrac{1}{12}\omega_{b0}^{2}t^{2}y(t)-\dfrac{i}{6}ty(t)h(t)\omega_{b0}-\dfrac{1}{12}h(t)\omega_{b0}^{2}t^{2}y(t)\nonumber\\
&+\dfrac{i}{90}ty(t)h(t)\omega_{b0}\bigl(h^{2}(t)+x(t)y(t)\bigr)-\dfrac{1}{720} t^{4}\omega_{b0}^{4}y(t)+\dfrac{i}{60}t^{3}\omega_{b0}^{3}y(t)h(t)\nonumber\\
&-\dfrac{i}{180}t^{3}\omega_{b0}^{3}y(t)h(t)-\dfrac{1}{30}t^{2}\omega_{b0}^{2}y(t)h^{2}(t)+\dfrac{1}{90}t^{2}\omega_{b0}^{2}y(t)\bigl(h^{2}(t)+x(t)y(t)\bigr)\nonumber\\
u_{bb}(t)\equiv & -it\omega_{\mathsf{B}}-h(t)-\dfrac{i}{6}tx(t)y(t)\omega_{b0}+\dfrac{i}{90}tx(t)y(t)\omega_{b0}\bigl(h(t)^{2}+x(t)y(t)\bigr) \nonumber\\
&-\dfrac{i}{180}t^{3}\omega_{b0}^{3}x(t)y(t)-\dfrac{1}{30}t^{2}\omega_{b0}^{2}h(t)x(t)y(t) .
\label{e54}
\end{align}
Thus, from Eq. \eqref{e40} the Floquet Hamiltonian is
\begin{align}
\bar{H}=(i\hbar/\tau)\mathpzc{E}(\tau),
\label{e55}
\end{align}
from which we can also obtain the eigensystem $\{\bar{\epsilon},|\bar{\epsilon}\rangle\}$. By having the Floquet Hamiltonian and using Eq. \eqref{e11} with $t_{0}=0$ the periodic operator $ P (t,t_{0})$ is as follows:
\begin{align}
P(t,0)=U_{\mathsf{S}}(t,0) e^{ it\bar{H}/\hbar},
\label{e57}
\end{align}
where $U_{\mathsf{S}}(t,0)$ is obtained by solving Eq. \eqref{ec1} numerically. To construct the master equation, the next step is to calculate the correlation functions of the baths. In doing so we should rewrite the interaction Hamiltonian between the system and the hot and the cold baths [Eqs. \eqref{e36} and (\ref{e37})] as in the following:
\begin{align}
H_{\mathsf{S}\mathsf{B}_{\mathsf{h}}} &=\sigma_{x}^{(\mathsf{h})} \otimes \textstyle{\sum_{k}} f_{k}(\hat{a}_{k}+\hat{a}_{k}^{\dagger})+\sigma_{y}^{(\mathsf{h})} \otimes \textstyle{\sum_{k}} if_{k}(\hat{a}_{k}-\hat{a}_{k}^{\dagger}) ,
\label{e64}\\
H_{\mathsf{S}\mathsf{B}_{\mathsf{c}}} &=\sigma_{x}^{(\mathsf{c})} \otimes \textstyle{\sum_{q}} g_{q}(\hat{b}_{q}+\hat{b}_{q}^{\dagger})+\sigma_{y}^{(\mathsf{c})} \otimes \textstyle{\sum_{q}} ig_{q}(\hat{b}_{q}-\hat{b}_{q}^{\dagger}) ,
\label{e66}
\end{align}
where
\begin{align}
& \sigma_{x}^{(\mathsf{h})}\equiv \dfrac{1}{2}\left( \vert 0\rangle\langle 1\vert +\vert 1\rangle\langle 0\vert \right) \qquad B_{1}^{(\mathsf{h})}= \textstyle{\sum_{k}} f_{k} \big(\hat{a}_{k}+\hat{a}_{k}^{\dagger}\big) , \nonumber\\
& \sigma_{y}^{(\mathsf{h})}\equiv \dfrac{i}{2}\left( \vert 0\rangle\langle 1\vert -\vert 1\rangle\langle 0\vert \right) \qquad B_{2}^{(\mathsf{h})}= \textstyle{\sum_{k}}if_{k}\big(\hat{a}_{k}-\hat{a}_{k}^{\dagger}\big),
\label{e65}\\
& \sigma_{x}^{(\mathsf{c})}\equiv \dfrac{1}{2}\left( \vert b\rangle\langle 1\vert +\vert 1\rangle\langle b\vert \right) \qquad B_{1}^{(\mathsf{c})}= \textstyle{\sum_{q}} g_{q} \big(\hat{b}_{q}+\hat{b}_{q}^{\dagger}\big) ,\nonumber\\
& \sigma_{y}^{(\mathsf{c})}\equiv \dfrac{i}{2}\left( \vert b\rangle\langle 1\vert -\vert 1\rangle\langle b\vert \right) \qquad B_{2}^{(\mathsf{c})}= \textstyle{\sum_{q}} ig_{q} \big(\hat{b}_{q}-\hat{b}_{q}^{\dagger}\big).
\label{e67}
\end{align}

According to Eq. \eqref{e23}, we also need to calculate the bath operators in the interaction picture, $\bm{B}(t)=e^{iH_{\mathsf{B}}t/\hbar}Be^{-iH_{\mathsf{B}}t/\hbar}$, which are obtained as
\begin{align}
& \bm{B}_{1}^{(\mathsf{h})}(t)= \textstyle{\sum_{k}} f_{k}\big( e^{-it\zeta_{k}} \hat{a}_{k} + e^{it\zeta_{k}} \hat{a}_{k}^{\dagger}\big) ,\nonumber\\
& \bm{B}_{2}^{(\mathsf{h})}(t)= \textstyle{\sum_{k}} if_{k}\big( e^{-it\zeta_{k}} \hat{a}_{k} - e^{it\zeta_{k}} \hat{a}_{k}^{\dagger}\big),
\label{e69}\\
& \bm{B}_{1}^{(\mathsf{c})}(t)=\textstyle{\sum_{q}} g_{q}\big( e^{-it\nu_{q}} \hat{b}_{q} + e^{it\nu_{q}} \hat{b}_{q}^{\dagger}\big) , \nonumber\\
& \bm{B}_{2}^{(\mathsf{c})}(t)= \textstyle{\sum_{q}} ig_{q}\big( e^{-it\nu_{q}} \hat{b}_{q} - e^{it\nu_{q}} \hat{b}_{q}^{\dagger}\big) .
\label{e70}
\end{align}
All $\Gamma_{ij}(x)$ coefficients can be obtained similarly. For example, 
\begin{align}
\Gamma^{(\mathsf{h})}_{11}(x)&= \textstyle{\int_{0}^{\infty}} ds\,e^{-isx} \big\langle \bm{B}^{(\mathsf{h})\dagger}_{1}(s) \bm{B}^{(\mathsf{h})}_{1}(0)\big\rangle =\int_{0}^{\infty}  ds\,e^{-isx} \textstyle{\sum_{k}} f_{k}^{2}\left[ \bar{n}(\zeta_{k}, \beta) e^{is\zeta_{k}} + [\bar{n}(\zeta_{k}, \beta)+1] e^{-is\zeta_{k}}\right] \nonumber\\
&= \textstyle{\sum_{k}} f_{k}^{2}\left\lbrace \bar{n}(\zeta_{k}, \beta)\int_{0}^{\infty}  ds\,e^{-i(x-\zeta_{k})s} +[ \bar{n}(\zeta_{k}, \beta)+1]\int_{0}^{\infty}  ds\,e^{-i(x+\zeta_{k})s}\right\rbrace \nonumber\\
&= \textstyle{\int_{\infty}^{\infty}} d\nu\, \textstyle{\sum_{k}} \delta (\nu -\zeta_{k}) f^{2}_{k}\left\lbrace \bar{n}(\nu ,\beta)\int_{0}^{\infty} ds\,e^{-i(x-\nu)s} +[\bar{n}(\nu ,\beta)+1]\int_{0}^{\infty} ds\,e^{-i(x+\nu)s}\right\rbrace \nonumber\\
&= \textstyle{\int_{-\infty}^{\infty}} d\nu\, J^{(\mathsf{h})}(\nu)\left\lbrace \bar{n}(\nu ,\beta)\left[ \pi \delta (x-\nu)-i\mathbbmss{P}\dfrac{1}{x-\nu}\right] + [\bar{n}(\nu ,\beta)+1]\left[ \pi \delta (x+\nu)-i\mathbbmss{P}\dfrac{1}{x+\nu}\right]\right\rbrace \nonumber\\
&=\pi \bar{n}(x ,\beta)\big[ J^{(\mathsf{h})}(x)-J^{(\mathsf{h})}(-x)\big] -i\mathbbmss{P}\textstyle{\int_{-\infty}^{\infty}} d\nu\, J^{(\mathsf{h})}(\nu)\left[ \dfrac{\bar{n}(\nu ,\beta)}{x-\nu}+\dfrac{\bar{n}(\nu ,\beta)+1}{x+\nu}\right] ,
\label{e71}
\end{align}
where we have introduced the spectral density function of the hot bath as
\begin{align}
J^{(\mathsf{h})}(x)= \textstyle{\sum_{k}} f^{2}_{k}\delta (x -\zeta_{k}),
\label{e72}
\end{align}
and also we have used the following formula:
\begin{align}
\textstyle{\int_{0}^{\infty}} ds\,e^{-i(x+\nu)s}=\pi \delta (x+\nu)-i\mathbbmss{P}\dfrac{1}{x+\nu} ,
\label{e73}
\end{align}
with $\mathbbmss{P}$ being the Cauchy principal value. Moreover, the following relations for the bosonic creation and annihilation operators have been used: 
\begin{gather}
\langle \hat{a}^{\dagger}_{k} \hat{a}_{l}\rangle =\delta_{kl} \, \bar{n}(\zeta_{k}, \beta),\\
\langle \hat{a}_{l} \hat{a}^{\dagger}_{k}\rangle =\delta_{kl}\, [\bar{n}(\zeta_{k}, \beta)+1],\\
\langle \hat{a}_{k} \hat{a}_{l}\rangle =0,\\ 
\langle \hat{a}_{k}^{\dagger} \hat{a}_{l}^{\dagger}\rangle =0,
\end{gather}
where $\beta$ is the inverse temperature of the bath. Finally according to Eq. \eqref{e28} by separating the real part from the imaginary part of the bath correlation function, we find the following expressions:
\begin{align}
& \gamma^{(\mathsf{h})}_{11}(x)=2\pi \bar{n}(x ,\beta) \big[ J^{(\mathsf{h})}(x)-J^{(\mathsf{h})}(-x)\big], \qquad \xi^{(\mathsf{h})}_{11}(x)=-\mathbbmss{P}\int_{-\infty}^{\infty} d\nu\, J^{(\mathsf{h})}({\nu}) \left[ \dfrac{\bar{n}(\nu ,\beta)}{x-\nu}+\dfrac{\bar{n}(\nu ,\beta)+1}{x+\nu}\right] ,\nonumber\\
& \gamma^{(\mathsf{h})}_{12}(x)=2 \mathbbmss{P}\int_{-\infty}^{\infty} d\nu J^{(\mathsf{h})}({\nu}) \left[ \dfrac{\bar{n}(\nu ,\beta)}{x-\nu}-\dfrac{\bar{n}(\nu ,\beta)+1}{x+\nu}\right], \qquad \xi^{(\mathsf{h})}_{12}(x)=\pi \bar{n}(x ,\beta)\left[ J^{(\mathsf{h})}(x)+J^{(\mathsf{h})}(-x)\right] , \nonumber\\
& \gamma^{(\mathsf{h})}_{21}(x)=2 \mathbbmss{P}\int_{-\infty}^{\infty} d\nu\, J^{(\mathsf{h})}({\nu}) \left[ \dfrac{\bar{n}(\nu ,\beta)+1}{x+\nu}-\dfrac{\bar{n}(\nu ,\beta)}{x-\nu}\right], \qquad \xi^{(\mathsf{h})}_{21}(x)=-\pi \bar{n}(x ,\beta)\left[ J^{(\mathsf{h})}(x)+J^{(\mathsf{h})}(-x)\right] , \nonumber\\
& \gamma^{(\mathsf{h})}_{22}(x)=2\pi \bar{n}(x ,\beta)\big[ J^{(\mathsf{h})}(x)-J^{(\mathsf{h})}(-x)\big], \qquad \xi^{(\mathsf{h})}_{22}(x)=-\mathbbmss{P}\int_{-\infty}^{\infty} d\nu\, J^{(\mathsf{h})}({\nu}) \left[ \dfrac{\bar{n}(\nu ,\beta)}{x-\nu}+\dfrac{\bar{n}(\nu ,\beta)+1}{x+\nu}\right] .
\label{e74}
\end{align}

For specificity, we choose the Ohmic spectral density function for the baths,
\begin{align}
J(x)=J_{0} x e^{-x^{2}/\omega^{2}_{\mathrm{cutoff}}},
\label{e75}
\end{align}
where $J_{0}$ is a model-dependent and also frequency-independent constant and $\omega_{\mathrm{cutoff}}$ is the bath cutoff frequency. Hence Eqs. (\ref{e74}) yield
\begin{align}
& \gamma_{ij}(x)=0,\qquad i\neq j \nonumber\\
& \gamma_{ii}(x)=
\begin{cases}
4\pi \bar{n}(x ,\beta) J(x) & \text{, $x\neq 0$} \\
4\pi J_{0}/(\hbar\beta) & \text{, $x=0$} , 
\end{cases}
\label{e76}\\
& \xi_{ij}(x)=0\qquad i\neq j \nonumber\\
& \xi_{ii}(x)=-2\mathbbmss{P} \textstyle{\int_{0}^{\infty}}  d\omega\,  G(x,\omega) ,
\label{e77}
\end{align}
where
\begin{align}
G(x,\omega)\equiv
\begin{cases}
J_{0}\omega e^{-\omega^{2}/\omega_{\mathrm{cutoff}}^{2}} \frac{(x-\omega) e^{\beta \hbar \omega}+x+\omega}{\left( e^{\beta \hbar \omega}-1\right)(x^{2}- \omega^2)} & \text{, $\omega \neq 0$} \\
2 J_{0}/(\hbar\beta x) & \text{, $\omega=0$} .
\end{cases}
\label{e78}
\end{align}

\begin{figure*}[tp]
\includegraphics[scale=0.19]{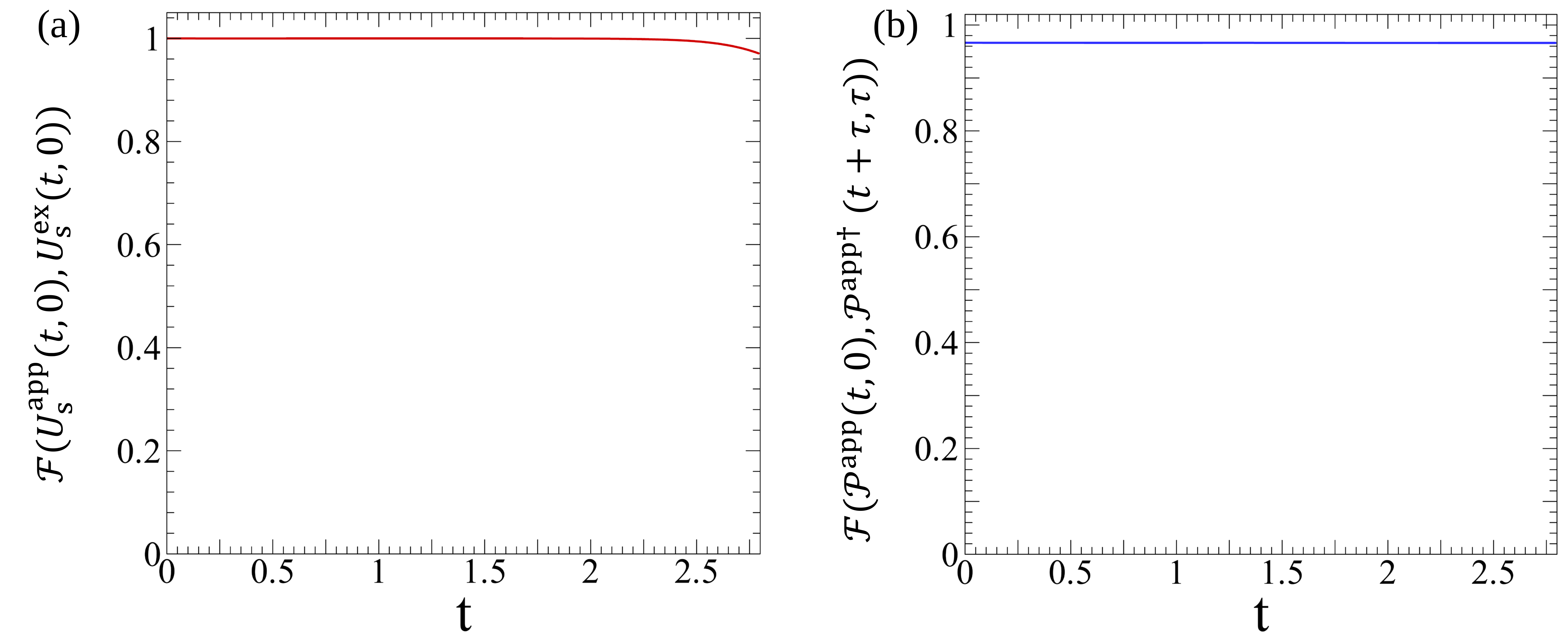}
\caption{(a) Fidelity between the exact and approximate solutions (Magnus + BCH expansions) of the evolution operator in the Schr\"{o}dinger picture and in the first period for the model discussed in Sec. \ref{sec:F-FL-results}. (b) Fidelity of $P(t,0)$ in two periods which demonstrates the periodic property of $P(t,0)$. All quantities are in natural units.}
\label{fig8}
\end{figure*}

\begin{remark}
If rather than taking the external field as in Eq. (\ref{e39}), we had chosen the field as
\begin{align}
V_{1}(t)=\mu \cos(\Omega t) \big( \vert 1\rangle\langle b\vert +\vert b\rangle\langle 1\vert \big),
\label{e58}
\end{align}
then Eqs. (\ref{e44}), (\ref{e52}), and (\ref{e53}) would be, respectively, replaced with
\begin{gather}
{\bm{V}}_{I}(t) =\mu \cos(\Omega t)\left( e^{it\omega_{1b}}\vert 1\rangle\langle b\vert +e^{-it\omega_{1b}}\vert b\rangle\langle 1\vert \right) ,
\label{e59}\\
\bm{\Lambda} (t) =x(t)\vert b\rangle\langle 1\vert +h(t)\vert b\rangle\langle b\vert -h(t)\vert 1\rangle\langle 1\vert +y(t)\vert 1\rangle\langle b\vert ,
\label{e60}\\
\mathpzc{E}(t) =v_{bb}(t)\vert b\rangle\langle b\vert + v_{b1}(t)\vert b\rangle\langle 1\vert +v_{1b}(t)\vert 1\rangle\langle b\vert +v_{11}(t)\vert 1\rangle\langle 1\vert -it\omega_{0}\vert 0\rangle\langle 0\vert -it \textstyle{\sum_{j=2}^{n-1}} \omega_{j}\vert j\rangle\langle j\vert, 
\label{e61}
\end{gather}
where the time-dependent functions $v_{bb}(t)$, $v_{b1}(t)$, $v_{1b}(t)$, and $v_{11}(t)$ have the same definitions as $u_{00}(t)$, $u_{0b}(t)$, $u_{b0}(t)$, and $u_{bb}(t)$, respectively, after the replacements $0\rightarrow b$ and $b\rightarrow 1$. Thus, the Floquet Hamiltonian $\bar{H}$, the periodic operator $P(t,0)$, and the Lindblad operators would similarly be determined through Eqs. \eqref{e55}, \eqref{e57}, and \eqref{e14}. 
\end{remark}

\subsection{Benchmarking the results}

As in Sec. \ref{sec:F-FL-results}, here we assume that $H_{0}$ is given by $(\epsilon_{0},\epsilon_{b},\epsilon_{1})= (0,2.5,3)$ and $\{|0\rangle=(1,0,0)^{T}, |1\rangle=(0,1,0)^{T}, |b\rangle=(0,0,1)^{T}\}$. The system is assumed to be initially in the state $\varrho_{\mathsf{S}} (0)=\vert 0\rangle_{\mathsf{S}}\langle 0\vert$, and is coupled to two thermal baths with $\beta_{\mathsf{c}} / \beta_{\mathsf{h}}=30/4$. We also assume $(\mu,\Omega=2\pi/\tau)=(0.1,0.25$). 

In order to obtain the Floquet Hamiltonian we first need to employ the Magnus + BCH expansions to obtain the operator $U_{\mathsf{S}}(t,0)$ as an exponential in the form of Eq. (\ref{expon}). Since these expansions are infinite series, in practice we need to truncate them, which in turn renders the obtained $U_{\mathsf{S}}(t,0)$, $\bar{H}$, and $P(t,0)$ to be approximate. As a result, one should investigate the validity of these approximations. To benchmark the approximate operators, here we have also calculated $U_{\mathsf{S}}(t,0)$ numerically through the $4$th order Runge-Kutta method \cite{lambert1991numerical}. Our figures-of-merit are the following fidelities \cite{PhysRevLett.87.177901}:  

\begin{table}
\caption{Parameters of the master equation \eqref{e29}. Here we consider two cases for the external field: (i) the field couples $\vert 0\rangle$ and $\vert b\rangle$ levels [Eq. (\ref{e39})], and (ii) the field couples $\vert 1\rangle$ and $\vert b\rangle$ states [Eq. (\ref{e58})]. All matrix representations are in the $\{|0\rangle, |1\rangle,|b\rangle\}$ eigenbasis of $H_{0}$ (as introduced in Sec. \ref{sec:F-FL-results}), and all quantities are in natural units.}
\label{tab-3}
\begingroup
\squeezetable{
\begin{ruledtabular}
\begin{tabular}{l|ccc} 
& $V_{0}$: \text{field couples $\vert 0\rangle$ and $\vert b\rangle$} & $V_{1}$: \text{field couples $\vert 1\rangle$ and $\vert b\rangle$} \\
\hline
$\bar{H}$ & $\left( \begin{matrix}
0.0317 & 0 & -0.2939+0.3347 i \\
0 & 3 & 0 \\
-0.2939-0.3347 i & 0 & 2.4683 
\end{matrix} \right)$ & $\left( \begin{matrix}
0 & 0 & 0 \\
0 & 2.9991 & 0.0250 \\
0 & 0.0250 & 2.5009 
\end{matrix} \right) $ \\
$\bar{\epsilon}$ & $\left\lbrace 3, 2.5472, -0.0472\right\rbrace $ & $\left\lbrace 3.0003, 2,4997, 0\right\rbrace $ \\
$\omega$ & $\left\lbrace \pm 0.4528, \pm 3.0472,\pm 2.5943, 0\right\rbrace $ & $\left\lbrace \pm 0.5006, \pm 3.0003,\pm 2.4997, 0\right\rbrace $ \\
$q$ & $\left[ -24,24\right] $ & $ \left[ -3,3 \right] $ \\
$H_{\mathrm{lamb}}^{(\mathrm{F}),(\mathrm{h})}$ & $\left( \begin{matrix}
-0.0145 & 0 & -0.0016+0.0018 i \\
0 & 0.0166 & 0 \\
-0.0016-0.0018 i & 0 & -0.0015 
\end{matrix} \right)$ & $\left( \begin{matrix}
-0.0190 & 0 & 0 \\
0 & 0.0162 & 0.0006 \\
0 & 0.0006 & 0.0035 
\end{matrix} \right) $ \\
$H_{\mathrm{lamb}}^{(\mathrm{F}) ,(\mathrm{c})}$ & $\left( \begin{matrix}
-0.0073 & 0 & 0.0243-0.0277 i \\
0 & 0.2392 & 0 \\
0.0243+0.0277 i & 0 & -0.2088 
\end{matrix} \right) $ & $\left( \begin{matrix}
0 & 0 & 0 \\
0 & 0.1912 & 0.0182 \\
0 & 0.0182 & -0.1711 
\end{matrix} \right) $ \\
\end{tabular}
\end{ruledtabular}
}
\endgroup
\end{table}
\begin{gather}
\mathpzc{F}\left[ U^{\mathrm{app}}_{\mathsf{S}}(t,0),U^{\mathrm{ex}}_{\mathsf{S}}(t,0)\right]=\dfrac{1}{3}\big| \mathrm{Tr}\big[ U^{\mathrm{app}}_{\mathsf{S}}(t,0)U_{\mathsf{S}}^{\mathrm{ex}\,\dagger}(t,0)\big] \big| ,
\label{e101}\\
\mathpzc{F}\left[ P^{\mathrm{app}}(t,0), P^{\mathrm{app}\,\dag}(t+\tau,\tau)\right]=\dfrac{1}{3}\big| \mathrm{Tr}\big[ P^{\mathrm{app}}(t,0), P^{\mathrm{app}\, \dag}(t+\tau,\tau) \big] \big|,
\label{e102}
\end{gather}
where $0\leqslant t\leqslant \tau$.
The first quantity measures the fidelity between the approximate and almost exact solutions for $U_{\mathsf{S}}(t,0)$. The second quantity measures periodicity of the approximate $P(t,0)$ [obtained according to Eq. \eqref{e57} where we have used only the Magnus expansion in order to calculate $U_{\mathsf{S}}(t,0)$], as expected in $P (t+n\tau,n\tau)=P (t+n\tau,0)=P(t,0)$ and the discussion after Eq. (\ref{e11}), for two periods. Figure \ref{fig8} a shows that the fidelity of the approximate and almost-exact unitaries $U_{\mathsf{S}}(t,0)$ is more than $0.97$ in $0\leqslant t\leqslant \tau$. Figure \ref{fig8} b demonstrates that the fidelity between the periodic unitary operators $P(t,0)$ in two periods is more than $0.96$. We remark that in numerical calculations of $S_{\alpha}(q,\omega)$, we have offset all values less than $10^{-3}$ to $0$. With these favorable results, we obtain the other relevant quantities of the model (the Floquet Hamiltonian $\bar{H}$, the Lamb-shift Hamiltonian $H^{(\mathrm{F})}_{\mathrm{lamb}}$, and the range of $q$s and $\omega$s) as in Table \ref{tab-3}. These results have then been used to obtain the results discussed in Sec. \ref{sec:F-FL-results}.

\section{Redfield master equation}
\label{sec:redfield}

In this section, we explain how the Redfield master equation for a multilevel quantum system can be obtained, and in particular we focus on the Jaynes-Cummings model discuss in this paper. 

\subsection{Time-independent system Hamiltonian: Redfield master equation}
\label{sec:Tindependent}

Consider a quantum system which is coupled to a radiation bath. The total system Hamiltonian is described by Eq. \eqref{e79}, where we consider the  time-independent case with $H_{\mathsf{S}}=H_{0}$, and the bath is comprised of bosons with wave vectors $\vec{k}$, frequencies $\nu_{k}$, and polarization $\lambda$ [see Eq. (\ref{e81})]. The interaction Hamiltonian is described as Eq. (\ref{e82}), where the couplings between the system and the bath are given by
\begin{align}
g_{k\lambda}^{(ij)} = -E_{k} \vec{\mu}_{ij}\cdot\hat{e}_{k\lambda} ,
\label{e83}
\end{align}
where $\hat{e}_{k\lambda}$ denotes the unit vector of $\lambda$ polarization for mode $k$, and $\vec{\mu}_{ij}=e\langle i\vert \vec{r}\vert j\rangle$ is the electric-dipole transition matrix elements, with $e$ being the electric charge \cite{Scully-book}. Here the quantity $ E_{k}$ has the dimension of an electric field and is defined as $E_{k}=\sqrt{\hbar \nu_{k}/(2\epsilon_{0} V)}$, where $V$ is the quantization volume and $\epsilon_{0}$ is the vacuum permittivity. We stress that in the derivation of the master equation we use Eq. \eqref{e73} for the integral $\textstyle{\int_{0}^{\infty}} ds\,e^{-i(x+\nu)s}$, including both delta function and the Cauchy principal value terms, contributing to the decoherence operators and the Lamb-shift Hamiltonian, respectively. We remind the following relation for transforming the interaction picture back to the Schr{\"o}dinger picture:
\begin{gather}
\bm{\varrho}_{\mathsf{S}}(t) =U_{\mathsf{S}}^{\dagger}(t,0)\varrho_{\mathsf{S}}(t)U_{\mathsf{S}}(t,0) ,
\label{e95} \\
\dfrac{d\varrho_{\mathsf{S}}(t)}{dt} =-\dfrac{i}{\hbar}\left[ H_{\mathsf{S}}(t),\varrho_{\mathsf{S}}(t)\right] +U_{\mathsf{S}}(t,0)\big[ \dfrac{d\bm{\varrho}_{\mathsf{S}}(t)}{dt} \big] U_{\mathsf{S}}^{\dagger}(t,0) ,
\label{e96}
\end{gather}
where $U_{\mathsf{S}}(t,0)$ is the unitary operator generated by $H_{\mathsf{S}}(t)=H_{0}$. After doing some algebra, we obtain the following Redfield master equation in the Schr{\"o}dinger picture:
\begin{align}
&\dfrac{d\varrho_{\mathsf{S}}(t)}{dt}= -\dfrac{i}{\hbar}\left[ H_{\mathsf{S}},\varrho_{\mathsf{S}}(t)\right] - \dfrac{1}{6\pi c^{3}\hbar \epsilon_{0}}\textstyle{\sum_{i}} \sum_{i'j'} \omega_{i'j'}^{3} \times \nonumber\\
&\times \big[ (\vec{\mu}_{i' i}\cdot\vec{\mu}_{i'j'})\bar{n}(\omega_{i'j'} ,\beta)\left( \vert i\rangle \langle j'\vert \varrho_{\mathsf{S}}(t)+\varrho_{\mathsf{S}}(t)\vert j'\rangle \langle i\vert \right)+(\vec{\mu}_{i j'}\cdot\vec{\mu}_{i'j'})[\bar{n}(\omega_{i'j'},\beta)+1]\left( \vert i\rangle \langle i'\vert \varrho_{\mathsf{S}}(t)+\varrho_{\mathsf{S}}(t)\vert i'\rangle \langle i\vert \right) \big] \nonumber\\
&+\dfrac{1}{6\pi c^{3}\hbar \epsilon_{0}} \textstyle{\sum_{ij}} \sum_{i'j'}\vec{\mu}_{i j}\cdot\vec{\mu}_{i'j'} \omega_{i'j'}^{3} \times \nonumber\\
& \times \big[\bar{n}(\omega_{i'j'},\beta) \left[ \vert i\rangle \langle j\vert \varrho_{\mathsf{S}}(t)\vert j'\rangle \langle i'\vert +\vert i'\rangle \langle j'\vert \varrho_{\mathsf{S}}(t)\vert j\rangle \langle i\vert \right]+[\bar{n}(\omega_{i'j'},\beta)+1] \left[ \vert j\rangle \langle i\vert \varrho_{\mathsf{S}}(t)\vert i'\rangle \langle j'\vert +\vert j'\rangle \langle i'\vert \varrho_{\mathsf{S}}(t)\vert i\rangle \langle j\vert \right] \big] \nonumber\\
&-\dfrac{i}{\hbar}\sum _{ij}\sum _{i'}\dfrac{\vec{\mu}_{ij}\cdot\vec{\mu}_{i'j}}{6\epsilon _{0} \pi ^{2} c^{3}} \big[ \big( \vert i\rangle \langle i'\vert \varrho _{\mathsf{S}}(t) -\varrho _{\mathsf{S}}(t)\vert i'\rangle \langle i\vert \big) \mathbbmss{P} \int _{0} ^{\infty} d\nu\, \dfrac{\nu ^{3}(\bar{n}(\nu ,\beta)+1)}{\omega _{i'j}-\nu} \big] \nonumber\\
&-\dfrac{i}{\hbar}\sum _{ij}\sum _{j'}\dfrac{\vec{\mu}_{ij}\cdot\vec{\mu}_{ij'}}{6\epsilon _{0} \pi ^{2} c^{3}} \big[ \big( \varrho _{\mathsf{S}}(t)\vert j'\rangle \langle j\vert - \vert j\rangle \langle j'\vert \varrho _{\mathsf{S}}(t)  \big) \mathbbmss{P} \int _{0} ^{\infty} d\nu\, \dfrac{\nu ^{3}\bar{n}(\nu ,\beta)}{\omega _{ij'}-\nu} \big] \nonumber\\
&+\dfrac{i}{\hbar}\sum _{ij}\sum _{i'j'}\dfrac{\vec{\mu}_{ij}\cdot\vec{\mu}_{i'j'}}{6\epsilon _{0} \pi ^{2} c^{3}} \big[ \big( \vert i\rangle \langle j\vert \varrho _{\mathsf{S}}(t)\vert j'\rangle \langle i'\vert - \vert i'\rangle \langle j'\vert \varrho _{\mathsf{S}}(t) \vert j\rangle \langle i\vert  \big) \mathbbmss{P} \int _{0} ^{\infty} d\nu\, \dfrac{\nu ^{3}\bar{n}(\nu ,\beta)}{\omega _{i'j'}-\nu} \big] \nonumber\\
&+\dfrac{i}{\hbar}\sum _{ij}\sum _{i'j'}\dfrac{\vec{\mu}_{ij}\cdot\vec{\mu}_{i'j'}}{6\epsilon _{0} \pi ^{2} c^{3}} \big[ \big( \vert j'\rangle \langle i'\vert \varrho _{\mathsf{S}}(t)\vert i\rangle \langle j\vert - \vert j\rangle \langle i\vert \varrho _{\mathsf{S}}(t) \vert i'\rangle \langle j'\vert  \big) \mathbbmss{P} \int _{0} ^{\infty} d\nu\, \dfrac{\nu ^{3}(\bar{n}(\nu ,\beta)+1)}{\omega _{i'j'}-\nu} \big] .
\label{eRd-full}
\end{align}
We have used this equation in Sec. \ref{sec:Rscenario}.

\subsection{Time-periodic system Hamiltonian: Floquet-Redfield master equation}
\label{sec:floquet}

Now we consider a quantum system in contact with a bath and also driven by an external periodic field. The model we use is similar to the one discussed in the precious section. To derive the master equation for such a system system, we need to combine both Redfield and Floquet approaches. According to Appendix \ref{app:floquet-lindblad}, to derive the master equation we need $\bm{H}_{\mathsf{int}}(t)$. By using Eqs. \eqref{e7} and \eqref{e11} we obtain
\begin{align}
\bm{H}_{\mathsf{int}}(t)&=- \textstyle{\sum_{i,j}} \sum_{k,\lambda} g_{k\lambda}^{(ij)} e^{i\bar{H}t/\hbar} P ^{\dagger}(t,0)\sigma_{ij}P(t,0)e^{-i\bar{H}t/\hbar}\otimes U_{\mathsf{B}}^{\dagger}(t,0)\hat{a}_{k,\lambda}U_{\mathsf{B}}(t,0)+ \mathrm{h.c.}, \nonumber\\
&= - \textstyle{\sum_{i,j}} \sum_{k,\lambda} \sum_{q}\sum_{\left\lbrace \omega\right\rbrace } g_{k\lambda}^{(ij)} e^{it(\omega +q\Omega -\nu_{k})}\sigma_{ij}(q,\omega)\otimes\hat{a}_{k,\lambda}+ \mathrm{h.c.},
\label{e86}
\end{align}
where 
\begin{gather}
\sigma_{ij}\equiv \vert i\rangle\langle j\vert,\\
U_{\mathsf{B}}^{\dagger}(t,0)\hat{a}_{k,\lambda}U_{\mathsf{B}}(t,0)=e^{-i\nu_{k}t}\hat{a}_{k,\lambda},\\
P ^{\dagger}(t,0)\sigma_{ij}P(t,0)= \textstyle{\sum_{q\in\mathbbmss{Z}}} \sigma_{ij}(q)e^{iq\Omega t} , \qquad \sigma_{ij}^{\dagger}(q)=\sigma_{ij}(-q) \\
\sigma_{ij}(q,\omega)= \textstyle{\sum_{\bar{\epsilon}-\bar{\epsilon}'=\hbar \omega}} \vert\bar{\epsilon}\rangle\langle\bar{\epsilon} \vert \sigma_{ij}(q)\vert\bar{\epsilon}'\rangle\langle\bar{\epsilon}'\vert , \qquad \sigma_{ij}^{\dagger}(q,\omega)=\sigma_{ij}(-q,-\omega) \\
\sigma_{ij}(t)= \textstyle{\sum_{q}} \sum_{\omega} e^{it(\omega +q\Omega)}\sigma_{ij}(q,\omega) .
\label{e89}
\end{gather}

By inserting $\bm{H}_{\mathsf{int}}(t)$ [Eq. \eqref{e86}] in Eq. \eqref{e9} and after some algebra, we obtain the master equation of the system in the interaction picture,
\begin{align}
\dfrac{d\bm{\varrho}_{\mathsf{S}} (t)}{dt}=&-\dfrac{1}{6\pi c^{3}\hbar \epsilon_{0}} \textstyle{\sum_{i,j}} \sum_{i',j'}\sum_{q,\omega}\sum_{q',\omega'}(\vec{\mu}_{ij} \cdot \vec{\mu}_{i'j'}) \times \nonumber\\
& \Big( e^{it(\omega -\omega'+(q-q')\Omega)} \big[(\omega'+q'\Omega)^{3} [\bar{n}(\omega'+q'\Omega ,\beta)+1]+\dfrac{i}{\pi}\mathbbmss{P} \int _{0} ^{\infty} d\nu\, \dfrac{\nu ^{3}(\bar{n}(\nu ,\beta)+1)}{\omega'+q'\Omega-\nu} \big] \sigma_{ij}(q,\omega)\sigma_{i'j'}^{\dagger}(q',\omega')\bm{\varrho}_{\mathsf{S}}(t) \nonumber\\
&+e^{-it(\omega -\omega'+(q-q')\Omega)} \big[(\omega'+q'\Omega)^{3} \bar{n}(\omega'+q'\Omega ,\beta)-\dfrac{i}{\pi}\mathbbmss{P} \int _{0} ^{\infty} d\nu\, \dfrac{\nu ^{3} \bar{n}(\nu ,\beta)}{\omega'+q'\Omega-\nu} \big] \sigma_{ij}^{\dagger}(q,\omega)\sigma_{i'j'}(q',\omega')\bm{\varrho}_{\mathsf{S}}(t) \nonumber\\
&- e^{it(\omega -\omega'+(q-q')\Omega)} \big[(\omega'+q'\Omega)^{3} \bar{n}(\omega'+q'\Omega ,\beta)+\dfrac{i}{\pi}\mathbbmss{P} \int _{0} ^{\infty} d\nu\, \dfrac{\nu ^{3} \bar{n}(\nu ,\beta)}{\omega'+q'\Omega-\nu} \big] \sigma_{ij}(q,\omega)\bm{\varrho}_{\mathsf{S}}(t)\sigma_{i'j'}^{\dagger}(q',\omega') \nonumber\\
&- e^{-it(\omega -\omega'+(q-q')\Omega)} \big[(\omega'+q'\Omega)^{3} [\bar{n}(\omega'+q'\Omega ,\beta)+1]-\dfrac{i}{\pi}\mathbbmss{P} \int _{0} ^{\infty} d\nu\, \dfrac{\nu ^{3}(\bar{n}(\nu ,\beta)+1)}{\omega'+q'\Omega-\nu}\big] \sigma_{ij}^{\dagger}(q,\omega)\bm{\varrho}_{\mathsf{S}}(t)\sigma_{i'j'}(q',\omega') \nonumber\\
&- e^{-it(\omega -\omega'+(q-q')\Omega)} \big[(\omega'+q'\Omega)^{3} \bar{n}(\omega'+q'\Omega ,\beta)-\dfrac{i}{\pi}\mathbbmss{P} \int _{0} ^{\infty} d\nu\, \dfrac{\nu ^{3} \bar{n}(\nu ,\beta)}{\omega'+q'\Omega-\nu} \big] \sigma_{i'j'}(q',\omega')\bm{\varrho}_{\mathsf{S}}(t)\sigma_{ij}^{\dagger}(q,\omega) \nonumber\\
&-e^{it(\omega -\omega'+(q-q')\Omega)}  \big[(\omega'+q'\Omega)^{3} [\bar{n}(\omega'+q'\Omega ,\beta)+1]+\dfrac{i}{\pi}\mathbbmss{P} \int _{0} ^{\infty} d\nu\, \dfrac{\nu ^{3}(\bar{n}(\nu ,\beta)+1)}{\omega'+q'\Omega-\nu} \big] \sigma_{i'j'}^{\dagger}(q',\omega')\bm{\varrho}_{\mathsf{S}}(t)\sigma_{ij}(q,\omega) \nonumber\\
&+e^{-it(\omega -\omega'+(q-q')\Omega)}  \big[(\omega'+q'\Omega)^{3} [\bar{n}(\omega'+q'\Omega ,\beta)+1]-\dfrac{i}{\pi}\mathbbmss{P} \int _{0} ^{\infty} d\nu\, \dfrac{\nu ^{3}(\bar{n}(\nu ,\beta)+1)}{\omega'+q'\Omega-\nu} \big] \bm{\varrho}_{\mathsf{S}}(t)\sigma_{i'j'}(q',\omega')\sigma_{ij}^{\dagger}(q,\omega) \nonumber\\
&+e^{it(\omega -\omega'+(q-q')\Omega)}  \big[(\omega'+q'\Omega)^{3} \bar{n}(\omega'+q'\Omega ,\beta)+\dfrac{i}{\pi}\mathbbmss{P} \int _{0} ^{\infty} d\nu\, \dfrac{\nu ^{3} \bar{n}(\nu ,\beta)}{\omega'+q'\Omega-\nu} \big] \bm{\varrho}_{\mathsf{S}}(t)\sigma_{i'j'}^{\dagger}(q',\omega')\sigma_{ij}(q,\omega) \Big) .
\label{e94}
\end{align}
 
To transform back into the Schr{\"o}dinger picture, we apply Eq. \eqref{e95}, where now $U_{\mathsf{S}}(t,0)$ is the unitary operator given by Eq. \eqref{e11}. Substituting Eq. \eqref{e94} on the RHS of Eq. \eqref{e96} then yields 
\begin{align}
&\dfrac{d\varrho_{\mathsf{S}} (t)}{dt} =-\dfrac{i}{\hbar}\left[ H_{\mathsf{S}}(t),\varrho_{\mathsf{S}}(t)\right] - \dfrac{1}{6\pi c^{3}\hbar \epsilon_{0}}\textstyle{\sum_{ij}} \sum_{i'j'}\sum_{q\omega}\sum_{q'\omega'}(\vec{\mu}_{ij} \cdot \vec{\mu}_{i'j'}) \nonumber\\
& \times\Big[ e^{it(q-q')\Omega} \big[ (\omega'+q'\Omega)^{3} [\bar{n}(\omega'+q'\Omega ,\beta)+1]+\dfrac{i}{\pi}\mathbbmss{P} \int _{0} ^{\infty} d\nu\, \dfrac{\nu ^{3}(\bar{n}(\nu ,\beta)+1)}{\omega'+q'\Omega-\nu} \big] P(t,0)\sigma_{ij}(q,\omega)\sigma_{i'j'}^{\dagger}(q',\omega') P ^{\dagger}(t,0)\varrho_{\mathsf{S}}(t) \nonumber\\
&+e^{-it(q-q')\Omega} \big[ (\omega'+q'\Omega)^{3} \bar{n}(\omega'+q'\Omega ,\beta)-\dfrac{i}{\pi}\mathbbmss{P} \int _{0} ^{\infty} d\nu\, \dfrac{\nu ^{3}\bar{n}(\nu ,\beta)}{\omega'+q'\Omega-\nu} \big] P(t,0)\sigma_{ij}^{\dagger}(q,\omega)\sigma_{i'j'}(q',\omega') P ^{\dagger}(t,0)\varrho_{\mathsf{S}}(t) \nonumber\\
&- e^{it(q-q')\Omega} \big[ (\omega'+q'\Omega)^{3} \bar{n}(\omega'+q'\Omega ,\beta)+\dfrac{i}{\pi}\mathbbmss{P} \int _{0} ^{\infty} d\nu\, \dfrac{\nu ^{3}\bar{n}(\nu ,\beta)}{\omega'+q'\Omega-\nu} \big] P(t,0)\sigma_{ij}(q,\omega) P ^{\dagger}(t,0)\varrho_{\mathsf{S}}(t)P(t,0)\sigma_{i'j'}^{\dagger}(q',\omega')  P ^{\dagger}(t,0)\nonumber\\
&- e^{-it(q-q')\Omega} \big[ (\omega'+q'\Omega)^{3} [\bar{n}(\omega'+q'\Omega ,\beta)+1]-\dfrac{i}{\pi}\mathbbmss{P} \int _{0} ^{\infty} d\nu\, \dfrac{\nu ^{3}(\bar{n}(\nu ,\beta)+1)}{\omega'+q'\Omega-\nu} \big] P(t,0)\sigma_{ij}^{\dagger}(q,\omega) P ^{\dagger}(t,0)\varrho_{\mathsf{S}}(t)P(t,0)\sigma_{i'j'}(q',\omega') P ^{\dagger}(t,0)\nonumber\\
&- e^{-it(q-q')\Omega} \big[ (\omega'+q'\Omega)^{3} \bar{n}(\omega'+q'\Omega ,\beta)-\dfrac{i}{\pi}\mathbbmss{P} \int _{0} ^{\infty} d\nu\, \dfrac{\nu ^{3}\bar{n}(\nu ,\beta)}{\omega'+q'\Omega-\nu} \big] P(t,0)\sigma_{i'j'}(q',\omega') P ^{\dagger}(t,0)\varrho_{\mathsf{S}}(t)P(t,0)\sigma_{ij}^{\dagger}(q,\omega) P ^{\dagger}(t,0)\nonumber\\
&-e^{it(q-q')\Omega} \big[ (\omega'+q'\Omega)^{3} [\bar{n}(\omega'+q'\Omega ,\beta)+1]+\dfrac{i}{\pi}\mathbbmss{P} \int _{0} ^{\infty} d\nu\, \dfrac{\nu ^{3}(\bar{n}(\nu ,\beta)+1)}{\omega'+q'\Omega-\nu} \big] P(t,0)\sigma_{i'j'}^{\dagger}(q',\omega') P ^{\dagger}(t,0)\varrho_{\mathsf{S}}(t)P(t,0)\sigma_{ij}(q,\omega) P ^{\dagger}(t,0) \nonumber\\
&+e^{-it(q-q')\Omega} \big[ (\omega'+q'\Omega)^{3} [\bar{n}(\omega'+q'\Omega ,\beta)+1]-\dfrac{i}{\pi}\mathbbmss{P} \int _{0} ^{\infty} d\nu\, \dfrac{\nu ^{3}(\bar{n}(\nu ,\beta)+1)}{\omega'+q'\Omega-\nu} \big] \varrho_{\mathsf{S}}(t)P(t,0)\sigma_{i'j'}(q',\omega')\sigma_{ij}^{\dagger}(q,\omega) P ^{\dagger}(t,0)\nonumber\\
&+e^{it(q-q')\Omega} \big[ (\omega'+q'\Omega)^{3} \bar{n}(\omega'+q'\Omega ,\beta)+\dfrac{i}{\pi}\mathbbmss{P} \int _{0} ^{\infty} d\nu\, \dfrac{\nu ^{3}\bar{n}(\nu ,\beta)}{\omega'+q'\Omega-\nu} \big] \varrho_{\mathsf{S}}(t)P(t,0)\sigma_{i'j'}^{\dagger}(q',\omega')\sigma_{ij}(q,\omega) P ^{\dagger}(t,0) \Big] .
\label{e97}
\end{align}

The above master equation contains some fast oscillating terms which may lead to nonpositivity of the density matrices. To alleviate this problem, we remove some fast oscillating terms by applying a \textit{partial secular approximation} which is given by the assumption $\Omega^{-1}\ll \mathpzc{T}_{R}$, where $\mathpzc{T}_{R}$ is the relaxation time of the driven system. This approximation neglects all terms where $q'\neq q$. Note that this approximation is different from that of Ref. \cite{tscherbul_partial_2015}. As a result, we obtain the following modified Floquet-Redfield master equation:
\begin{align}
\dfrac{d\varrho_{\mathsf{S}} (t)}{dt}=& -\dfrac{i}{\hbar}\left[ H_{\mathsf{S}}(t),\varrho_{\mathsf{S}}(t)\right] -\dfrac{1}{6\pi c^{3}\hbar \epsilon_{0}}\sum_{ij}\sum_{i'j'}\sum_{q}\sum_{\omega \omega'}(\vec{\mu}_{ij} \cdot \vec{\mu}_{i'j'}) \times \nonumber\\
& \Big[ L_{ij}(q,\omega ;t)L^{\dagger}_{i'j'}(q,\omega';t)\varrho_{\mathsf{S}}(t) \big[ (\omega'+q\Omega)^{3} [\bar{n}(\omega'+q\Omega ,\beta)+1]+\dfrac{i}{\pi}\mathbbmss{P} \int _{0} ^{\infty} d\nu\, \dfrac{\nu ^{3}(\bar{n}(\nu ,\beta)+1)}{\omega'+q\Omega-\nu}\big]   \nonumber\\
&+ L^{\dagger}_{ij}(q,\omega ;t)L_{i'j'}(q,\omega';t)\varrho_{\mathsf{S}}(t) \big[(\omega'+q\Omega)^{3} \bar{n}(\omega'+q \Omega ,\beta)-\dfrac{i}{\pi}\mathbbmss{P} \int _{0} ^{\infty} d\nu\, \dfrac{\nu ^{3}\bar{n}(\nu ,\beta)}{\omega'+q\Omega-\nu}   \big]     \nonumber\\
&- L_{ij}(q,\omega ;t)\varrho_{\mathsf{S}}(t)L^{\dagger}_{i'j'}(q,\omega';t) \big[(\omega'+q\Omega)^{3} \bar{n}(\omega'+q \Omega ,\beta)+\dfrac{i}{\pi}\mathbbmss{P} \int _{0} ^{\infty} d\nu\, \dfrac{\nu ^{3}\bar{n}(\nu ,\beta)}{\omega'+q\Omega-\nu}  \big]   \nonumber\\
&- L^{\dagger}_{ij}(q,\omega ;t)\varrho_{\mathsf{S}}(t)L_{i'j'}(q,\omega';t) \big[ (\omega'+q\Omega)^{3} [\bar{n}(\omega'+q \Omega ,\beta)+1]-\dfrac{i}{\pi}\mathbbmss{P} \int _{0} ^{\infty} d\nu\, \dfrac{\nu ^{3}(\bar{n}(\nu ,\beta)+1)}{\omega'+q\Omega-\nu} \big]     \nonumber\\
&- L_{i'j'}(q,\omega';t)\varrho_{\mathsf{S}}(t)L^{\dagger}_{ij}(q,\omega ;t) \big[ (\omega'+q\Omega)^{3} \bar{n}(\omega'+q \Omega ,\beta)-\dfrac{i}{\pi} \mathbbmss{P} \int _{0} ^{\infty} d\nu\, \dfrac{\nu ^{3}\bar{n}(\nu ,\beta)}{\omega'+q\Omega-\nu} \big]    \nonumber\\
&- L^{\dagger}_{i'j'}(q,\omega';t)\varrho_{\mathsf{S}}(t)L_{ij}(q,\omega ;t) \big[ (\omega'+q\Omega)^{3}  [\bar{n}(\omega'+q \Omega ,\beta)+1]+\dfrac{i}{\pi} \mathbbmss{P} \int _{0} ^{\infty} d\nu\, \dfrac{\nu ^{3}[\bar{n}(\nu ,\beta)+1]}{\omega'+q\Omega-\nu} \big]      \nonumber\\
&+ \varrho_{\mathsf{S}}(t)L_{i'j'}(q,\omega';t)L^{\dagger}_{ij}(q,\omega ;t) \big[ (\omega'+q\Omega)^{3}  [\bar{n}(\omega'+q \Omega ,\beta)+1]-\dfrac{i}{\pi} \mathbbmss{P} \int _{0} ^{\infty} d\nu\, \dfrac{\nu ^{3}[\bar{n}(\nu ,\beta)+1]}{\omega'+q\Omega-\nu} \big]       \nonumber\\
&+ \varrho_{\mathsf{S}}(t)L^{\dagger}_{i'j'}(q,\omega' ;t)L_{ij}(q,\omega ;t) \big[ (\omega'+q\Omega)^{3} \bar{n}(\omega'+q\Omega ,\beta)+\dfrac{i}{\pi} \mathbbmss{P} \int _{0} ^{\infty} d\nu\, \dfrac{\nu ^{3}\bar{n}(\nu ,\beta)}{\omega'+q\Omega-\nu} \big]      \Big],
\label{e98-}
\end{align}
where
\begin{gather}
L_{ij}(q,\omega ;t)= P(t,0)\sigma_{ij}(q,\omega) P ^{\dagger}(t,0) .
\end{gather}
We have used this equation in Sec. \ref{sec:FRscenario}.

\section{Computing the Lamb-shift correction in the Redfield equation}
\label{sec:Lamb-app}

In both Eqs. (\ref{eRd-full}) and (\ref{e98-}) we have included the Lamb-shift terms. These terms may give rise to divergence. It has been known that by by introducing a finite cutoff frequency $W<\infty$ one can make these corrections finite. In fact, the physical reason for the introduction of the cutoff frequency is related to the long-wavelength approximation ($\omega|\vec{r}|/c\ll1$, where $|\vec{r}|$ is of the order of the atom size) \cite{book:Cohen-Tannoudji-QED}. After introducing $W$ we need the following integral for the computation of the Lamb-shift corrections:
\begin{align}
\mathbbmss{P} \int_{0}^{W} d\nu\, \dfrac{\nu ^{3}[\bar{n}(\nu ,\beta)+1]}{\omega -\nu}=\mathbbmss{P} \int_{0}^{W} d\nu\, \dfrac{\nu ^{3} \bar{n}(\nu ,\beta)}{\omega -\nu}+\mathbbmss{P} \int_{0}^{W} d\nu\, \dfrac{\nu ^{3}}{\omega -\nu}.
\label{divInt_ap}
\end{align}
Note that the second term of the RHS can also be recast as
\begin{align}
\mathbbmss{P} \int_{0}^{W} d\nu\, \dfrac{\nu ^{3}}{\omega -\nu}=-\int_{0}^{W} d\nu\, \nu ^{2} -\omega \int_{0}^{W} d\nu\,  \nu - \omega ^{2}\int_{0}^{W} d\nu + \mathbbmss{P} \int_{0}^{W} d\nu \dfrac{\omega ^{3}}{\omega -\nu} .
\label{divInt1_app}
\end{align}

A careful quantum-electrodynamical (QED) investigation reveals that the first term of Eq. \eqref{divInt1_app} is fully compensated and cancelled out by other terms emerged from the QED treatment \cite{book:Cohen-Tannoudji}. As in Appendix \ref{sec:lm}, consider a charged particle (an atom or a molecule) coupled to a radiation field. Based on Eq. (\ref{q8}) in the ``$\vec{E} \cdot \vec{r}$'' representation, we have an extra term $\varepsilon_{\mathrm{dipole}}$ corresponding to a dipole self-energy of the particle. In particular, in the Jaynes–Cummings model by considering a cutoff frequency this dipole self-energy term is simply a constant energy value and thus it is usually discarded in the literature \cite{Rokaj_2018}. However, interestingly, it has been known that this term contributes an extra term in the Lamb-shift correction which is the same as the first term of the RHS of Eq. \eqref{divInt1_app}, now with the opposite sign. That is, the dipole self-energy contribution in the energy shift completely cancels out the effect of the first term of the RHS of Eq. \eqref{divInt1_app}. Furthermore, as argued in Remark \ref{remark-} of Appendix \ref{sec:lm}, in the low-intensity regime, the appearance of $-H_{I2}$ term in the Hamiltonian contributes an opposite term in the Lamb-shift correction, which cancels out the effect of the the second term of the RHS of Eq. (\ref{divInt1_app}). 

\begin{remark}
Note that for low-energy photons one can ignore the interaction between the spin magnetic moment of the charged particle and the magnetic field of the radiation field \cite{book:Cohen-Tannoudji-QED}.
\end{remark}

\twocolumngrid
\end{widetext}
%


\end{document}